\documentclass[letter,11pt]{article}
\pdfoutput=1
\usepackage{jheppub} 
\usepackage{etoolbox}
\makeatletter
\patchcmd{\maketitle}{\@fpheader}{}{}{}
\makeatother
\usepackage[T1]{fontenc}
\usepackage{bbm}
\usepackage{physics}
\usepackage{ulem}
\usepackage[dvipsnames]{xcolor} 
\usepackage{wrapfig}
\usepackage{amsmath}
\usepackage{algorithm}
\usepackage[noend]{algpseudocode}
\usepackage{mathrsfs}
\usepackage{dsfont}
\usepackage{mathtools}

\newcommand{\ads}{AdS$_3\,\,$}
\newcommand{\Z}{\mathbb{Z}}
\newcommand{\footref}[1]{Footnote~\ref{#1}}
\newcommand{\varthetaconst}{\vartheta \left[\begin{matrix} \mathbf{a}\\ \mathbf{b}\end{matrix}\right]}
\newcommand{\spin}[4]{\left[ \begin{matrix} #1 & #2 \\ #3 & #4 \end{matrix} \right]}
\newcommand{\abw}{\{a_i\}, \{b_i\}, \{w_i\}}
\newcommand{\Kabw}{ | \{a_i\}, \{b_i\}, \{w_i\} \rangle }
\newcommand{\Babw}{ \langle \{a_i\}, \{b_i\}, \{w_i\} | }
\newcommand{\Kabwp}{ | \{a_i'\}, \{b_i'\}, \{w_i'\} \rangle }

\newcommand{\vac}{ {\rm vac} }
\newcommand{\M}{\mathbb{M}}
\title{\boldmath Establishing strongly-coupled 3D AdS quantum gravity with Ising dual using all-genus partition functions }

\author[a]{Chao-Ming Jian,}
\author[b]{Andreas W. W. Ludwig,}
\author[a,c]{Zhu-Xi Luo,}
\author[a,d]{Hao-Yu Sun}
\author[e,f]{and Zhenghan Wang}

\affiliation[a]{Kavli Institute for Theoretical Physics, University of California, Santa Barbara, CA 93106-4030, USA}
\affiliation[b]{Department of Physics, University of California, Santa Barbara, CA 93106-9530, USA}
\affiliation[c]{Department of Physics and Astronomy, University of Utah, Salt Lake City, UT 84112-0830, USA}
\affiliation[d]{Center for Theoretical Physics and Department of Physics, University of California, Berkeley, CA 94720-7300, USA}
\affiliation[e]{Microsoft Station Q, Santa Barbara, CA 93106-6105, USA}
\affiliation[f]{Department of Mathematics, University of California, Santa Barbara,
CA 93106-3080, USA}

\abstract{We study 3D pure Einstein quantum gravity with negative cosmological constant, in the regime where the AdS radius $l$ is of the order of the Planck scale. Specifically, when the Brown-Henneaux central charge $c=3l/2G_N$ ($G_N$ is the 3D Newton constant) equals $c=1/2$, we 
establish duality  between 3D gravity  
and 2D Ising conformal field theory
by matching gravity and conformal field theory partition functions for AdS spacetimes with general asymptotic boundaries. This duality was suggested by a genus-one calculation of Castro et al. [Phys. Rev. D {\bf 85}, 024032 (2012)]. Extension beyond genus-one requires new mathematical results based on 3D Topological Quantum Field Theory; these turn out to uniquely select the $c=1/2$ theory among all those with $c<1$, extending the previous results of Castro et al.. \\ Previous work suggests the reduction of the calculation of the gravity partition function to a problem of summation over the orbits of the mapping class group action on a ``vacuum seed''. But whether or not the summation is well-defined for the general case was unknown before this work. Amongst all theories with Brown-Henneaux central charge $c<1$, the sum is finite and unique {\it only} when $c=1/2$, corresponding to a dual Ising conformal field theory on the asymptotic boundary.
}

\begin{document}
\maketitle
\flushbottom

\section{Introduction and Summary of Results}
\label{sec:intro}

A way of looking at 3D pure gravity with negative cosmological constant using partition functions and modular properties was initiated in work by Dijkgraaf et al. \cite{FareyTail}, Witten \cite{Witten}, and Maloney and Witten \cite{MaloneyWitten}. 
In a pioneering paper \cite{Castro}, Castro et al. argued that two well-known Conformal Field Theories (CFT) in two-dimensional space time, i.e., Ising and tricritical Ising minimal models \cite{BPZ,FQS}, are dual to pure Einstein quantum gravity in three-dimensional spacetime with negative cosmological constant, i.e., in Anti-de Sitter spacetime (AdS$_3$).\footnote{In the same paper similar arguments are also presented for certain versions of theories of higher spin quantum gravity. See also \footref{foot:higher}.}
These are theories of strongly-coupled gravity where the AdS radius $l$ is of the order of the Planck scale. The arguments provided by Castro et al. in support of these dualities at the corresponding values of the Brown-Henneaux \cite{BrownHenneaux} central charges $c=3l/2G_N$ ($G_N$ is the 3D Newton constant) consisted in demonstrating a match between the gravity partition function of Euclidean \ads spacetime when its asymptotic boundary is a 2D torus, with the torus partition function of the corresponding 2D minimal model CFT.

To be specific, one can think of the finite-temperature partition function of 
pure Einstein gravity in Euclidean \ads as being written as a path integral. The latter is 
formally a sum
over every smooth 3-manifold $X$ whose asymptotic boundary is a torus $T^2$,
\begin{equation}
\label{eq:integral}
Z_{\text{grav}}(\tau,\bar{\tau})=\int_{\partial X=T^2} \mathcal{D} g_{\mu\nu}~e^{-c \  S_E[g_{\mu\nu}]},
\end{equation}
with $\tau$ the conformal structure parameter of the boundary 
torus, the Brown-Henneaux
central charge $c=3l/2G_N$ playing the role of the inverse gravity coupling constant (large in the semi-classical regime). $S_E[g_{\mu\nu}]$ is the Einstein-Hilbert action with $g_{\mu\nu}$ the complete Riemannian metric tensor on $X$.
One will need to both sum over all different geometries
of the bulk 3-manifold $X$ with the same equivalence class of conformal structures (i.e., the same conformal class) on the boundary torus, as well as integrate over all different boundary metrics connected by small diffeomorphisms, i.e., those isotopic to the identity. The {\it full} gravitational path integral can then be written as
\begin{equation}
Z_{\text{grav}}=\sum_{X \ ({\rm where} \  \partial X = T^2)} Z(X,\tau),
\label{eq:sumM}
\end{equation}
where $Z(X,\tau)$ denotes the contribution from the sum over all metrics related by small diffeomorphisms on a particular 
3-manifold $X$ with a fixed
conformal structure parameter
$\tau$ on the asymptotic torus boundary, while the summation over $X$ means summing over different $\tau$'s in the same conformal class.

In the semi-classical (large $c$) limit, the smooth 3-manifolds $X$ contributing to the path integral turn out to be only those which admit classical solutions, i.e., which are saddle
points\footnote{One main conclusion 
of \cite{MaloneyWitten} is that in the weak-coupling/semiclassical regime, one has to include geometries corresponding to \textit{complex} saddle points of $S_E[g]$ in order to have a Hilbert space interpretation of the gravity theory. However, since here we are only concerned with the strongly coupled regime, we are not bound by these considerations.}
of the Einstein-Hilbert action $S_E[g]$, and only solid tori are commonly considered, see \cite{MaldacenaStrominger, FareyTail,MaloneyWitten}.
Following the logic pursued in previous work \cite{Castro,FareyTail,Witten} on this problem, the gravitational path integral can then be thought of as being organized as a sum over classical solutions, along with a {\it full} treatment of {\it all} quantum fluctuations around each saddle point.
For the case of solid tori $X$, different saddles correspond to inequivalent ways of filling in the bulk $X$ of the boundary torus $T^2$, and are related to each other by $SL(2,\mathbb{Z})$ modular transformations. The gravity partition function (\ref{eq:sumM}) can then be obtained as the sum of inequivalent images of a certain ``vacuum seed''
partition function
in Euclidean \ads under the action of $SL(2,\mathbb{Z})$.
Physically, this ``vacuum seed'' describes the gravitational partition function of thermal \ads where the spatial cycle of the boundary torus $T^2$ is contractible in the bulk, whereas the cycle of Euclidean time is not. This corresponds to a particular solid torus $X$, for example see Figure \ref{fig:ads}. As argued in \cite{Castro}, the gravitational ``vacuum seed''
partition function can be obtained exactly by using the remarkable and fundamental results of Brown and Henneaux \cite{BrownHenneaux}; this is reviewed in Section \ref{LabelSectionThevacuumseed} below, 
and the result summarized in the next paragraph.
After action on the ``vacuum seed'' gravitational partition function with a non-trivial modular transformation, the spatial cycle may no longer be contractible
in the bulk while the temporal cycle may now be; in that case the corresponding gravitational partition function describes physically that of a BTZ black hole \cite{BTZ}.
The sum over modular transformations in  $SL(2,\mathbb{Z})$ 
appearing in (\ref{eq:sumM}) can also be seen  to 
originate from general coordinate invariance, 
independent of invoking semi-classical notions such as saddle points,
because non-trivial modular transformations correspond to large diffeomorphisms,
not continuously connected to the identity; in the gravitational path
integral for the ``vacuum seed'' partition function, on the other hand, these large
diffeomorphisms are thought to be excluded. (This complementary point of view was also stressed in \cite{Castro}.)

As was argued in \cite{Castro}, with certain assumptions the gravitational ``vacuum  seed'' partition function turns out to be precisely
equal to the vacuum character of the dual CFT. 
Furthermore, owing to the fact that the vacuum character of rational CFTs is invariant under a certain finite index subgroup of the modular group
$SL(2,\mathbb{Z})$ \cite{Bantay, NgSch}, 
the modular sum
in (\ref{eq:sumM}) over the infinite group $SL(2,\mathbb{Z})$ of modular transformations reduces in fact to the sum over a finite number of right cosets of that finite index subgroup in $SL(2,\mathbb{Z})$ when the Brown-Henneaux central charge $c$ is equal to that of a unitary conformal minimal model CFT \cite{BPZ,FQS}.
For Brown-Henneaux central charge $c=1/2$, the resulting finite sum was shown in \cite{Castro} to be proportional to the partition function of the 2D Ising CFT on the torus $T^2$.

Based on the above analysis of solid tori $X$, Castro et al. argued in \cite{Castro} that amongst all \cite{BPZ,FQS} the unitary Virasoro minimal models with central charge $c<1$, only the Ising and tricritical Ising CFTs are dual to pure Einstein gravity at the corresponding values of the Brown-Henneaux central charge.\footnote{Some of the $\mathcal{W}_N$ minimal models,
were also conjectured in \cite{Castro} to be possibly dual to higher-spin gravity theories instead of being dual to pure Einstein gravity. This is due to the existence of an extended chiral conformal algebra, generated by conserved currents possessing (conformal) spins with values ranging from $3$ up to $N$. These currents generalize the spin-2 stress-energy tensor $T_{\mu\nu}$
which generates ``pure graviton'' excitations in the pure Einstein gravity discussed in Section \ref{LabelSectionThevacuumseed}, and lead to a ``truncated version'' of higher spin Vasiliev gravity, 
the latter containing generalized graviton excitations of arbitrary integer spin (see e.g. \cite{Vasiliev,Gopakumar}). As it is well known, the presence of extended chiral conformal algebras can lead to multiple modular invariants in 2D CFTs, but by extending the ``vacuum seed'' to  the vacuum representation of the extended chiral algebra, a single modular invariant can be built, and generalizations to higher spin gravity of the Virasoro arguments leading to Ising and Tricritical Ising are possible as described in \cite{Castro} based on genus-one considerations.\label{foot:higher}}
A possible gravitational explanation of this observation could be as follows: It turns out that amongst all unitary Virasoro minimal CFTs with central charge $c<1$, only the Ising and tricritical Ising CFTs satisfy the condition that the conformal weights $h$ of all non-trivial primary states are larger than $c/24$. In CFTs with large central charge $c$, this inequality describes a necessary condition that a primary state of conformal weight $h$ can be interpreted as being dual to a black hole \cite{Strominger}. 
Assuming that this condition is still valid in the strong-coupling regime where $c$ is not large, Ising and tricritical Ising would be the only unitary minimal model CFTs with $c<1$ in which all primary states can be interpreted as being dual to black holes. All other $c<1$ unitary minimal model CFTs would then contain, in addition to black holes, other primary {\it matter} fields, and these CFTs could thus not be dual to pure Einstein gravity. (We will come back in
Appendix \ref{sec:discussion} to the interpretation of primary states in the Ising CFT as states dual to black holes in strongly-coupled Einstein gravity, by suggesting a possible expression for their Bekenstein-Hawking entropy.)

The focus of the present paper is pure Einstein quantum gravity on Euclidean 3-manifolds $X$ whose asymptotic boundaries $\partial X$ are {\it higher-genus} Riemann surfaces. This arises physically because 3-manifolds $X$ whose boundaries are Riemann surfaces of higher genus $g \geq 2$, are known \cite{Aminneborg, Brill, Krasnov1, Krasnov2, Yin1, Giombi, Skenderis} to be the Euclidean spacetimes corresponding to multi-boundary wormholes in Lorentzian signature. $X$ is commonly 
restricted to handlebodies, and we will follow this assumption here; more complicated saddles such as non-handlebodies 
were studied
in \cite{Yin2} in the semi-classical regime. We plan to come back to this issue in future work. 
There is a variety of interesting and important physical questions related to such multi-boundary wormholes (see, e.g., \cite{Multibdry} for a relatively recent discussion), and a complete description of the duality between 3D quantum gravity and the associated 2D CFT at the asymptotic boundary must include all those spacetimes. In other words, any proposed duality must also be valid in any such multi-boundary wormhole spacetime. For that reason, it is important to investigate the duality between quantum gravity in \ads and the CFT on the asymptotic boundary at higher genus $g\geq 2$. 

For the gravitational partition function at general genus $g$, there is again 
formally a sum
over geometries of the smooth
handlebody $X$ and 
over
its boundary geometries
\begin{equation}
Z_{\text{grav}}(\Omega, \bar{\Omega})
=\int_{\partial X = \Sigma_g} {\cal D}g_{\mu\nu} \ e^{-c~S_E[g_{\mu\nu}]},
\label{eq:sumMGenusg}
\end{equation}
where the ``period matrix'' $\Omega$, a $g\times g$-dimensional symmetric complex matrix, completely parametrizes the conformal structure of the genus $g$ Riemann surface $\Sigma_g$ constituting
the boundary of $X$. The gravitational path integral can then again be written in the form
\begin{equation}
Z_{\text{grav}}=\sum_{X \ ({\rm where} \ \partial X = \Sigma_g)} \ Z(X,\Omega),
\label{eq:SumXGenusg}
\end{equation}
where $Z(X,\Omega)$ stands for the contribution from the sum over all metrics connected by small diffeomorphisms on a particular smooth handlebody $X$ with a fixed 
period matrix
$\Omega $ on its asymptotic boundary $\Sigma_g$, while the summation over $X$ means summing over different 
period matrices $\Omega$ in the same conformal class.

Different Euclidean saddles can be constructed by specifying which cycles of the Riemann surface $\Sigma_g$ are contractible in the interior of 
the 3-manifold
$X$, and such cycles are mapped into each other under the action of the mapping class group (MCG) $\Gamma_g$ of the Riemann surface $\Sigma_g$. To compute the gravitational path
integral in (\ref{eq:SumXGenusg}), our strategy is analogous
to the torus case: We again start with the contribution from 
a certain gravitational ``vacuum seed'' partition function
$Z_{\text{vac}}(\Omega, \bar{\Omega})$
corresponding to the trivial saddle and perform a modular sum to write the complete partition function in (\ref{eq:SumXGenusg}) in the following more explicit form
\begin{equation}
Z_{\text{grav}}(\Omega, \bar{\Omega}) =
\sum_{\gamma\in\Gamma_c\backslash\Gamma_g} Z_{\text{vac}}(\gamma\Omega, \bar{\gamma}\bar{\Omega}).
\label{eq:ModularSumHigherGenus}
\end{equation}
Here, $\Gamma_c$ denotes the subgroup of the mapping class group $\Gamma_g$
of the Riemann surface 
$\Sigma_g$ (the latter being an infinite group) which leaves the 
``vacuum seed partition function''
$Z_{\text{vac}}(\Omega, \bar{\Omega})$ invariant, and $\Gamma_c\backslash\Gamma_g$
is the right coset space; it is over this coset space that the sum in
(\ref{eq:ModularSumHigherGenus}) is performed. 
Whether this sum has an infinite or a finite number of terms depends in general (a): on the value of the Brown-Henneaux central charge, and (b): on the genus $g$. When the sum is infinite, there is no natural procedure to associate
a value to it.\footnote{A natural regularization scheme would require a probability measure on the (infinite) mapping class group that is also invariant under ``translations'' (i.e., under group multiplications).  A group with such a translation-invariant measure that is further finitely additive
(the measure of a finite disjoint union of sets is the sum of the measures of these sets) is called amenable. All mapping class groups are non-amenable as they contain non-abelian free groups as subgroups.  Subgroups of amenable groups are amenable and non-abelian free groups are known to be non-amenable. It follows that there are no natural regularization schemes to sum over mapping class groups in this sense. However, this theorem does not apply to summations over cosets such as the regularized sum considered for genus one in \cite{MaloneyWitten}. It is not clear how to generalize their treatment to higher genus at the current stage.\label{foot:regularization}} 
In this paper, we show that for all theories of pure Einstein gravity in \ads with Brown-Henneaux central charge $c<1$, this sum is finite and unique {\it only} when $c=1/2$, corresponding to the dual CFT at the asymptotic boundary to be the Ising CFT. Therefore we argue that in the strong-coupling regime of Brown-Henneaux central charge $c=3l/2G_N<1$, pure Einstein gravity is only dual to a 2D CFT if $c=1/2$.

We arrive at this conclusion by extending the results obtained for 
genus one
by Castro et al. \cite{Castro}. Recall that, as mentioned above, Castro et al. argued solely based on genus-one considerations that the only 
2D CFTs with central charge $c<1$ that can be dual to pure Einstein gravity in \ads at the corresponding Brown-Henneaux central charges are the Ising and the Tricritical Ising CFTs of central charges $c=1/2$ and $c=7/10$, respectively.
The results we obtain in the present paper, based on consideration of {\it arbitrary} genus $g$, are two-fold:

(i) {\it For Brown-Henneaux central charge $c=1/2$.} 
After first identifying the gravitational
genus-$g$ ``vacuum seed'' partition function, we observe that the orbit of the vacuum seed under the MCG action is dictated by a projective representation $\rho_g$ of the MCG $\Gamma_g$ that is identical to the projective representation induced by the holomorphic conformal blocks of the 2D Ising CFT. We then show, using 
the properties
of $\rho_g$, that the action of the MCG $\Gamma_g$ on the vacuum seed  generates an orbit that is always a finite set for any genus $g$ and, hence, leads only to a {\it finite} sum in (\ref{eq:ModularSumHigherGenus}). We further prove that this projective representation $\rho_g$ is {\it irreducible}, which, by Schur's Lemma,
leads to the conclusion that the finite sum in (\ref{eq:ModularSumHigherGenus}) for the gravitational partition function is unique, and is precisely proportional to the partition function of the 2D Ising CFT.\footnote{The physical significance of the factor of proportionality is not entirely clear at this point.} 
The key mathematical results that we prove in this paper and that underlie our physics conclusions on the quantum gravity partition function at $c=1/2$ 
are: (1) The representation $\rho_g$, when viewed as a mapping from
the MCG
$\Gamma_g$ to a unitary group, has a {\it finite} image set for 
any genus $g$, and (2)
the projective representation $\rho_g$ of the
MCG $\Gamma_g$ is always {\it irreducible} for any genus $g$. These results are obtained by exploiting the connection between the 2D Ising CFT and the 3D Ising topological quantum field theory (TQFT).
We first provide a simplified discussion on these results in Section \ref{sec:two}
for the genus-two case, and continue with the discussion of the general genus $g$ case in Section \ref{sec:more}.

(ii) {\it For Brown-Henneaux central charge $c=7/10$.} While the genus-one considerations by Castro et al. \cite{Castro} 
would permit the conclusion
that pure Einstein gravity in \ads at $c=7/10$ is dual to the 2D Tricritical Ising CFT at the asymptotic boundary, their arguments do not carry over to higher genus $g\geq 2$.
(As discussed above, consideration of arbitrary genus is necessary for a complete description of a duality.) 
We arrive at this conclusion by considering the 
3D TQFT related to the 2D Tricritical Ising CFT at $c=7/10$. We show that, at Brown-Henneaux central charge $c=7/10$, the sum occurring in the $g\geq 2$ gravitational partition function  \eqref{eq:ModularSumHigherGenus} has an infinite number of terms and cannot be naturally regularized, as explained in \footref{foot:regularization}. A detailed discussion will be provided in Section \ref{sec:difficulty}.

The remainder of the paper is organized as follows. In Section \ref{sec:torus}, we review the torus case. Section \ref{sec:two} presents a discussion of the genus-two case, while Section \ref{sec:more} presents a complete discussion and proof for general genus $g$, which is independent of the previous section and is more mathematically involved.
The difficulty in extending to the Tricritical Ising case is discussed in more detail at the end of Section \ref{sec:more}. Several Appendices spell out various details. In the last appendix \ref{sec:discussion}, the duality is used to compute the gravitational entropy and we find a resemblance to the topological correction to the entanglement entropy occurring in the context of topological phases of matter.

\section{Gravitational partition function with torus asymptotic boundary}
\label{sec:torus}
The simplest Euclidean smooth 3-manifold $X$ that contributes to the sum in \eqref{eq:sumM} is that of thermal AdS$_3$, topologically a solid torus. It is described in the semi-classical limit ($c\gg 1$) by the following metric
\begin{equation}
ds^2=l^2\left(d\rho^2+\cosh^2\rho~dt_E^2+\sinh^2\rho~d\phi^2\right),
\label{eq:tads}
\end{equation}
where $\phi\sim\phi+2\pi$ denotes a spatial cycle which is contractible in the bulk of $X$, and the Euclidean time $t_E$ parametrizes a non-contractible cycle. Defining $z=-t_E+i\phi$, the complex coordinate $z$ parametrizes points on the asymptotic boundary ($\rho \to \infty$) of $X$, and it is periodically identified according to
\begin{equation}
    z\sim z+2\pi i n\sim z+2\pi i m\tau,~ m,n\in\mathbb{Z},
\end{equation}
where the first identification is automatic (due to the periodicity of $\phi$), while the second is to construct the thermal AdS$_3$
space-time, and $\tau$ is the complex parameter specifying the conformal structure of the boundary torus. Large diffeomorphisms, i.e., elements of the MCG, act on this conformal structure parameter as
\begin{equation}
    \tau\rightarrow \gamma\cdot\tau =\frac{a\tau+b}{c\tau+d},\quad \gamma=\left(\begin{matrix} a & b\\ c & d\end{matrix}\right)\in SL(2,\mathbb{Z}).
\end{equation}
Note that the large diffeomorphisms do not change the conformal structure on the boundary torus. Therefore, all conformal structure parameters
$\gamma\cdot\tau$ with $\gamma \in SL(2,\mathbb{Z}))$ (in other words, all $\tau$'s related to each other by the MCG) specify the same conformal structure. Each of $\gamma \tau$ gives a classical Euclidean solution to the Einstein's 
equation \cite{MaldacenaStrominger}, i.e., a valid saddle point of (\ref{eq:integral}). These may or may not be different saddle points depending on the choice of $\gamma$. For example, the combination $a=0,$ $b=1,$ $c=-1,$ $d=0$ realizes a modular $S$ transformation, which maps $\tau\mapsto -1/\tau$ and the resultant saddle is the Euclidean BTZ black hole \cite{BTZ}. It is related to thermal \ads by exchanging the spatial and temporal cycles, consistent with the defining feature of a Euclidean 
BTZ black hole - the existence of a a temporal cycle contractible in the bulk. It was shown in \cite{MaloneyWitten} that the only smooth solutions to the equation of motion with torus boundary conditions are the ones above, but not all these solutions labeled by $\gamma$ are inequivalent. Specifically, an overall sign flip of $a, b, c, d$ does not change the saddle, neither does a constant integer shift $(a,b)\rightarrow (a,b)+n (c,d)$ generated by the modular $T$ transformation. Physically, the latter observation corresponds to the fact that adding a contractible cycle to a non-contractible cycle leaves the non-contractible cycle still non-contractible.
We denote the subgroup of $SL(2,\Z)$ generated by $T$ by $\Gamma_\infty$. 
So in the semi-classical regime, different saddles are labeled by different right cosets of $\Gamma_{\infty}$ in $SL(2,\mathbb{Z})$, or equivalently by integers $(c,d)$ corresponding to solid tori $M_{c,d}$. Notice that all solid tori $M_{c,d}$ share the same hyperbolic metric \eqref{eq:tads}, because by a famous theorem of Sullivan \cite{Sullivan,McMullen,Finiteness}, for a fixed conformal class of 
the asymptotic boundary, the bulk is a {\it unique} smooth and infinite-volume hyperbolic 3-manifold, with a rigid complete metric. 

\begin{figure}
    \centering
    \includegraphics[width=0.85\textwidth]{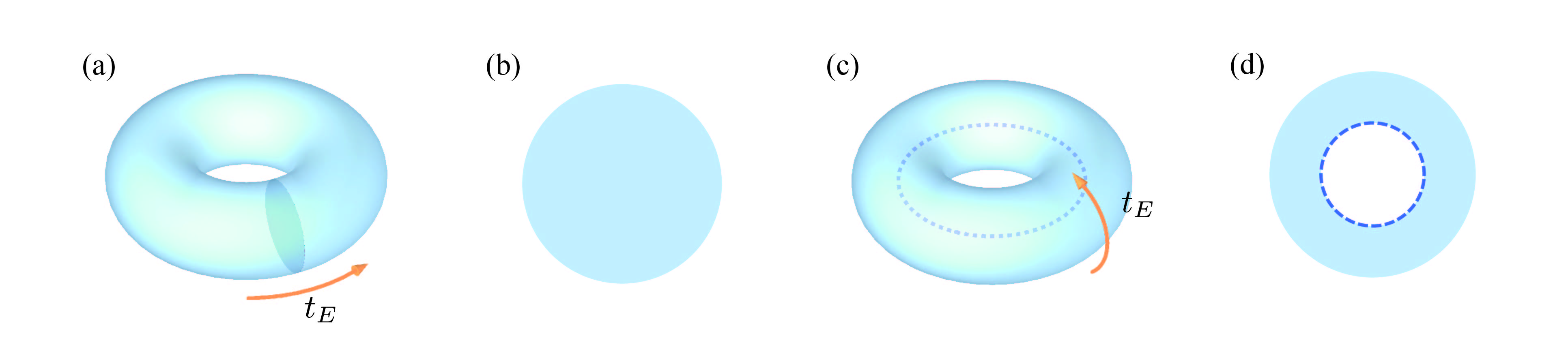}
    \caption{ 
    From left to right: (a) Geometry for Euclidean thermal AdS$_3$, with time going along the longitudinal direction; (b) The constant time slice of (a) is a disk; (c) Geometry for Euclidean BTZ black hole, with the event horizon being the dashed line in the core of the solid torus; (d) Constant time slice of (c) is an annulus,where the inner boundary is the event horizon.}
    \label{fig:ads}
\end{figure}

These saddle-point Euclidean spacetimes $M_{c,d}$ can be obtained from the corresponding Lorentzian ones via analytical continuation, which amounts to taking the Schottky double of its Lorentzian $t=0$ constant time slice \cite{Krasnov1, Krasnov2}. 
The Schottky double of a surface is essentially two copies of the surface glued along their boundaries, i.e., a closed surface. (For a surface without a boundary, the Schottky double is two disconnected copies of the surface, with all moduli replaced by their complex conjugates in the second copy.) In Figure \ref{fig:ads}, we depict the examples of Euclidean
thermal \ads with $(c,d)=(0,1)$ and the Euclidean BTZ black hole $(c,d)=(1,0)$, as well as their constant time slices. Both are non-rotating\footnote{For a definition see Appendix \ref{app:selection}.} and possess an equal time $t=0$ surface with a $\Z_2$ time-reversal symmetry.

It turns out that in the strongly coupled regime, $\Gamma_\infty$ is enhanced to a larger group $\Gamma_c$ (a ``new gauge symmetry'') \cite{Castro}, which is a finite index subgroup of $SL(2,\mathbb{Z})$. Hence the inequivalent manifolds $X$ are then labeled by right cosets
$\gamma \in \Gamma_c\backslash SL(2,\mathbb{Z})\equiv\Gamma$
and one can write
\begin{equation}
Z_{\text{grav}}(\tau,\bar{\tau})=\sum_{\gamma\in\Gamma} Z_{\text{vac}}(\gamma \tau, \gamma \bar{\tau}),
\end{equation}
where $Z_{\text{vac}}$ is the partition function of the 
``vacuum seed'', by which we here mean  here that of thermal AdS$_3$ spacetime.

\subsection{Vacuum seed}
\label{LabelSectionThevacuumseed}

To compute $Z_{\text{vac}}(\tau,\bar{\tau})$, one needs to evaluate in the path integral (\ref{eq:integral}) the contribution from metrics that are continuously connected to thermal AdS$_3$. In this subsection (and only here), we temporarily resort to Lorentzian signature for convenience. These metrics differ from that of empty AdS$_3$, the Lorentzian counterpart of the Euclidean thermal AdS$_3$, by small diffeomorphisms that preserve the Brown-Henneaux boundary conditions
\begin{equation}
ds^2\sim l^2\left[d\rho^2+\frac{1}{4}e^{2\rho} (-dt^2+d\phi^2)+O(\rho^0)\right]
\end{equation}
at large $\rho$. 

The classical phase space of the theory is the same as the configuration space of all classical excitations that are continuously connected to the global \ads ground state metric \eqref{eq:tads}. Brown and Henneaux \cite{BrownHenneaux} 
observed\footnote{See also, e.g., the compact review in \cite{Strominger}.} that the phase space charges $H(\zeta_n)$ corresponding to such diffeomorphisms $\zeta_n$ satisfy the Virasoro algebra with central charge $c=3l/2G_N$.
Acting on the ground state with these charge operators, one obtains the boundary graviton states, whose norms must be positive. Upon performing canonical quantization as proposed in \cite{Castro, BrownHenneaux}, these charge operators are promoted to operators representing the generators of the Virasoro algebra,
\begin{equation}
L_n\equiv H[\zeta_n],\quad\bar{L}_n\equiv H[\bar{\zeta}_n].
\end{equation}

Then the vacuum state is annihilated by $L_0$ and $\bar{L}_0$, as well as other Virasoro lowering operators. This state corresponds semi-classically to empty Lorentzian AdS$_3$. The conformal symmetry constrains the theory strongly, and the boundary gravitons are described by the states obtained by acting with chains of Virasoro raising operators on the vacuum, i.e., these are the descendant states $L_{-n_1}\cdots L_{-n_k}|0\rangle$, with $n_i>1$. Our desired partition function $Z_{\text{vac}}$ is then the generating function that counts these states. 

In the strongly coupled regime of Brown-Henneaux central charge
$c<1$, the requirement of unitarity constrains the central charge to
the values $c=1-6/m(m+1)$ corresponding to the `Virasoro minimal model' CFTs, where $m$ is an integer larger than two. 
Furthermore, eliminating the null states gives the vacuum (identity)
character of an irreducible highest-weight representation of the Virasoro algebra (see for example \cite{BigYellowBook}),
\begin{equation}
\label{eq:torusvac}
Z_{\text{vac},\,g=1}=\text{Tr}_{\text{vac}} \ q^{L_0}\bar{q}^{\bar{L}_0}=\abs{\chi_{1,1}(\tau)}^2,
\quad {\rm where} \ q=e^{2\pi i \tau}.
\end{equation}
Here the subscript $r,s$ of a general character $\chi_{r,s}$ denotes indices of the Kac table that label all possible
irreducible representations of the Virasoro algebra at this central charge
(where $r=1, 2, ...,m-1$ and $s=1, 2, ...,m$), and\footnote{The character is known \cite{RochaCaridi,Cappelli} to take the form
$\eta(\tau)q^{-(1-c)/24} \chi_{r,s}=$
$\sum_{k \in \mathbb{Z}} \left (q^{h_{r+ 2km, s}} - q^{h_{r+2km, -s}}\right)
\equiv S$. Using the identity $h_{r+l m, s+l (m+1)}= h_{r,s}$ for all $l\in \mathbb{Z}$, which
follows from \eqref{eq:KacConformalWeights} by inspection, one can bring the
expression above into the form
$S= q^{h_{r,s}}$
$+\sum_{k=1}^\infty  \left[q^{h_{r+2k m, s}} + q^{h_{r, s+ 2k (m=1)}}\right]$
$-\left[ \left ( \sum_{k=0}^\infty q^{h_{r+(2k+1) m,-s + (m+1)}} \right)
+\left(\sum_{k=1}^\infty q^{h_{r,-s+2k(m+1)}}\right)\right]$. After expressing this as a single sum over $l$ from $l=1$ to
$\infty$, where $l=2k$ for even $l$, and $l=(2k+1)$ for odd $l$, the sum $S$ is easily seen to
yield the result presented in the following equation.}

\begin{eqnarray}
&&\chi_{r,s}=
\\ \nonumber
&&=\frac{q^{(1-c)/24}}{\eta(\tau)}\left[q^{h_{r,s}}+
\sum\limits_{l=1}^\infty (-1)^l \left(
q^{h_{r+l m,s (-1)^l+(m+1)[1-(-1)^l]/2}}
+
q^{h_{r,s (-1)^l+ l(m+1) + (m+1) [1-(-1)^l]/2}}
\right)\right], \qquad \quad
\label{eq:KacCharacterChi-r-s}
\end{eqnarray}
with $\eta(\tau)
=q^{1/24}\prod_{n=1}^{\infty}(1-q^n)$ the Dedekind eta function and the highest weight $h_{r,s}$ given by 
\begin{equation}
h_{r,s}=\frac{[(m+1)r-ms]^2-1}{4m(m+1)}. 
\label{eq:KacConformalWeights}
\end{equation}

\subsection{Modular sum and duality to the Ising CFT}
The simplest minimal model is the Ising CFT with $c=1/2$. There are three irreducible representations of the Virasoro algebra satisfying $h_{1,1}=0,$  $h_{2,1}=1/2$ and $h_{1,2}=1/16$. 
The partition function of the theory is simply the diagonal modular invariant
\begin{equation}
Z_{\text{Ising}}(\tau,\bar{\tau})=|\chi_{1,1}(\tau)|^2+|\chi_{1,2}(\tau)|^2+|\chi_{2,1}(\tau)|^2,
\end{equation}
where the three summands are conformal characters of the identity, energy and spin operators, respectively \cite{BigYellowBook}. These characters can also (for Ising)  be expressed in terms of the Riemann or Jacobi theta function, as reviewed in Appendix \ref{app:Ising}, equation \eqref{eq:TorusChiTheta}. 

On the other hand, the gravitational partition function is obtained by summing all images of the vacuum character under 
$\Gamma \equiv \Gamma_c\backslash SL(2,\mathbb{Z})$, the right coset space of $\Gamma_c$ in $SL(2,\mathbb{Z})$, as
\begin{equation}
Z_{\text{grav}}(\tau,\bar{\tau})=\sum_{\gamma\in\Gamma} |\chi_{1,1}(\gamma\tau)|^2,
\label{eq:IsingSum}
\end{equation}
where $\Gamma_c$ is the set of all ``pure gauge transformations'' of the vacuum, which are
defined to be those
elements of $SL(2,\mathbb{Z})$ that act trivially on
the modulus of the vacuum character,
$|\chi_{1,1}|$:
\begin{equation}
\label{eq:chi11}
\Gamma_c=\{\gamma\in SL(2,\mathbb{Z})\large{\mid} \  \abs{\chi_{1,1}(\gamma\tau)}=|\chi_{1,1}(\tau)|\}.
\end{equation}
This is a finite index subgroup as proven in \cite{Bantay}, so the summation in \eqref{eq:IsingSum} has a finite number of terms, unlike
the $c>1$ Farey-tail cases that were discussed in Refs \cite{FareyTail, MaloneyWitten}. We will see in later sections that the finiteness property, seen here at genus one,
extends (for Ising) to the case of higher genus. Starting from
the ``vacuum seed''
$|\chi_{1,1}|^2$, \eqref{eq:torusvac}, and repeatedly acting on it with the generators 
\begin{equation}
S=\left(\begin{matrix} 0 & -1\\ 1 & 0 \end{matrix}\right),\quad T=\left(\begin{matrix} 1 & 1 \\ 0 & 1\end{matrix}\right)
\end{equation}
of $SL(2,\mathbb{Z})$,
one finds 24 inequivalent contributions, 
which sum up to 
\begin{equation}
\label{eq:torusindex}
Z_{\text{grav}}=8 Z_{\text{Ising}}.
\end{equation}
The physical meaning of this constant factor of 8 is at present unclear, while its mathematical meaning, along with extra new results on $\Gamma_c$ that go beyond those presented in \cite{Castro},
are collected in Appendix \ref{app:kernel}. Therefore we see the equality of the partition functions of pure Einstein gravity in \ads at Brown-Henneaux central charge $c=3l/2G_N=1/2$ and that of the Ising CFT, at genus one.

\section{Gravitational partition functions with genus-2 asymptotic boundaries}
\label{sec:two}

Now we generalize the discussion of the duality between Euclidean \ads and 
Virasoro minimal model CFTs
to genus two. The current section is more ``physical'' or intuitive, 
compared to Section \ref{sec:more} which discusses the case for arbitrary genus and will be more mathematically involved. We will focus on the $c=3l/2G_N=1/2$ theory and present its gravitational partition function as well as its relation to the Ising CFT in Section \ref{sec:TwoArgue}, followed by a review of the relevant mathematical concepts in Section \ref{sec:math}. 

\subsection{Gravitational partition function}\label{sec:TwoArgue}

Similar to the genus-one case, the key assumption in the computation of the 
gravitational partition function is that the path integral is equal to the contribution from classical saddle points and the full set of quantum fluctuations around them, irrespective of the fact that the Brown-Henneaux central charge is now of order one. 

\begin{figure}[htbp]
\includegraphics[width=0.95\textwidth]{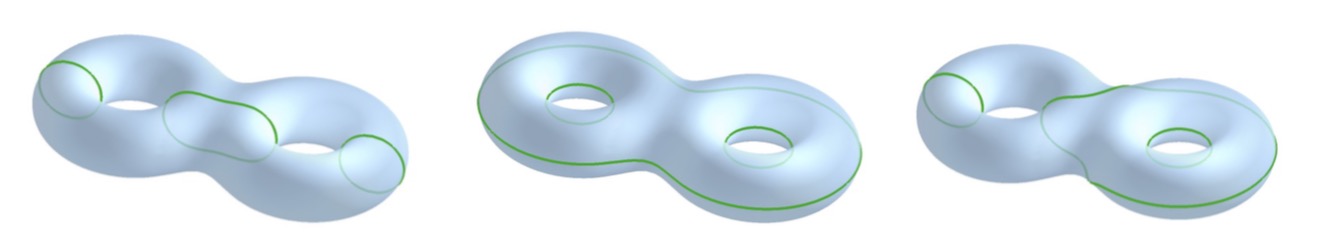}
\caption{Different Euclidean saddles with $\Z_2$
time-reflection symmetry. They analytically continue to three copies of thermal \ads (left), three-sided wormhole (middle), and one thermal \ads plus a BTZ black hole (right). The areas encircled by the green lines are the sets of fixed points of the action of the $\Z_2$ symmetry.}
\label{fig:saddle}
\end{figure}

As briefly reviewed in the last section, the analytical continuation from Lorentzian to  Euclidean signature basically amounts to taking a Schottky double. When the Lorentzian geometry contains three asymptotic regions, its constant time slice is a pair of pants. The boundary of the corresponding Euclidean spacetime is thus obtained (following the notion of the Schottky double, mentioned above) by gluing two pairs of pants together, thereby obtaining a genus-two Riemann surface. Different ways of gluing give distinct saddles and correspond to different choices of contractible cycles in the bulk. In Figure \ref{fig:saddle}, we sketch three bulk geometries that possess a $\Z_2$ time-reflection symmetry \cite{Multibdry}. 
The left one depicts the case which corresponds to three disconnected thermal \ads spacetimes in Lorentzian signature. The green circles label the interfaces between the two pairs of pants. The middle panel describes the Euclidean version of the three-sided wormhole. The right figure is the case with one copy of thermal \ads and a BTZ black hole. Different bulk saddles can be transformed into each other by the action of
the mapping class group 
(whose definition will be reviewed in Section \ref{sec:math}).

The full partition function can thus be written as the modular sum of one of the saddles, namely as that of the vacuum saddle without black holes,
\begin{equation}
\label{eq:ModularSum}
Z_{\text{grav}}(\Omega,\bar{\Omega})=\sum_{\gamma\in \Gamma} Z_{\text{vac}} (\gamma\Omega,\bar{\gamma}\bar{\Omega}),
\end{equation}
where $\Gamma= \Gamma_c\backslash\Gamma_g$
is the right coset space of the mapping class group (MCG) $\Gamma_g$ of the Riemann surface $\Sigma_g$ with respect to $\Gamma_c$, the symmetry group that leaves $Z_{\text{vac}}$ 
invariant. - Here $g=2$. The $2\times 2$-dimensional complex, symmetric period matrix $\Omega$ is a higher-genus generalization of the modular parameter $\tau$ in genus one, whose definition is presented in Section \ref{sec:math} below. The conformal structure on the asymptotic boundary is specified by the period matrix $\Omega$. All period matrices $\Omega$ related to each other by the MCG correspond to the {\it same} conformal structure.

\subsubsection{Vacuum seed}
\label{Sec:Genus2_vacuum_Seed}

We are interested in the case where the bulk gravity is a genus-two handlebody, which can be viewed as three solid cylinders that meet at a cup and a cap (each being a ``3-ball'' - the interior of a 2-dimensional sphere), compare e.g., Figure \ref{fig:cylinders}. We will choose the notation for the elementary cycles depicted in Figure \ref{fig:handlebody} below.
\begin{figure}[htbp]
\centering
\includegraphics[width=0.6\textwidth]{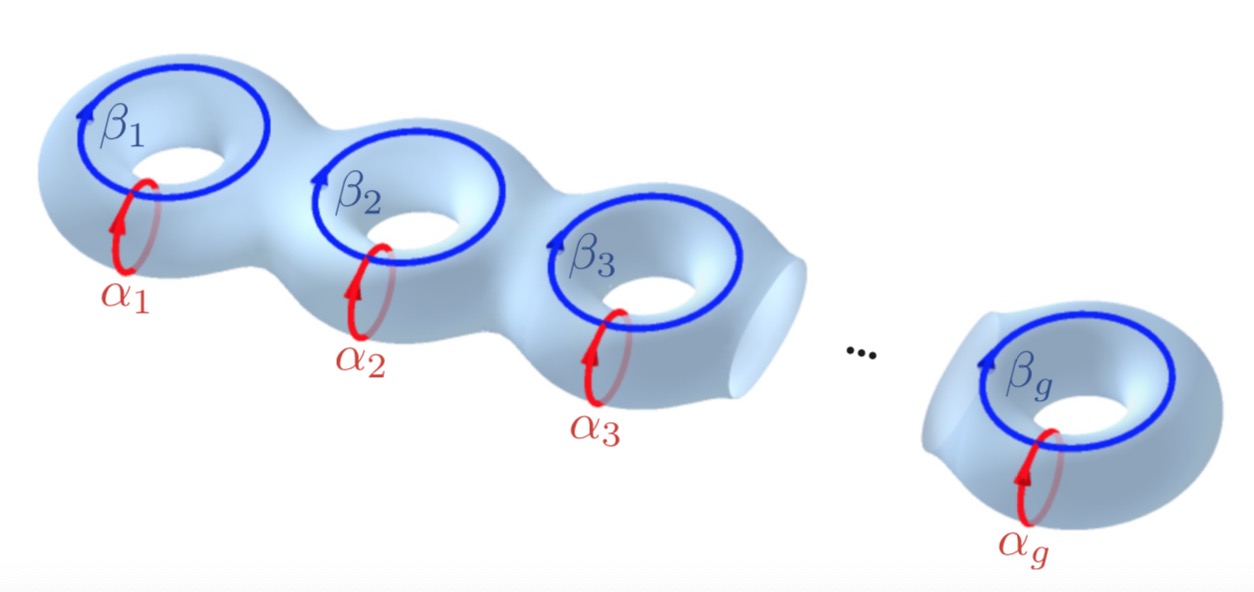}
\caption{The canonical homology basis for $\Sigma_g$.}
\label{fig:handlebody}
\end{figure}
The vacuum sector $Z_\text{vac}$ dominates the full partition function in the low-temperature limit, which we define to be the limit where the three solid cylinders are long and thin, like in Figure \ref{fig:cylinders}. (This is analogous to the genus-one case, where in the low-temperature limit, the dominant geometry is the one whose boundary torus has a longitude much larger than its meridian.) In this limit, a natural local coordinate system can be chosen, such that a constant time slice is a disjoint union of three disks, i.e., the cross sections of the three solid cylinders (see Figure \ref{fig:cylinders}), while the time direction is along the longitudinal direction of the cylinders.\footnote{From a TQFT point of view, this corresponds to the case where only the trivial anyons propagate in the long cylinders. The relationship with TQFT is discussed briefly in Appendix \ref{app:Ising} and will be generally described in Section \ref{sec:more}.}
Such a topology analytically continues to three copies of thermal AdS$_3$. Namely, all the $\alpha$-cycles in Figure \ref{fig:handlebody} need to be contractible in the bulk. 
\begin{figure}[hbtp]
\centering
\includegraphics[width=0.45\textwidth]{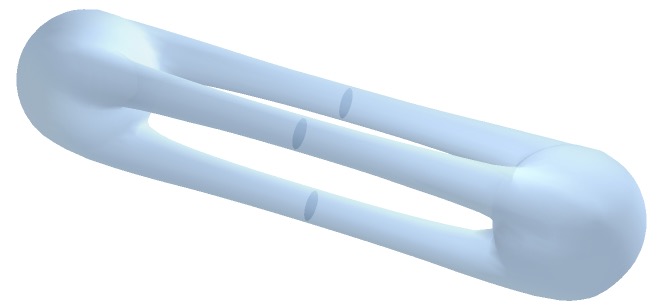}
\caption{The low-temperature or long-cylinder limit of the genus-two geometry.}
\label{fig:cylinders}
\end{figure}

For a bulk geometry with a higher-genus asymptotic boundary, we believe that the 
association with the Brown-Henneaux central charge $c=3l/2G_N$ is still valid.\footnote{In their original paper \cite{BrownHenneaux}, given the global AdS$_3$ metric with $\rho$ being the radial direction,
$$ ds^2=-\left(\frac{\rho^2}{l^2}\right)dt^2+\left(\frac{l^2}{\rho^2}\right)d\rho^2+\rho^2d\phi^2, $$ after quotienting it by some discrete subgroup of the isometry group of global AdS$_3$, off-diagonal entries of the new metric need to satisfy the asymptotic conditions
$$ g_{t\rho}\sim\mathcal{O}(1/\rho^3),\quad g_{t\phi}\sim\mathcal{O}(1),\quad g_{\rho\phi}\sim\mathcal{O}(1/\rho^3), $$
in order to produce two copies of Virasoro algebras with central charge $c$ on the boundary. In principle, these conditions can be checked here using the Fefferman-Graham metric for asymptotic AdS$_{d+1}$, constructed basically by shooting geodesics inwards from the boundary \cite{Fefferman1,Fefferman2}: $ds^2=g_{\rho\rho}d\rho^2+g_{ij}dx^idx^j$ $=\displaystyle{\frac{l^2d\rho^2}{4\rho^2}+\frac{1}{\rho}\tilde{g}_{ij}(\rho,x)dx^idx^j}$, where the $d$-dimensional metric $\tilde{g}_{ij}(\rho,x)$ is the $\rho$-dependent Euclidean boundary metric.} 
Recall in the genus-one case, the boundary torus describes the time evolution of graviton states living on the boundary of a disk. When the Brown-Henneaux central charge is
$c=1/2$, these states correspond to the quantum states of the 2D Ising CFT in the vacuum sector $\chi_{1,1}$ (and $\bar{\chi}_{1,1}$). For genus two and in the local coordinate system where a constant time slice consists of three disjoint disks (see Figure \ref{fig:cylinders}), the boundary graviton states live on the boundary of each disk. Hence locally, the boundary graviton states correspond to three copies of $\chi_{1,1}$ states (and $\bar{\chi}_{1,1}$ states). Globally, the former should correspond to states in the vacuum conformal block of the Ising CFT at genus two (the analogue of the $\chi_{1,1}$ sector for genus one), which we denote by
$\chi_{\text{vac}}$. Therefore, we assume $Z_{\text{vac}}$ to be of the same form as the partition function of the Ising vacuum conformal block. This assumption is a natural extension of results in \cite{Yin1, Bootstrap}. In the large-$c$ and the pinching limit of the genus-2 asymptotic boundary, Ref. \cite{Yin1} calculated the vacuum seed of \ads to order$1/c^2$. 
This was then shown to match exactly with the partition function of the vacuum conformal block of a 2D large-$c$ CFT \cite{Bootstrap}. Naturally, we expect this match to hold to all orders of $1/c$, thereby justifying the assumption.

The full partition function of the 2D Ising CFT theory on a Riemann surface of arbitrary genus 
was worked out in Ref. \cite{DVV} using a $\mathbb{Z}_2$ orbifold of the free compactified boson theory, and in Ref. \cite{AGMV} using a single, non-interacting Majorana fermion. 
In the formulation using the Majorana fermion, a choice of boundary conditions (spin structure) has to be imposed.
The contribution from each choice of boundary conditions or spin structure can be written as the norm of the regularized determinant of the corresponding chiral Dirac operator. The determinant can further be separated into two factors, one being the Riemann theta function of the corresponding spin structure (whose definition will be reviewed in Section \ref{sec:math}), while the other is independent of spin structures and only a function of the metric. In what follows, the former will be denoted as the classical contribution to the partition function, and the latter will be called the quantum contribution. (Note this has a different meaning from the ``quantum'' used to describe gravitational theories which are beyond semiclassical regime.
The word ``quantum'' here stems from the fact that this universal factor accounts for the quantum fluctuations of the boson fields in the $\mathbb{Z}_2$ orbifold.) For more details about the quantum contribution, we refer to Appendix \ref{app:Ising}. In fact, not only the full partition of the 2D Ising CFT, but each of the conformal blocks also factorizes into a classical and a quantum piece. 
Given the identification of the gravitational vacuum seed and the vacuum conformal block of the 2D Ising CFT, discussed above, we can write $Z_{\vac} = Z_{\vac}^{\rm cl} Z_{\vac}^{\rm qu}$ (where ``cl'' stands for classical and ``qu'' stands for quantum). In the following discussion, we will be interested in how the different sectors or conformal blocks in the theory transform into each other under the mapping class group. For this purpose, it is enough to temporarily ignore the overall quantum factor that is the same for all conformal blocks and focus on the classical contribution of the gravitational vacuum seed 
\begin{equation}
\label{eq:Zvac}
Z_{\text{vac}}^{\text{cl}}(\Omega,\bar{\Omega})=|\chi_{\text{vac}}^{\text{cl}}(\Omega)|^2=\frac{1}{16}\left|\sum_{b_1,b_2\in\{0,1/2\}}\vartheta^{1/2}\spin{a_1=0}{a_2=0}{b_1}{b_2}(0|\Omega)\right|^2,
\end{equation}
where $|\chi_{\text{vac}}^{\text{cl}}(\Omega)|^2$ is the classical contribution to the vacuum conformal block of the 2D Ising CFT,
and where
$\vartheta$ denotes the conventional Riemann theta function (see \eqref{eq:DefRiemannThetaFunction}
of Section \ref{sec:math} for a review of relevant notations).
Here, $a_{1,2}$ should be viewed as the two components of the characteristic vector ${\bf a}=(a_1,a_2)$ appearing in the theta function
for the genus-2 case. Similarly, $b_{1,2}$ are the two components of the characteristic vector ${\bf b}=(b_1,b_2)$. The number of components  of these 
characteristic vectors is given by the genus $g$ in general.
We explain in the following the specific choice of theta functions $\vartheta$ appearing in the above expression.

We know that along a contractible cycle,
the boundary condition for a fermion has to be anti-periodic.\footnote{This is the natural boundary  condition for fermions since they anti-commute. See also for example \cite{AGMV, BigYellowBook, Multibdry}. Periodic boundary conditions for fermions would imply a singularity inside the cycle, often called a $\Z_2$-vortex, or Majorana fermion zero mode.} Since, as discussed above,
in the gravitational vacuum seed
all the $\alpha$-cycles in Figure \ref{fig:handlebody} need to be contractible, 
all the corresponding boundary conditions on the (Majorana) fermion
along those cycles need to be anti-periodic. Consequently, the top characteristic vector of the theta functions that are relevant for the vacuum sector is zero, i.e., $a_1=a_2=0.$ Furthermore, the vacuum sector must be a equal-weight summation over both even and odd fermion number parities along every $\beta$-cycle. This means $Z_{\text{vac}}^{\text{cl}}$ has to be the modulus square of a equal-weight linear combination of the square root of Riemann theta functions that appear in equation \eqref{eq:Zvac}, as displayed in that same equation. 

The above form \eqref{eq:Zvac} for $Z_{\text{vac}}^{\text{cl}}$ is also analogous to the classical contribution to the vacuum seed on the {\it torus}. The latter, as reviewed in \eqref{eq:TorusChiTheta} of Appendix \ref{app:Ising}, is the equal-weight sum of the 
square roots of all theta functions whose characteristic 
`vector' ${\bf a}= (a_1)$ is zero. 
On the torus, this has a natural Hamiltonian interpretation that exists due to a global notion of time (leading to a clean
separation of 1D space and 1D time), which is absent at higher genus.\footnote{Namely, at genus
one there are four (one of them vanishing) holomorphic partition functions, $\chi_{\sigma_x,\sigma_{t_E}}\equiv Tr_{{\cal H}_{\sigma_x}} (\sigma_{t_E})^F q^{L_0}$, where $F$ denotes the fermion parity operator, $\sigma_x, \sigma_{t_E}=\pm 1$. Here $\mp\sigma_x$
denotes spatial (anti-)periodicity whereas $\mp\sigma_{t_E}$ denotes Euclidean temporal (anti-)periodicity. These holomorphic partition functions are proportional to 
$\vartheta^{1/2} \left[\begin{matrix} a\\ b\end{matrix}\right]$
with $a=(1-\sigma_x)/2$ and $b=(1-\sigma_{t_E})/2$  One then sees from (\ref{eq:TorusChiTheta}) of Appendix \ref{app:Ising} that the torus vacuum character $\chi_{1,1}$ is proportional to the sum of the square-roots of theta functions with $a=0$, summed over $b=0$ and $b=1/2$. The sum appearing in (\ref{eq:Zvac}) is the natural generalization of this genus-one expression to genus two.
}

In the pinching limit where the bulk 
`cylinder' connecting the two tori pinches off \cite{DVV}, i.e.,
where $\Omega_{12}, \Omega_{21} \to 0$, 
the (classical) vacuum seed partition function 
\eqref{eq:Zvac} reduces to $|\chi_{1,1}^{\text{cl}}(\tau_1) \ \chi_{1,1}^{\text{cl}}(\tau_2)|^2$, which is the product of the classical parts of the two torus vacuum seeds $|\chi_{1,1}(\tau_1) |^2$ and $| \chi_{1,1}(\tau_2)|^2$ of two tori with modular parameters $\tau_1$ and $\tau_2$.

One can check that \eqref{eq:Zvac} is invariant under a genus-two generalization of $\Gamma_\infty$, see Appendix \ref{app:generators}. This is a subgroup of the genus-two mapping class group $\Gamma_g$,\footnote{Basic facts about this genus-two generalization of $\Gamma_\infty$ will be discussed in Appendix \ref{app:generators} where this group is referred to
as $\Gamma_{\infty}^{[2]}$.} generated \cite{Yin1} by integer shifts of matrix elements of the period matrix $\Omega$, as well as the $SL(2,\Z)$ transformation that acts 
on $\Omega$ by conjugation
$\Omega\mapsto A\Omega A^T$. 
The genus-$2$ generalization of the group $\Gamma_\infty$
is the classical symmetry of the vacuum seed at large $c$, and it is enhanced in the case of strong coupling ($c<1$) to the previously mentioned group $\Gamma_c$, a subgroup of $\Gamma_{g=2}$
which is larger than $\Gamma_\infty$. This new ``gauge symmetry'' will be relevant in the modular sum as it turns out to be a finite-index subgroup.

As a consistency check of \eqref{eq:Zvac}, in the low-temperature or long-cylinder limit depicted in Figure \ref{fig:cylinders}, the leading contribution to $Z_{\text{vac}}^{\text{cl}}$ needs to be equal to that of the total classical contribution 
to the full Ising partition function at genus two in the same limit, as explained in Appendix \ref{app:AlgCurve}. The long-cylinder limit can be taken in the following way: The genus-two Riemann surface can be described as
a hyperelliptic curve, which is the set of solutions to the following 
equation (see for example \cite{CardyMaloneyMaxfield} for a recent discussion of this)
\begin{equation}
\label{eq:curve}
y(z)^2= \prod_{k=1}^3 \frac{z-u_k}{z-v_k}.
\end{equation}
Such a surface is a two-sheeted branched cover of the Riemann sphere, the points on which are parametrized by $z$, and the two sheets are labeled by the choice of the root $y$ which solves \eqref{eq:curve}. There is a $\Z_2$ ``replica symmetry'' generated by $y\rightarrow -y$, which physically corresponds to the time reversal symmetry
discussed above in Figure \ref{fig:saddle}, and the corresponding text. 
The covering map has $2\times 3=6$ branch points $(u_k, v_k)$. Monodromy of $z$ around one of the six branch points shifts $y\rightarrow -y$ and moves from one sheet to the other. The locations of the branch points span the moduli space\footnote{This is a $g=2$ coincidence, for general genus $g$ the moduli space $\mathcal{M}_{g}$ of a Riemann surface $\Sigma_g$ has real dimension $6g-6$, while the number of real branches is $2g+2$.} 
of the Riemann surface. Consequently, the period matrix can be expressed in terms of the branch points \cite{CoserTagliacozzoTonni}, and the long-cylinder limit corresponds to
taking $u_k-v_k$ to be small for $k=1,2,3,$.  
To obtain the vacuum seed partition function, the resulting period matrix $\Omega$ is inserted into \eqref{eq:Zvac}.
For the case of the long-cylinder limit at general genus $g$, one simply replaces the number $3$ appearing in \eqref{eq:curve} by $g+1$, and proceeds in an analogous fashion. 

A final remark is that, for our gravitational vacuum seed partition function $Z_{\text{vac}}$ to be identical with that of the vacuum conformal block of the boundary CFT (up to some constant factor), we further need to discuss the cup and cap regions, where three cylinders join. We argue that the three-point correlation functions that describe the graviton scattering processes in the gravity theory 
match those in the boundary conformal theory\footnote{At genus one, a related but somewhat different two-to-one scattering process in AdS$_3$ between the bulk duals of the light primaries $O$ and $\chi$, whose conformal weights less than $c/12$, is described in \cite{Kraus}. Their CFT three-point function is found to have the {\it same form} as the gravitational scattering amplitude between their gravitational duals in the BTZ background, with a proportionality factor only dependent on the saddle geometry. Although this process is not necessarily in pure gravity, this result is in support of our argument about the form of $Z_{\text{vac}}$.}.

\subsubsection{Genus two modular sum}
\label{sec:sum}

With the above expression for the vacuum seed, we now perform the sum over the images of the action with the MCG  (``modular sum'') as in \eqref{eq:ModularSum} 
at $g=2$.\footnote{Instead of $Z_{\text{vac}}^{\text{cl}}$, we use the full quantum conformal block $Z_{\text{vac}}=|\chi_{\text{vac}}|^2$ in the modular sum. The quantum contribution and issues related to it, are discussed in Appendices \ref{app:generators} and \ref{app:Ising}.
}. We will first provide the numerical results, and then give a mathematical argument for the finiteness of the modular sum. Independently, we will present
later in Section \ref{sec:finiteness_modular_sum}
another simple proof from a TQFT perspective for arbitrary genus.

As reviewed in Section \ref{sec:math}, the subgroup of the
mapping class group which acts non-trivially on the period matrix is $Sp(4,\Z)$. The generators of $Sp(2g,\Z)$ are reviewed in Appendix \ref{app:generators}. By acting repeatedly with the two generators of $Sp(4,\Z)$ on the vacuum seed partition function, we find 3840 inequivalent contributions with the aid of {\it Mathematica}.\footnote{This set is invariant under the action of Torelli group introduced in Section \ref{sec:math} below. The Torelli group acts by multiplying
$\left\{\vartheta[1/2, 1/2; 1/2, 1/2](\Omega)\cdot\vartheta^{*}[1/2, 1/2; 1/2, 1/2](\Omega)\right\}^{1/2}$ by a minus sign, which can be explicitly verified in the pinching limit using the formalism in \cite{DVV} and straightforwardly carries over to the general case away from that limit.
Only this specific theta function product is affected by the Torelli group action, because it is related to the sector (conformal block)
$\chi_{\sigma\psi\sigma}$ (in the language of Appendix \ref{app:Ising}), where there is a fermion $\psi$ in the middle of the genus-two handlebody (denoted by $b$ in Figure \ref{fig:anyon}, i.e., $b\to \psi$),
which acquires a negative sign upon the Dehn twist along the separating curve.}
These modular images sum up to 384 times the partition function $Z_{\rm Ising}$ of the 2D Ising CFT at genus two of Appendix \ref{app:Ising}):
\begin{equation}
\label{eq:384}
    Z_{\text{grav}}=384 Z_{\text{Ising}}.
\end{equation}
The factor $10=3840/384$ is simply the dimension of the conformal block basis, or simply the number of linearly independent Riemann theta functions. The physical meaning of the constant factor $384$ in (\ref{eq:384}) is unclear at this point.

We emphasize that all the above arguments are gravitational ones that solely come from the three-dimensional bulk. In the remainder of this section, we support the above computation by a mathematical explanation for the finiteness of the summation in the partition function (as in \eqref{eq:ModularSum}).

In the Ising case, there exists\footnote{This is a generalization of the mathematical result in \cite{Wright} 
which is explained in Appendix F.} a short exact sequence for any genus $g$,
\begin{equation}
\label{eq:short}
1\rightarrow \rho_g(D_g) \rightarrow \rho_g (\Gamma_g) \rightarrow Sp(2g,\mathbb{Z}_2)\rightarrow 1,
\end{equation}
where $\rho_g(\Gamma_g)$ is the image group of the mapping class group $\Gamma_g$ represented as matrices in the basis of Riemann theta functions, $D_g$ is the subgroup of the mapping class group that acts trivially on $H_1(\Sigma_g,\Z_2)$, and $\rho_g(D_g)$ is the corresponding image group. The latter turns out to always be a subgroup of $\Z_8^N$, where $N$ is a finite positive integer.

Since $\rho_g(D_g)$ is abelian
(see Appendix \ref{app:propertyf}), \eqref{eq:short} gives a central extension of $Sp(2g,\mathbb{Z}_2)$. Such central extensions are classified by the second cohomology group $H^2(Sp(2g,\Z_2),~\rho_g(D_g))$:
For every group element $h\in Sp(2g,\Z_2)$ and
$n\in \rho_g(D_g)$, there is an element $(n, h)$ in $\rho_g(\Gamma_g)$, satisfying the group multiplication $(n_1,h_1)\cdot (n_2,h_2) = \left(\omega(h_1, h_2)n_1 n_2,h_1h_2\right),$ where $\omega(h_1, h_2)$ is a 2-cocycle with $\rho_g(D_g)$ coefficients.
Alternatively, one can interpret the above short exact sequence in terms of projective representations. Irreducible representations of the mapping class group $\Gamma_g$ correspond to the irreducible projective representations of $Sp(2g,\mathbb{Z}_2)$, where the projective phases are given by $\rho_g(D_g)$. 

Since $Z_{\text{vac}}$ involves taking the 
modulus square of the vacuum character, the overall phases of $\rho_g(D_g)$ will not matter. We can simply focus on the summation over elements of  $Sp(2g,\Z_2)$ that act non-trivially on the absolute values of the theta functions. At genus $g=2$, $Sp(4, \Z_2)$ turns out to be equal to the permutation group $S_6$ and contains $6!=720$ elements. Due to the short exact sequence \eqref{eq:short}, the image group of $\Gamma_g$ is clearly finite. 

In Section \ref{sec:more}, we will present an alternative simple proof for the finiteness of $\rho_g(\Gamma_g)$ that works for arbitrary genus, from a topological field theory perspective.

\subsection{Review of the relevant concepts}
\label{sec:math}
We first describe the homology of orientable, finite-type two-dimensional surfaces $\Sigma_g$ of genus $g$. When $\Sigma_g$ is compact, its homology  groups are free, with dim$H_0(\Sigma_g)=1$, dim$H_1(\Sigma_g)=2g$, dim$H_2(\Sigma_g)=1$. One can choose a canonical homology basis $\alpha_i, \beta_i$ with $1\leq i\leq g$ for $H_1(\Sigma_g)$ as in Figure \ref{fig:handlebody}. Any closed curve on $\Sigma_g$ generates a homology class, which can be uniquely decomposed into the classes generated by $\alpha_i, \beta_i$. They are normalized with respect to the algebraic intersection number $J(C_1, C_2)$ between two simple closed curves $C_1$ and $C_2$, by 
\begin{equation}
\label{eq:intersection}
J(\alpha_i,\alpha_j)=J(\beta_i,\beta_j)=0,\quad J(\alpha_i,\beta_j)=-J(\beta_i,\alpha_j)=\delta_{ij}.
\end{equation}

There are $g$ pairs of holomorphic and anti-holomorphic one-forms on $\Sigma_g$, 
denoted by $\{\omega_i, \bar{\omega}_i\}$ $(i=1,\cdots,g)$, which satisfy the normalization condition
\begin{equation}
\label{eq:normalization}
    \oint_{\alpha_i} \omega_j=\delta_{ij}.
\end{equation}
The period matrix defined by
\begin{equation}
    \oint_{\beta_i} \omega_j=\Omega_{ij}
\end{equation}
is then a $g\times g$ complex symmetric matrix, with a positive-definite imaginary part.\footnote{An 
alternative normalization for $\Omega$ more suitable for computation is considered in Appendix \ref{app:AlgCurve}.} 
Analogous equations as above hold for the anti-holomorphic
counterparts $\bar{\omega}_i$ and $\bar{\Omega}_{ij}$. 
The period matrix $\Omega$ generalizes the modular parameter $\tau$ for the 
torus, completely parametrizing the conformal structure of $\Sigma_g$. Note that a conformal structure of $\Sigma_g$ can be specified by different period matrices that are related to each other by the mapping class group.\footnote{The moduli space $\mathcal{M}_g$, the space of 
conformal structures of $\Sigma_g$, 
has real dimension $6g-6$. The {\it Torelli map} from $\mathcal{M}_g$ to the space of $\Omega$'s quotiented by the mapping class group
$\Gamma_g$ is injective, intuitively because the latter has real dimension $g(g+1)$, so the parametrization is complete.}

The mapping class group (MCG) $\Gamma_g$ of a genus-$g$ Riemann surface $\Sigma_g$ is the group of all isotopy classes of orientation preserving diffeomorphisms of $\Sigma_g$.
It is generated by Dehn twists around the cycles $C$
of $\Sigma_g$. A Dehn twist acts by excising a tubular neighborhood of $C$ inside $\Sigma_g$, twisting the latter by $2\pi$, and then gluing it back to the rest of the surface. There are two generators for each handle, and one for each closed curve linking the holes of two neighboring handles. 

$\Gamma_g$ leaves the intersections (\ref{eq:intersection}) invariant, thus acting on the canonical homology basis by $Sp(2g,\mathbb{Z})$ transformations. The $Sp(2g, \mathbb{Z})$ transformations act on the period matrix by
\begin{equation}
\gamma=\left(\begin{matrix}  A & B\\  C & D\\ \end{matrix}\right)\in Sp(2g, \mathbb{Z}), \quad \gamma: \Omega\rightarrow  (A\Omega+B)(C\Omega+D)^{-1},
\label{eq:sp2z}
\end{equation}
where $A, B, C, D$ are $g$ by $g$ matrices. 
At genus $g=2$, the minimal number of generators of $Sp(4,\mathbb{Z})$ 
is two \cite{Bender}; these are reviewed in Appendix \ref{app:generators}.
For $Sp(2g,\mathbb{Z})$ with $g\geq 3$ the minimal number of
generators is three \cite{Lu}. 

\begin{figure}[htbp]
\centering
\includegraphics[width=0.75\textwidth]{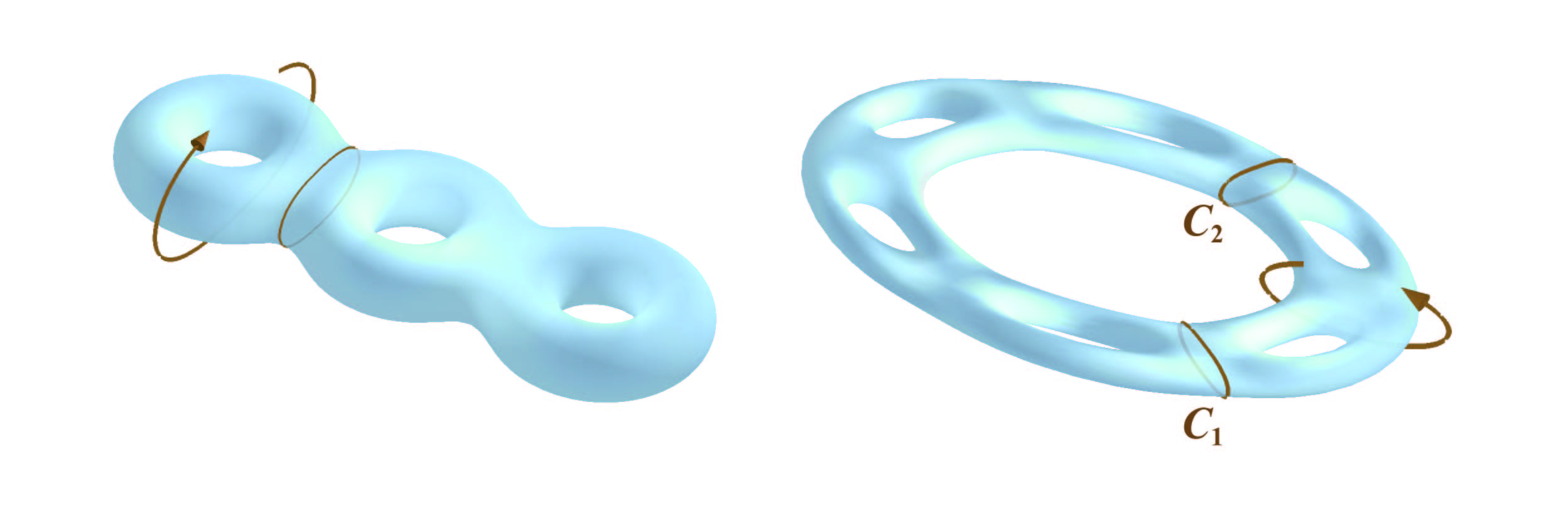}
\caption{Generators of the Torelli group $\mathcal{I}_g$. Left: Dehn twist along a separating curve. Right: the bounding pair map.}
\label{fig:Torelli}
\end{figure}

Some elements of $\Gamma_g$ act trivially on the canonical homology basis, leaving it invariant. These elements are diffeomorphisms homotopic to the identity and they form a normal subgroup of $\Gamma_g$, known as the Torelli group $\mathcal{I}_g$ \cite{Hatcher,Primer} . For genus two, $\mathcal{I}_g$ is infinitely generated by Dehn twists around the separating curve, i.e., the curve that separates the genus two surface into two tori. For $g\geq 3$, besides the ones that twist around the separating curves, there exists another type of generator, called the ``bounding pair map''.
A bounding pair map is the composition of a twist along a non-separating curve $C_1$ and an inverse twist along another non-separating curve $C_2$ which is disjoint from $C_1$ but represents the same homology class as $C_1$. 
So $C_1\cup C_2$ separates $\Sigma_g$ 
into two subsurfaces having $C_1\cup C_2$ as their common boundary. These two kinds of generators are shown in Figure \ref{fig:Torelli}.

In summary, we have the following non-splitting short exact sequence,
\begin{equation}
1\rightarrow \mathcal{I}_g \rightarrow \Gamma_g \rightarrow Sp(2g, \mathbb{Z})\rightarrow 1.
\label{eq:SES}
\end{equation}

Riemann or Siegel theta functions, which depend on two $g$-dimensional row vectors $\mathbf{a}, \mathbf{b} \in \mathbb{R}^g$ called characteristics, are defined by the following infinite sum \cite{Mumford, Fay, Igusa}, 
\begin{equation}
\label{eq:DefRiemannThetaFunction}
\vartheta\left[\begin{matrix} \mathbf{a}\\ \mathbf{b}\end{matrix}\right] (\mathbf{z}|\Omega)\equiv\sum_{\mathbf{n}\in \Z^g} \exp\left( i\pi (\mathbf{n}+\mathbf{a})\cdot \Omega\cdot (\mathbf{n}+\mathbf{a})+2\pi i(\mathbf{n}+\mathbf{a})\cdot (\mathbf{z}+\mathbf{b})\right),
\end{equation}
where $\mathbf{z}\in \mathbb{R}^g$ is a $g$-dimensional vector.

In this paper, we will be interested in 
and limit our discussion to the Ising case 
described by a single Majorana fermion species, 
where the characteristic vectors
are $\mathbf{a}, \mathbf{b} \in (\frac{1}{2}\mathbb{Z})^g$. In this case there is, associated with each theta function, the notion of a {\it spin-structure} 
of characteristics $\left[\begin{matrix} \mathbf{a}\\ \mathbf{b}\end{matrix}\right]$, denoting a $2\times g$-matrix. The spin structure is called to be even or odd depending on whether $4\mathbf{a}\cdot\mathbf{b}$ is even or odd, respectively. This can be seen from the following identity
\begin{equation}
\vartheta \left[\begin{matrix} \mathbf{a}\\ \mathbf{b}\end{matrix}\right] (-\mathbf{z}|\Omega)=(-1)^{4\mathbf{a}\cdot\mathbf{b}}\varthetaconst (\mathbf{z}|\Omega).
\label{eq:ThetaSpinStructure}
\end{equation}
Additionally, due to the identity 
\begin{equation}
\vartheta \left[\begin{matrix} \mathbf{a}+\mathbf{n}\\ \mathbf{b}+\mathbf{m}\end{matrix}\right] (\mathbf{z}|\Omega)=e^{2\pi i\mathbf{a}\cdot\mathbf{m}}\varthetaconst (\mathbf{z}|\Omega),
\end{equation}
where $\mathbf{m},$ $\mathbf{n}\in \mathbb{Z}^g$, it is enough to only consider $\mathbf{a}, \mathbf{b}\in (\frac{1}{2}\mathbb{Z}_2)^g$. 
At genus $g$, there are $2^{g-1}(2^g-1)$ odd spin structures and $2^{g-1}(2^g+1)$ even ones. The theta functions $\vartheta(\Omega)$ always vanish for odd spin structures, which is obvious from \eqref{eq:ThetaSpinStructure}.

Riemann theta functions are also weight-1/2 modular forms. From now on we will denote $\vartheta(0|\Omega)$ by $\vartheta(\Omega)$ for convenience. When their argument $\Omega$ is acted on by
$\gamma=\left(\begin{matrix} A & B \\ C & D\end{matrix}\right)\in Sp(2g,\mathbb{Z})$, they transform as \cite{Igusa, AGMV}:
\begin{equation}
\label{eq:AGMV}
    \vartheta\left[\begin{matrix} \mathbf{a'}\\ \mathbf{b'}\end{matrix}\right]\left(\gamma\Omega\right)=\epsilon(\gamma) e^{-i\pi \phi(\mathbf{a},\mathbf{b})}\det\left(C\Omega+D\right)^{1/2} \vartheta\left[\begin{matrix} \mathbf{a}\\ \mathbf{b}\end{matrix}\right](\Omega),
\end{equation}
where
\begin{equation}
\label{eq:trans}
 \left\{\begin{matrix} \mathbf{a'}\\ \mathbf{b'}\end{matrix}\right\}= \left(\begin{matrix} D & -C \\ -B & A\end{matrix}
\right)\left\{\begin{matrix} \mathbf{a}\\ \mathbf{b}\end{matrix}\right\}+\frac{1}{2}\left\{\begin{matrix} {\left(CD^T\right)_d}\\ {\left(AB^T\right)_d}\end{matrix}\right\},
\end{equation}
and
\begin{equation}
    \phi\left(\mathbf{a},\mathbf{b}\right)=  \left(\mathbf{a}D^TB\mathbf{a}+\mathbf{b}C^TA\mathbf{b}\right)-\left[2\mathbf{a}B^TC\mathbf{b}+\left(\mathbf{a}D^T-\mathbf{b}C^T\right)\left(AB^T\right)_d\right],
\end{equation}
which is $\Omega$-independent.

In \eqref{eq:trans},
$\left\{\begin{matrix} \cdot \\ \cdot \end{matrix}\right\}$ 
means concatenating two $g$-dimensional row vectors into a single $2g$-dimensional column vector,
i.e.,
$\left\{\begin{matrix}\mathbf{a}  \\ \mathbf{b}  \end{matrix}\right\}$ 
$\equiv\left [\begin{matrix}\mathbf{a}^T  \\ \mathbf{b}^T
\end{matrix}\right]$
where 
$\cdot^T$ denotes the matrix transpose, whereas $(\cdot)_d$ denotes the $g$-dimensional row vector whose entries are the diagonal elements of the $g\times g$ matrix appearing inside the parentheses $(\,)$. The subtle phase $\epsilon(\gamma)$ is always an eighth root of unity independent of $\mathbf{a}$ and $\mathbf{b}$, and incidentally, if $\gamma=I_{2g}\,\,\text{mod}\,\,2$, then $\epsilon^2(\gamma)=e^{\pi i\text{Tr}(D-1)/2}$.

We note that the action of the group $Sp(4,\mathbb{Z})$ on the Riemann theta functions at genus $g=2$ defines a 10-dimensional projective, not a linear representation. The explicit forms of the matrix representations of the (two) generators of the group are displayed in Appendix \ref{app:generators}.

\section{Gravitational partition functions with boundaries of arbitrary genus}
\label{sec:more}

In this section, we discuss the full gravitational partition function at Brown-Henneaux central charge
$c=1/2$ with an asymptotic boundary being a Riemann surface of arbitrary genus following the same strategy as in the genus-2 case. The full gravitational partition function $Z_{\text{grav}}$ at Brown-Henneaux
central charge $c=1/2$ with a genus-$g$ asymptotic boundary $\Sigma_g$ is again formulated as a sum over the contributions from different saddle points which are all related to the ``vacuum seed'' contribution $Z_{\text{vac}}$ by the action of the 
mapping class group $\Gamma_g$ of the asymptotic boundary $\Sigma_g$. Given the period matrix $\Omega$ that specifies the conformal structure on  the asymptotic boundary $\Sigma_g$, we should write the full gravitational partition function as
\begin{align}
Z_{\text{grav}}(\Omega,\bar{\Omega})=\sum_{\gamma\in \Gamma} Z_{\text{vac}} (\gamma\Omega,\bar{\gamma}\bar{\Omega}),
\label{eq:Zgrav_MultiGenus}
\end{align}
where $\Gamma=\Gamma_c\backslash\Gamma_g $ is the right coset space of the mapping class group $\Gamma_g$ by its subgroup $\Gamma_c$ that leaves the vacuum seed invariant. In this sum, the term with trivial $\gamma$ represents the contribution from the vacuum sector (as known as the ``vacuum seed") while other terms present the contributions from other saddle points. 

In the following, we will first argue in Section \ref{sec:genus_g_vacuum_seed} that the vacuum seed $Z_{\text{vac}} (\Omega,\bar{\Omega})$ 
at Brown-Henneaux central charge $c=1/2$ can be identified with the vacuum conformal block of the 2D Ising CFT on the asymptotic boundary $\Sigma_g$ with the same period matrix $\Omega$. Then, we will show that the $\Gamma_g$-orbit $\{Z_{\text{vac}} (\gamma\Omega,\bar{\gamma}\bar{\Omega}) | \gamma\in \Gamma_g\}$ of the vacuum seed, which appears in (\ref{eq:Zgrav_MultiGenus}), is dictated by the projective representation $\rho_g$ of the MCG $\Gamma_g$ induced by the holomorphic conformal blocks of the 2D Ising CFT on $\Sigma_g$. Subsequently, we will prove in Section \ref{sec:finiteness_modular_sum} a mathematical result stating that $\rho_g$, viewed as a mapping from $\Gamma_g$ to a unitary group, has a {\it finite} image set ${\rm im}(\rho_g)$, which has 
the immediate consequence that the sum $\sum_{\gamma\in \Gamma_g}$ in (\ref{eq:Zgrav_MultiGenus}) is {\it finite}. Furthermore, in Section \ref{sec:irreducibility}, we will prove another mathematical result stating 
that the MCG representation $\rho_g$ is {\it irreducible}. Using the irreducibility of $\rho_g$, we can show that the finite sum in (\ref{eq:Zgrav_MultiGenus}) for the full gravitational partition function is precisely proportional to the partition function of the 2D Ising CFT on the asymptotic boundary $\Sigma_g$. In Section \ref{sec:relation}, we establish duality between 3D AdS quantum gravity at Brown-Henneaux central charge $c=1/2$ and 
2D Ising CFT. 
There, we will also further comment on our arguments for the gravitational vacuum seed $Z_{\text{vac}} (\Omega,\bar{\Omega})$. In Section \ref{sec:difficulty}, we will discuss, from the
perspective of the higher-genus partition function, the fundamental difficulty in extending the duality to the case with Brown-Henneaux 
central charge $c=7/10$. 

\subsection{Vacuum seed}
\label{sec:genus_g_vacuum_seed}
Similar to the discussion of the genus-2 asymptotic boundary, to identify the vacuum seed, namely the gravitational partition function contributed by the vacuum sector, we start with a handlebody $X$ with a genus-$g$ asymptotic boundary $\partial X= \Sigma_g$. The classical saddle point geometry on such a handlebody $X$ is asymptotically AdS$_3$  \cite{Aminneborg,Brill,Skenderis,Krasnov1,Krasnov2,Yin1,Giombi}. As stated in Section \ref{Sec:Genus2_vacuum_Seed}, we believe that the asymptotic behavior of the geometry ensures that the Brown-Henneaux central charge $c=3l/2G_N$ is still applicable even if the boundary genus $g$ is larger than 1. In the following, we will always focus on the case with 
Brown-Henneaux central charge $c=1/2$. 

\begin{figure}[htbp]
\centering
\includegraphics[width=0.5\textwidth]{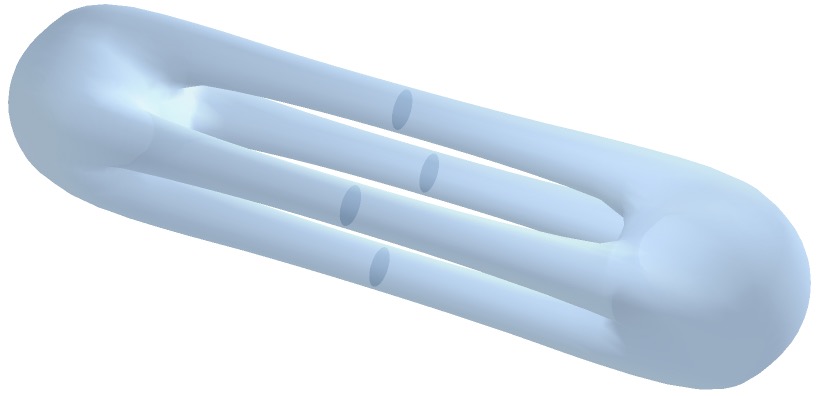}
\caption{A handlebody with a genus-$3$ asymptotic boundary. Each shaded disk shown should be associated 
with a Hilbert space of the boundary graviton states which form the representation of the Virasoro algebra with $c=1/2$.}
\label{fig:cylinders_genusG}
\end{figure}

As far as topology goes, the genus-$g$ handlebody $X$ can be viewed as two 3-balls
(the interiors of two 2-dimensional spheres) 
connected by $g+1$ solid cylinders. A genus-$3$ example is shown in Figure \ref{fig:cylinders_genusG}.\footnote{In this paper, we only study handlebodies in 3 dimensions. A genus-$g$ (3-dimensional) handlebody means a handlebody with a genus-$g$ 2-dimensional boundary.} Similar to the genus-2 discussion, we believe that the vacuum seed $Z_{\text{vac}}$ should dominate the (full) gravitational partition function on the 3-manifold $X$ in the limit where the boundary period matrix $\Omega$ is chosen such that, for each of the solid cylinder regions, the boundary circumference is much shorter than the length of the cylinder. In such a limit, it is natural to consider a (local) coordinate system such that the Euclidean time direction is along the longitudinal direction of each solid cylinder region. The Hilbert space of quantum gravity states should then be associated to a constant-time slice, which is a disjoint union of the cross sections of each of the solid cylinders, namely the disjoint union of $g+1$ disks. For example, for $g=3$, the Hilbert space of quantum gravity states should 
be associated with a disjoint union of 4 disks as shown in Figure \ref{fig:cylinders_genusG}.

Recall that in the discussion of the case with a genus-1 asymptotic boundary, the quantum gravity states defined on a single disk are the boundary graviton states that form the irreducible
(identity) representation of the Virasoro algebra with the corresponding Brown-Henneaux central charge $c$. For $c=1/2$ in particular, the boundary graviton states on a single disk are in one-to-one correspondence with quantum states of the 2D Ising CFT within the $|\chi_{1,1}|^2$ sector. 

Coming back to the genus-$g$ handlebody, we now need to assign a Hilbert space to the 
disjoint union of $g+1$ disks. We naturally expect the Hilbert space to be identified as the tensor product of $g+1$ copies of boundary graviton states obtained in the genus-1 discussion. In this picture, each solid cylinder region physically describes the time evolution of the boundary graviton states. 

So far, we have have been discussing the solid cylinder regions of the handlebody. Each of the 3-ball regions in the handlebody glues together all of the solid cylinders. Physically, each of them should describe the scattering process of $g+1$ boundary graviton states. Since the boundary graviton states are in one-to-one correspondence with the quantum states of the 2D Ising CFT, we further make the proposal
that the vacuum seed, $Z_{\text{vac}} (\Omega,\bar{\Omega})$, is identical to the vacuum conformal block of the 2D Ising CFT on the asymptotic boundary $\Sigma_g$ with period matrix
$\Omega$\footnote{In the vacuum conformal block of the 2D Ising CFT, the states propagating along the boundary of the solid cylinder regions all belong to the irreducible representation of the Virasoro algebra (and its anti-holomorphic copy) associated with $|\chi_{1,1}|^2$.}, which we naturally expect to 
factorize into holomorphic and the anti-holomorphic pieces,
i.e., 
\begin{align}
Z_{\text{vac}} (\Omega,\bar{\Omega}) =  \chi_{\text{vac}} (\Omega)\bar{\chi}_{\text{vac}} (\bar{\Omega}),
\label{eq:Zvac_MultiGenus}
\end{align}
where $\chi_{\text{vac}}(\Omega)$ and $\bar{\chi}_{\text{vac}}(\bar{\Omega})$ are the respective holomorphic and anti-homolorphic vacuum conformal blocks of the 2D Ising CFT on the genus-$g$ surface $\Sigma_g$ with period matrix $\Omega$. 
In fact, our proposed form of the vacuum seed is simply a natural 
extension of results in \cite{Yin1} and \cite{Bootstrap}. To be more specific,  \cite{Yin1} calculates the vacuum seed of the pure 3D AdS gravity with a genus-2 asymptotic boundary in the large-$c$ limit and also in the degeneration limit of the boundary. The result is obtained to the order $1/c^2$. \cite{Bootstrap} shows that the 
vacuum conformal block of a 2D large-$c$ CFT matches exactly with the result of \cite{Yin1} to all the orders calculated. Naturally, such a matching is expected to hold to all orders of $1/c$. Hence, (\ref{eq:Zvac_MultiGenus}) is a reasonable assumption when we take $c=1/2$. In addition, we will also see in the following subsections that a vacuum seed of the form of (\eqref{eq:Zvac_MultiGenus}) 
does yield a sensible expression for the full gravitational partition function through the modular sum  (\ref{eq:Zgrav_MultiGenus}).

\subsection{Finiteness of the Modular Sum}
\label{sec:finiteness_modular_sum}
To perform the modular sum (\ref{eq:Zgrav_MultiGenus}), we need to ensure that the summation over the set of right cosets, $\Gamma= \Gamma_c\backslash\Gamma_g$, is finite. $\Gamma_g$ is the MCG of the asymptotic boundary $\Sigma_g$ and $\Gamma_c$ is the subgroup of $\Gamma_g$ that leaves the vacuum seed $Z_{\text{vac}} (\Omega, \bar{\Omega})$ invariant. The finiteness of the set $\Gamma$ is mathematically equivalent to the finiteness of
the orbit of the vacuum seed $Z_{\text{vac}}$ under the MCG action, namely the finiteness of
the set $\{Z_{\text{vac}}\left(\gamma\Omega,\bar{\gamma}\bar{\Omega}\right) | \gamma \in \Gamma_g \}$. In Section \ref{sec:genus_g_vacuum_seed}, we have argued that the vacuum seed $Z_{\text{vac}}$ is given by the product of the holomorphic and anti-holomorphic vacuum conformal blocks of the 2D Ising CFT. Therefore, the MCG orbit of the vacuum seed $Z_{\text{vac}}$ is dictated by the $\Gamma_g$ 
action on the conformal blocks of the 2D Ising CFT on $\Sigma_g$. 

$\Sigma_g$ is a genus-$g$ Riemann surface. 
Considering only the holomorphic vacuum conformal block $\chi_{\text{vac}}$, the 2D Ising CFT has a total of $N_g = 2^{g-1}(2^{g}+1)$ holomorphic conformal blocks on $\Sigma_g$. They form an $N_g$-dimensional vector space which admits a $\Gamma_g$ action:
\begin{align}
 \chi_i(\gamma \Omega)  = \sum_{i'=1}^{N_g} \Big(\rho_g(\gamma) \Big)_{ii'}  \chi_{i'}(\Omega).
 \label{eq:MCG_action_on_conformal_blocks}
\end{align}
Here $\gamma \in \Gamma_g$, where $\chi_i(\Omega)$ with $i=1,2,..,N_g$ denote the $N_g$ different holomorphic conformal blocks of the 2D Ising CFT on the surface $\Sigma_g$, and $\rho_g (\gamma) \in U(N_g)$ is an $N_g \times N_g$ unitary matrix that depends on $\gamma$ (but not on the period matrix $\Omega$). In fact, $\rho_g$ is a projective representation of the MCG $\Gamma_g$:
For any $\gamma,\gamma' \in \Gamma_g$, $\rho_g (\gamma)\rho_g (\gamma')$ is equal to $\rho_g (\gamma\gamma')$ up to a $U(1)$ phase. The $\Gamma_g$ action on the anti-holomorphic conformal blocks of the 2D Ising CFT is naturally given by the complex-conjugated version of (\ref{eq:MCG_action_on_conformal_blocks}). Therefore, we will only discuss the representation $\rho_g$ that dictates the $\Gamma_g$ action on the holomorphic conformal blocks in the following discussion. 

When viewed as a map from $\Gamma_g$ to $U(N_g)$, $\rho_g$ has an image set $\rho_g(\Gamma_g) \equiv \{ \rho_g (\gamma)| \gamma\in \Gamma_g \}$ which is a subset of $U(N_g)$. In the following, we will prove that $\rho_g(\Gamma_g) $ is a finite set. Combining (\ref{eq:Zvac_MultiGenus}) and (\ref{eq:MCG_action_on_conformal_blocks}), it is straightforward to see that the finiteness of the set $\rho_g(\Gamma_g)$ directly implies the finiteness of the MCG orbit $\{Z_{\text{vac}}\left(\gamma\Omega,\bar{\gamma}\bar{\Omega}\right) | \gamma \in \Gamma_g \}$ and, consequently, leads to the conclusion that the modular sum (\ref{eq:Zgrav_MultiGenus}) is finite.

We will prove the finiteness of $\rho_g(\Gamma_g)$ by contradiction. Let's assume that $\rho_g(\Gamma_g)$ is an infinite set. First, we show that this assumption leads to the consequence that $\{\Tr\rho_g (\gamma) | \gamma\in \Gamma_g \}$ also has to be an infinite set. Since $\rho_g (\gamma) \in U(N_g)$, $|\Tr\rho_g (\gamma) | \leq N_g$. To show that $\{\Tr\rho_g (\gamma) | \gamma\in \Gamma_g \}$ is an infinite set, it is sufficient to show that, for any small number $\epsilon>0$, we can either find (i) a pair of elements $\gamma,\gamma' \in \Gamma_g$ such that $0< |\Tr \rho_g (\gamma) - \Tr \rho_g (\gamma')|< \epsilon$ or (ii) an element $\gamma'' \in \Gamma_g$ such that $ N_g - \epsilon <|\Tr\rho_g(\gamma'') | < N_g$. First, we start with a sufficiently small $\epsilon' >0$. Since $U(N_g)$ is a compact space, the assumption that $\rho_g(\Gamma_g)$ is an infinite set guarantees the existence of a pair of elements $\gamma,\gamma'\in \Gamma_g$ such that 
$0<\norm{\rho_g(\gamma) -\rho_g(\gamma')} <\epsilon'$ where $\norm{\cdot}$ represents the Frobenius norm.\footnote{The Frobenius norm $\norm{A}$ of a matrix $A$ is defined as the square root of the sum of the absolute squares of its elements, namely $\norm{A} = \sqrt{\Tr(A^\dag A)}$.} $\rho_g(\gamma)$ is not identical to $\rho_g(\gamma')$. But we still need to distinguish two situations depending on whether $\rho_g(\gamma)$ and $\rho_g(\gamma')$ differ by only a $U(1)$ phase or not. In first situation where $\rho_g(\gamma)$ differs from $\rho_g(\gamma')$ by a $U(1)$ phase, the sufficiently small $\epsilon'$ can guarantee that $0<|\Tr\rho_g(\gamma) - \Tr\rho_g(\gamma')|<\epsilon$. Hence, we find the pair of elements $\gamma, \gamma'$ described in (i). In second situation where $\rho_g(\gamma)$ is not proportional to $\rho_g(\gamma')$, we notice $\rho_g(\gamma^{-1}\gamma') $, which is equal to $\rho_g(\gamma)^{-1}\rho_g(\gamma')$ up to a $U(1)$ phase, is then not proportional to the identity operator. Then, with $\gamma''=\gamma^{-1}\gamma'$, $|\Tr\rho_g(\gamma'')|<N_g$. However, with a sufficiently small $\epsilon'$, $\rho_g(\gamma'')$ can be arbitrarily close to the identity operator up a $U(1)$ phase. Therefore, we have $ N_g - \epsilon <|\Tr\rho_g(\gamma'') | < N_g$. Hence, we find the element $\gamma''$ described in (ii). Now, we can conclude that the assumption that $\rho_g(\Gamma_g)$ is an infinite set has a consequence that $\{\Tr\rho_g (\gamma) | \gamma\in \Gamma_g \}$ also has to be an infinite set.

In the remainder of this subsection, we will show that $\{\Tr\rho_g (\gamma) | \gamma\in \Gamma_g \}$ in fact cannot be an infinite set and, hence, that the assumption that $\rho_g(\Gamma_g)$ is an infinite set is incorrect. 

For any $\gamma\in \Gamma_g$, $\Tr\rho_g (\gamma)$ can be interpreted as a partition function of the 3D Ising topological quantum field theory (TQFT). The 3D Ising TQFT is closely related to the 2D Ising CFT. In particular, the 3D Ising TQFT assigns a $N_g$-dimensional Hilbert space to the genus-$g$ surface $\Sigma_g$ whose basis vectors are in one-to-one correspondence with the holomorphic conformal blocks of the 2D Ising CFT on $\Sigma_g$ \cite{MooreSeiberg}. The details of this correspondence will be reviewed in the next subsection. A $\Gamma_g$ action $\gamma$ on the genus-$g$ surface $\Sigma_g$ induces a unitary transformation within the 3D Ising TQFT Hilbert space which is exactly given by $\rho_g(\gamma)$. $\Tr\rho_g(\gamma)$ can be interpreted as the 3D Ising TQFT partition function $Z_{\rm iTQFT} (M_\gamma)$ evaluated on the mapping torus $M_\gamma \equiv \frac{[0,1]\times \Sigma_g}{(0,x)\sim (1,\gamma(x))}$. The mapping torus $M_\gamma$ is a 3-manifold obtained from gluing the two $\Sigma_g$ boundary components of the Cartesian product $[0,1]\times \Sigma_g$ with an MCG action $\gamma$ performed on one of the $\Sigma_g$ components. 
For a general 3-manifold $M^3$, the 3D Ising TQFT partition can be expressed as \cite{Walker1991, Kirby1991}
\begin{align}
Z_{\rm iTQFT} (M^3) = \sum_{ \text{spin structure }\zeta} e^{\frac{2\pi i}{16} \mu(M^3,\zeta)},
\label{eq:iTQFT_partition_function}
\end{align}
where $\sum_{\zeta}$ represents the summation over all spin structures $\zeta$ on $M^3$ and $\mu(M^3,\zeta)$ is Rokhlin's $\mu$-invariant\footnote{For $(M^3,\zeta)$, it is defined as the signature of the intersection form of any smooth compact spin 4-manifold with the spin boundary $(M^3,\zeta)$.} of the 3-manifold $M^3$ with the spin structure $\zeta$. The invariant
$\mu(M^3,\zeta)$ is defined modulo 16 and is always an even integer. For a general 3-manifold $M^3$, the number of spin structures on $M^3$ is equal to $| H^1(M^3,\mathbb{Z}_2)| = | H_1(M^3,\mathbb{Z}_2)|$. Here, we are viewing $H^1$ and $H_1$ as groups. $|\cdot |$ means the order of the group in this context. For the mapping torus $M_\gamma$ of $\Sigma_g$, we can consider the following long exact sequence (see, e.g., Example 2.48 of \cite{Hatcher2}),
\begin{align}
...\rightarrow 
H_n (\Sigma_g, \mathbb{Z}_2) 
\xrightarrow{~} 
H_n (\Sigma_g,\mathbb{Z}_2) 
\rightarrow 
H_{n} (M_{\gamma}, \mathbb{Z}_2)
\rightarrow 
H_{n-1} (\Sigma_g, \mathbb{Z}_2) \rightarrow ...,
\label{eq:long_exact_sequence}
\end{align}
which implies an upper bound on the number of spin structures on $M_\gamma$ that only depends on $g$ but not $\gamma$:
\begin{align}
|H^1(M_{\gamma}, \mathbb{Z}_2)| =  | H_{1} (M_{\gamma}, \mathbb{Z}_2) | \leq | H_{1} (\Sigma_g, \mathbb{Z}_2)|\times |H_{0} (\Sigma_g, \mathbb{Z}_2)|.
\label{eq:spin_structure_counting_inequality}
\end{align}
The inequality above is a direct consequence of the $H_1 (\Sigma_g,\mathbb{Z}_2) 
\rightarrow 
H_{1} (M_{\gamma}, \mathbb{Z}_2)
\rightarrow 
H_{0} (\Sigma_g, \mathbb{Z}_2)$ part of the long exact sequence (\ref{eq:long_exact_sequence}).\footnote{With the map $H_{1} (M_{\gamma}, \mathbb{Z}_2)
\rightarrow 
H_{0} (\Sigma_g, \mathbb{Z}_2)$ in (\ref{eq:long_exact_sequence}) viewed as a linear map between vector spaces (over $\mathbb{Z}_2$), the sum of the dimensions of its kernel and its image is equal to the dimension of $H_{1} (M_{\gamma}, \mathbb{Z}_2)$. The image of the linear map $H_{1} (M_{\gamma}, \mathbb{Z}_2)
\rightarrow 
H_{0} (\Sigma_g, \mathbb{Z}_2)$, as a vector space over $\mathbb{Z}_2$, has a dimension less than or equal to the dimension of $H_{0} (\Sigma_g, \mathbb{Z}_2)$. From the fact that (\ref{eq:long_exact_sequence}) is exact, the kernel of the map $H_{1} (M_{\gamma}, \mathbb{Z}_2)
\rightarrow 
H_{0} (\Sigma_g, \mathbb{Z}_2)$ has the same dimension as the image of the map $H_1 (\Sigma_g,\mathbb{Z}_2) 
\rightarrow 
H_{1} (M_{\gamma}, \mathbb{Z}_2)$ whose dimension is smaller than or equal to the dimension of the vector space $H_1 (\Sigma_g,\mathbb{Z}_2) $. Therefore, the dimension of the vector space $H_{1} (M_{\gamma}, \mathbb{Z}_2)$ is not greater than the sum of the dimensions of $H_{0} (\Sigma_g, \mathbb{Z}_2)$ and of $H_1 (\Sigma_g,\mathbb{Z}_2) $, which implies (\ref{eq:spin_structure_counting_inequality}).
} Therefore, according to (\ref{eq:iTQFT_partition_function}), for any $\gamma\in\Gamma_g$, 
\begin{align} 
& \Tr\rho_g(\gamma) =Z_{\rm iTQFT}(M_{\gamma})  \nonumber \\
&~ \in \left\{ \sum_{n=0,2,4,...,14} a_n e^{ \frac{2\pi i}{16} n} ~\Bigg|~ a_n \in \mathbb{Z},~0 \leq a_n \leq | H_{1} (\Sigma_g, \mathbb{Z}_2)|\times |H_{0} (\Sigma_g, \mathbb{Z}_2)| \right\}.
\end{align}
Notice that the set given in the second line a finite set. Therefore, $\{\Tr\rho_g (\gamma) | \gamma\in \Gamma_g \}$ cannot be an infinite set, which is in contradiction to the consequence of the assumption that $\rho_g(\Gamma_g)$ is an infinite set. Now, we can conclude that $\rho_g(\Gamma_g)$ has to be a finite subset of $U(N_g)$. It follows that the modular sum (\ref{eq:Zgrav_MultiGenus}) is finite. 

This proof of the finiteness of the modular sum (\ref{eq:Zgrav_MultiGenus}) relies on the expression of the vacuum seed $Z_{\rm vac}$ (\ref{eq:Zvac_MultiGenus}) that we argued for in Section \ref{sec:genus_g_vacuum_seed}. In fact, as long as the vacuum seed $Z_{\rm vac}$ can be written as a product of a holomorphic and an anti-holomorphic conformal block of the 2D Ising CFT (or even as a sum of products of this type), the proof given in this subsection is still applicable and the modular sum (\ref{eq:Zgrav_MultiGenus}) is still finite.

\subsection{Irreducibility of the MCG representation and the modular sum}
\label{sec:irreducibility}
With the modular sum (\ref{eq:Zgrav_MultiGenus}) proven to be finite, the full gravitational partition function $Z_{\text{grav}}(\Omega, \bar{\Omega})$ is then, by construction, invariant under any $\Gamma_g$ action on the asymptotic boundary $\Sigma_g$. Since a MCG action generally transforms the holomorphic (anti-holomorphic) vacuum conformal blocks of the 2D Ising CFT into a linear superposition of all holomorphic (anti-holomorphic) conformal blocks, we expect the modular sum (\ref{eq:Zgrav_MultiGenus}), together with the vacuum seed (\ref{eq:Zvac_MultiGenus}), to yield
\begin{align}
  Z_{\text{grav}} (\Omega, \bar{\Omega}) = \sum_{i,i'=1}^{N_g} B_{ii'} \bar{\chi}_{i'}(\bar{\Omega})  \chi_i(\Omega),
  \label{eq:Bilinear}
\end{align}
where $B$ is a 
$N_g \times N_g$ matrix. The invariance of $ Z_{\text{grav}} (\Omega, \bar{\Omega})$ under the 
action of the MCG implies that
\begin{align}
    \rho_g(\gamma)^{\dag} B \rho_g(\gamma) = B,
    \label{eq:M_modular_condition}
\end{align}
for any $\gamma \in \Gamma_g$. 
Importantly, as we will prove later in this subsection, the projective representation $\rho_g$ of the MCG $\Gamma_g$ is {\it irreducible}. As a consequence, by  Schur's lemma,
$B$ has to be proportional to the identity matrix to satisfy (\ref{eq:M_modular_condition}). Therefore, the full gravitational partition function satisfies
\begin{align}
    Z_{\text{grav}} (\Omega, \bar{\Omega}) \propto \sum_{i=1}^{N_g}  \bar{\chi}_{i}(\bar{\Omega})  \chi_i(\Omega).
    \label{eq:Zgrav_diagonal_modular_invariance}
\end{align}
In the following, we will present the proof of the irreducibility of the MCG representation $\rho_g$.

\begin{figure}[htbp]
\centering
\includegraphics[width=0.8\textwidth]{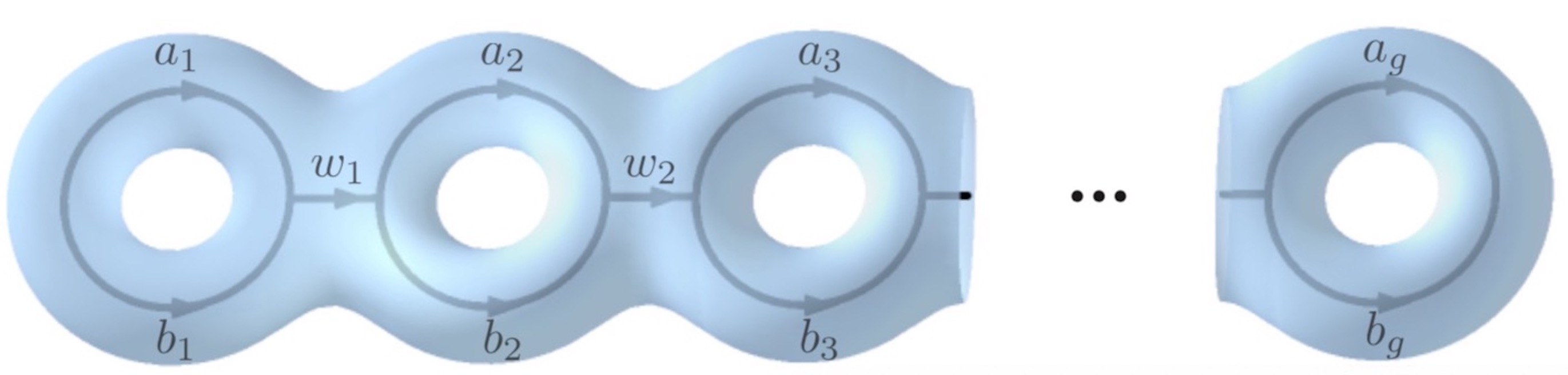}
\caption{$a_{1,2,...,g}~,~b_{1,2,...,g}~,~w_{1,2,...,g-1}\in\{1,\sigma,\psi\}$. Fusion rules need to be applied at each trivalent vertex for the fusion diagram to be admissible.}
\label{fig:ConformalBlock}
\end{figure}

First, we review the connections between the 2D Ising CFT and 3D Ising TQFT that will be useful for the proof of the irreducibility of the representation $\rho_g$. On the genus-$g$ surface $\Sigma_g$, there are $N_g$ holomorphic conformal blocks in the 2D Ising CFT and there are $N_g$ orthogonal quantum states in the 3D Ising TQFT. Each of the holomorphic conformal blocks 
has a corresponding TQFT quantum state and vice versa. Each of holomorphic conformal blocks and its corresponding TQFT quantum state can be represented by an admissible fusion diagram as shown in Figure \ref{fig:ConformalBlock}. Each line in the fusion diagram is labeled by $1$, $\sigma$ or $\psi$. That is to say, in Figure \ref{fig:ConformalBlock}, all the labels $a_{1,2,...,g}$, $b_{1,2,...,g}$ and $w_{1,2,..,g-1}$ take values 
in the set $\{1, \sigma, \psi\}$. The labels $\{1, \sigma, \psi\}$ should be viewed as the labels for the primary fields in the 2D (chiral) Ising CFT and, equivalently, also as the labels for the anyons (or objects or particles) in the 3D Ising TQFT. 
Note that the lines in the fusion diagrams are also directed. In general, a directed line carrying an anyon label $a$ is equivalent to the line with the opposite direction and with the label $\bar{a}$, namely the label for the anti-particle of $a$. The directions of all the lines in Figure \ref{fig:ConformalBlock} are chosen merely as a convention. In fact, in 3D Ising TQFT, each of $1$, $\sigma$ and $\psi$ is its own antiparticle.  Therefore, it should not cause confusion even if we don't specify the directions of the lines in a fusion diagram in 
in the discussion below. Also, $1$ represents the trivial anyon in the 3D Ising TQFT and the trivial 
(identity) primary operator in the 2D Ising CFT. In the fusion diagram, a line labeled by $1$ can also be erased. Only a so-called admissible fusion diagram corresponds to a holomorphic conformal block or a TQFT quantum state on $\Sigma_g$. For the fusion diagram in Figure \ref{fig:ConformalBlock} to be admissible in the 2D Ising CFT or the 3D Ising TQFT, we first need to require $a_1 = b_1$ and $a_g = b_g$. Moreover, an admissible fusion diagram also requires each trivalent vertex to be admissible. Each trivalent vertex has two incoming (outgoing) lines and one outgoing (incoming) line. If the anyons $a$ and $b$ labeling the two incoming (outgoing) lines have a fusion product $a\times b$ that contains the anyon $c$ labeling the one outgoing (incoming) line, the trivalent vertex is admissible. The full set of fusion rules of the 3D Ising TQFT (or the 2D Ising CFT) is given by 
\begin{align}
& 1\times 1= 1,~~~1\times \sigma = \sigma ,~~~ 1\times \psi = \psi, \nonumber \\
& \psi\times 1 = \psi,~~~\psi \times \sigma = \sigma,~~~\psi \times \psi =1,  \\
&\sigma \times 1 = \sigma,~~~\sigma \times \psi = \sigma,~~~ \sigma \times \sigma = 1+ \psi \nonumber.
\end{align}
One can directly show (see below)
that there are $N_g$
admissible fusion diagrams (with different anyon labels $a_{1,2,..,g}$, $b_{1,2,..,g}$ and $w_{1,2,..,g-1}$) of the form shown in Figure \ref{fig:ConformalBlock}, where
$N_g=2^{g-1}(2^{g-1}+1)$
is the dimension of the representation
$\rho_g$ of the MCG discussed above.

We will denote the Ising TQFT quantum state (and its correspond Ising-CFT conformal block) 
by the corresponding fusion diagram labels. For example, the Ising TQFT quantum state associated to the fusion diagram shown in Figure \ref{fig:ConformalBlock} will be denoted as $\Kabw$. Physically, in the language of 3D TQFT, one can think of an admissible fusion diagram as describing the world lines of anyon. Therefore, 
in the discussion below, 
we will also refer to a fusion diagram as an anyon diagram. The correspondence between the state $\Kabw$ and its fusion diagram can be understood as follows. The state $\Kabw$ on $\Sigma_g$ can be viewed as generated by the 3D Ising TQFT path integral on a genus-$g$ handlebody $H_g$ such that $\partial H_g = \Sigma_g$, and such that
the corresponding fusion diagram (or anyon diagram) is embedded in the core of $H_g$ (in the same configuration as shown in Figure \ref{fig:ConformalBlock}). In particular, there is a ``special'' state $|\vac\rangle \equiv | \{a_i=1\},\{b_i=1\},\{w_i=1\}\rangle $ with all of the labels on the fusion diagram set to be $1$. 
The state 
$|\vac\rangle$ can be viewed as the result of the Ising TQFT path integral on the handlebody $H_g$ without an anyon diagram inside 
(remember that anyon lines labeled by $1$ can be erased). The so-defined TQFT state $|\vac\rangle$ corresponds to the holomorphic vacuum conformal block $\chi_{\vac} (\Omega)$ of the 2D Ising CFT.

\begin{figure}
    \centering
    \includegraphics[width=0.2\textwidth]{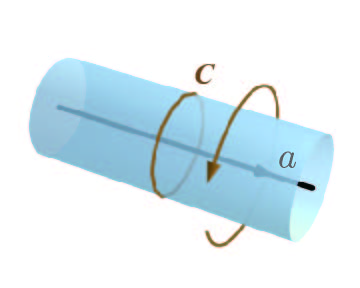}
    \caption{The Dehn twist along the non-contractible loop $C$ only yields a $U(1)$ phases $e^{i2\pi h_a}$ that depends on the label $a$. }
    \label{fig:single_Dehn}
\end{figure}

Because of the correspondence between the 
holomorphic conformal blocks of the 2D Ising CFT
and the states on $\Sigma_g$ of the 3D Ising TQFT, the MCG $\Gamma_g$ acts on the states $\Kabw$ via the same representation $\rho_g$. The $\Gamma_g$ action on the Ising-TQFT states can also be understood as follows. In the picture where the Ising-TQFT states are generated by the Ising TQFT path integral on a handlebody $H_g$ with an anyon diagram, the MCG action on $\Sigma_g = \partial H_g$ should be extended to the whole handlebody $H_g$. Such an extended action of $\Gamma_g$ 
deforms the anyon diagram inside $H_g$. 
The deformed anyon diagram can be rewritten in terms of a linear superposition of anyon diagrams of the original shape shown in Figure  \ref{fig:ConformalBlock} with different anyon labels. That is to say that when a state $\Kabw$ is acted 
on by an element $\gamma\in\Gamma_g$ of the MCG, the resulting state is in general a superposition of many states with different anyon labels in their fusion diagrams:
\begin{align}
& \rho_g (\gamma) \Kabw  \nonumber \\
& = \sum_{a_i', b_i', w_i'=\{1, \sigma,\psi\}} 
\langle \{a_i'\}, \{b_i'\}, \{w_i'\} |\rho_g (\gamma) \Kabw 
\ \ ~| \{a_i'\}, \{b_i'\}, \{w_i'\}  \rangle.
\end{align}
A particularly simple case is when the MCG action is a Dehn twist $\nu_C$ along a loop $C$ that 
is threaded by a single anyon line labeled by $a$ (as is shown in Figure \ref{fig:single_Dehn}). Such a Dehn twist does not change the shape of the anyon diagram, the action $\rho_g(\nu_C)$ only yields extra $U(1)$ phase $e^{i2\pi h_a}$ on the state represented by the anyon diagram, where $h_a$ depends on the anyon label $a$:
\begin{align}
    h_1 = 0, ~~ h_\sigma = 1/16,~~ h_\psi = 1/2.
\end{align}
Here $h_a$ can be viewed as the conformal weight of the primary field labeled by $a$ in the 2D (chiral) Ising CFT. Also, in the 3D Ising TQFT language, we can view $e^{i2\pi h_a}$ as the topological spin of the anyon labeled by $a$.

In the following, we will show that $\rho_g$ is an irreducible projective representation of the MCG $\Gamma_g$. In fact, the irreducibility of $\rho_g$ is equivalent to the statement that the $\mathbb{C}$-linear matrix algebra $\mathbb{C}[\rho_g]$ generated by $\rho_g\left(\Gamma_g\right)$ (through addition and matrix multiplication) is identical to the
{\it full} 
matrix algebra $\M_{N_g}$ of all $N_g \times N_g$ complex matrices, namely $\mathbb{C}[\rho_g] \cong \M_{N_g}$. Obviously,  $\mathbb{C}[\rho_g] \subseteq \M_{N_g}$. Therefore, what we need to prove is that $\M_{N_g} \subseteq \mathbb{C}[\rho_g]$. The strategy 
of the proof 
is to explicitly construct all the operators of the form 
$\Kabw \  \langle \{a_i'\}, \{b_i'\}, \{w_i'\} |$  
within $\mathbb{C}[\rho_g]$.

\begin{figure}
    \centering
    \includegraphics[width=0.7\textwidth]{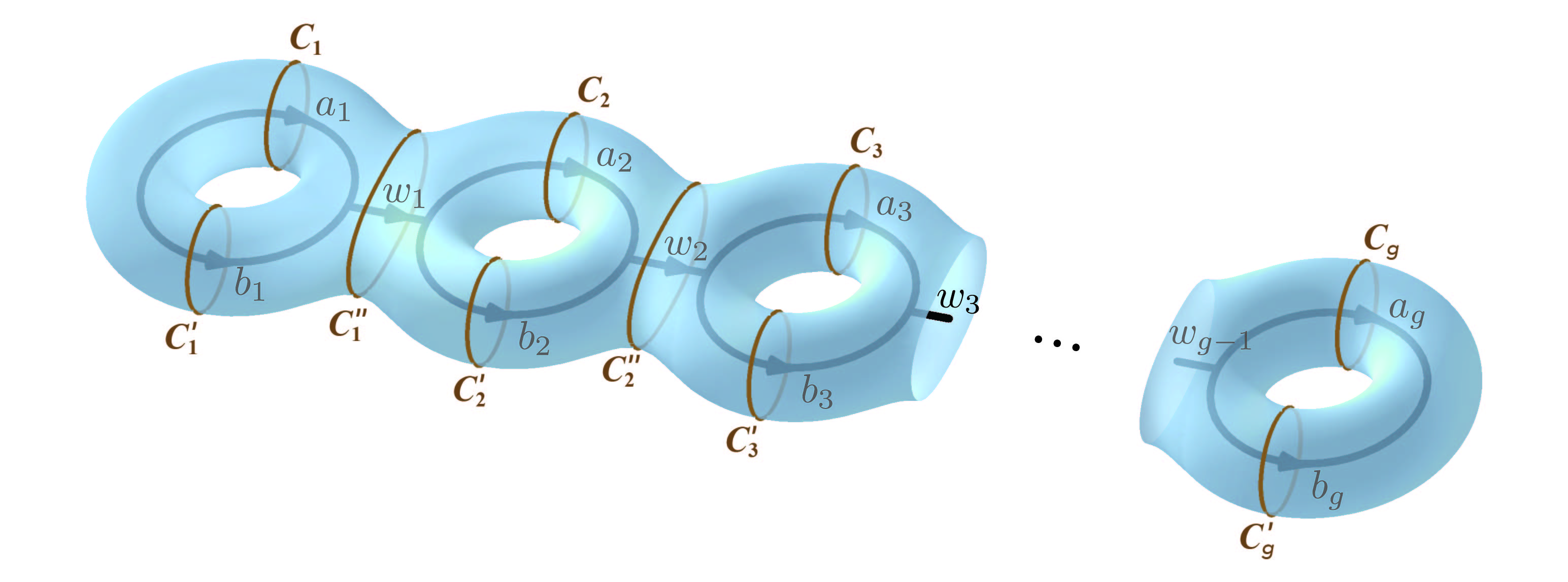}
    \caption{The Dehn twists along the loops $C_{1,2,..,g}$, $C'_{1,2,..,g}$ and $C''_{1,2,..,g-1}$ can distinguish all of the states $\Kabw$}
    \label{fig:DehnG}
\end{figure}

We will first construct the projection operators $\Kabw \Babw$, which will be denoted as $P_{\abw}$ in the following discussion, for any state $\Kabw $. For this purpose, we can focus on the set of non-intersecting loops $C_{1,2,..,g}$, $C'_{1,2,..,g}$ and $C''_{1,2,..,g-1}$ shown in Figure \ref{fig:DehnG}. The Dehn twists $\nu_{C_i}$, $\nu_{C_i'}$ and $\nu_{C_i''}$ along each of these loops commute with each other. A state $\Kabw $ with a fixed set of labels $a_{1,2,..,g}$, $b_{1,2,..,g}$ and $w_{1,2,..,g-1}$ is a simultaneous eigenstate of all such Dehn twists:
\begin{align}
& \rho_g(\nu_{C_j}) \Kabw = e^{i2\pi h_{a_j}} \Kabw,~~~~~{\rm for}~j=1,2,...,g \nonumber \\
& \rho_g(\nu_{C'_j}) \Kabw = e^{i2\pi h_{b_j}} \Kabw,~~~~~{\rm for}~j=1,2,...,g \\
& \rho_g(\nu_{C''_j}) \Kabw = e^{i2\pi h_{w_j}} \Kabw,~~~~~{\rm for}~j=1,2,...,g-1. \nonumber 
\end{align}
Since 
$e^{i2\pi h_1}$, $e^{i2\pi h_\sigma}$, $e^{i2\pi h_\psi}$ are all different, one can use the set of Dehn twists $\nu_{C_i}$, $\nu_{C_i'}$ and $\nu_{C_i''}$ to fully distinguish all the states $\Kabw$. Building on this, we can construct the following projection operators 
associated with any loop 
$C$ and an anyon label $1$, $\sigma$, or $\psi$ within $\mathbb{C}[\rho_g]$:
\begin{align}
    & P_1 (C) \equiv \frac{1}{16} \sum_{n=1}^{16} \rho_g\left(\nu_{C}^n\right),~~~~  \nonumber \\
    & P_\sigma (C) \equiv \left(1-e^{2\pi i/8}\right)^{-1}\left(\mathds{1} - \rho_g\left(\nu_{C}^2\right)  \right),~~~~ \\
    & P_\psi (C) = \mathds{1} - P_1 (C) - P_\sigma (C), \nonumber
\end{align}
where $\nu_C \in \Gamma_g$ represents the Dehn twist along the loop $C$ and $\mathds{1}$ represents the $N_g \times N_g$ identity matrix. Choosing $C$ to be $C_j$, $C_j'$ or $C_j''$, we see that 
\begin{align}
& P_a (C_j)  \Kabw = \delta_{a,a_j} \Kabw,~~~~~{\rm for}~j=1,2,...,g \nonumber \\
& P_b (C_j') \Kabw = \delta_{b,b_j} \Kabw,~~~~~{\rm for}~j=1,2,...,g \\
& P_w (C_j'')  \Kabw = \delta_{w,w_j}  \Kabw,~~~~~{\rm for}~j=1,2,...,g-1, \nonumber 
\end{align}
where 
$a_i,b_i,w_i\in \{1,\sigma,\psi\}$. 
Any projection operator $P_{\abw}$ 
onto a given state $\Kabw$ can then be written as a product of $P_a (C_j)$, $P_b (C_j')$ and $P_w (C_j'')$. Therefore, 
all projection operators $P_{\abw}$ belong 
to $\mathbb{C}[\rho_g]$.

Next, we will show that all the operators of the form $|\vac \rangle \Babw $ can be constructed within $\mathbb{C}[\rho_g]$. Upon 
inspection we observe 
that in any admissible fusion diagram of the form shown in Figure \ref{fig:ConformalBlock}, the labels $w_i$ for $i=1,2,...,g-1$ can only take values $1$ or $\psi$. We will first focus on the case with $w_i = 1$ for all
$i=1,2,...,g-1$. In this case, an admissible diagram further requires $a_i = b_i$ for all $i=1,2,...,g$. Therefore, the relevant states in this case are of the form $|\{a_i\}, \{b_i =a_i\}, \{w_i=1\}\rangle$, which will be denoted by
$|\{a_i\}\rangle$ in short hand in the following discussion. The anyon diagram of $|\{a_i\}\rangle$, after we have erased all the lines carrying label $1$, is simply a disjoint union of anyon loops labeled by $a_{1,2,...,g}$. To construct an operator of the form $|\vac \rangle \langle \{a_i\}|$ in $\mathbb{C}[\rho_g]$, it is sufficient to find an MCG element $\gamma$ such that $\langle \vac | \rho_g(\gamma) |\{a_i\} \rangle \neq 0$ which allows us to write $|\vac \rangle \langle \{a_i\}| $ as $| \vac \rangle \langle \vac | \rho_g(\gamma) | \{a_i\}\rangle\langle \{a_i\}|$ up to a non-zero multiplicative constant. Remember that we have already constructed the operators
$| \vac \rangle \langle \vac |$ and $| \{a_i\}\rangle\langle \{a_i\}|$ within $\mathbb{C}[\rho_g]$. Therefore, we only need to find the suitable MCG element $\gamma$. 

In principle, the choice of $\gamma$ can depend on the state $| \{a_i\}\rangle$. Interestingly, we can show that there is a specific MCG element $\gamma_0 \in \Gamma_g$ that works for all $| \{a_i\}\rangle$. 
The MCG element $\gamma_0$ can be identified as follows. Consider the disjoint union of two copies 
of a genus-$g$ handlebody $H_g$ and $H_g'$ whose  boundaries are given
by two identical copies $\Sigma_g$
and $\Sigma_g'$ of the {\it same} Riemann surface, i.e.,
$\Sigma_g = \partial H_g$ and $\Sigma_g' = \partial H_g'$. In general, we can perform a MCG action $\gamma \in\Gamma_g$ on $\Sigma_g'$ and then glue it to $\Sigma_g$. This procedure glues the two genus-$g$ handlebodies $H_g$ and $H_g'$ into a single closed 3-manifold that depends on the choice of $\gamma$. There exists an element $\gamma_0$ such that the resulting closed 3-manifold is
the 3-sphere $S^3$. We will show that $\langle \vac | \rho_g(\gamma_0) |\{a_i\} \rangle \neq 0$ for any states $|\{a_i\} \rangle$. Again, consider the setup with two copies $H_g$ and $H'_g$ of the genus-$g$ 
handlebody. Performing the 3D Ising-TQFT path integral on $H_g$ (without any anyon diagram) yields the state $|\vac\rangle$ on its boundary $\partial H_g=\Sigma_g$. Now, we embeded the anyon diagram of $| \{a_i\}\rangle$, which is a collection of disjoint anyon loops labeled by $a_{1,2,..,g}$ respectively, in $H_g'$. The TQFT path integral on $H_g'$ then yields the state $|\{a_i\} \rangle$ on its boundary $\partial H_g' =\Sigma_g'$. When $\Sigma'_g$ is acted on by
$\gamma_0$ and then glued to $\Sigma_g$, we obtain a 3D Ising TQFT path integral on $S^3$ together with the anyon diagram that was originally embedded in $H_g'$. The result of such a path integral is exactly $\langle \vac | \rho_g(\gamma_0) |\{a_i\} \rangle $. Since the anyon diagram involved here is a disjoint union 
of anyon loops labeled by $a_{1,2,..,g}$ respectively, the Ising TQFT path integral on $S^3$ with such anyon diagrams is definitely non-vanishing. Therefore, 
\begin{align}
    \langle \vac | \rho_g(\gamma_0) |\{a_i\} \rangle \neq 0,
\end{align}
for any choice of $a_{1,2,...,g}$. Consequently, we can conclude that the operators of the form $|\vac\rangle \langle \{a_i\}| $ all belong to $\mathbb{C}[\rho_g]$. By Hermitian conjugation, the any operator of the form  $|\{a_i\}\rangle \langle \vac| $ also belongs to $\mathbb{C}[\rho_g]$.

\begin{figure}
    \centering
    \includegraphics[width=0.7\textwidth]{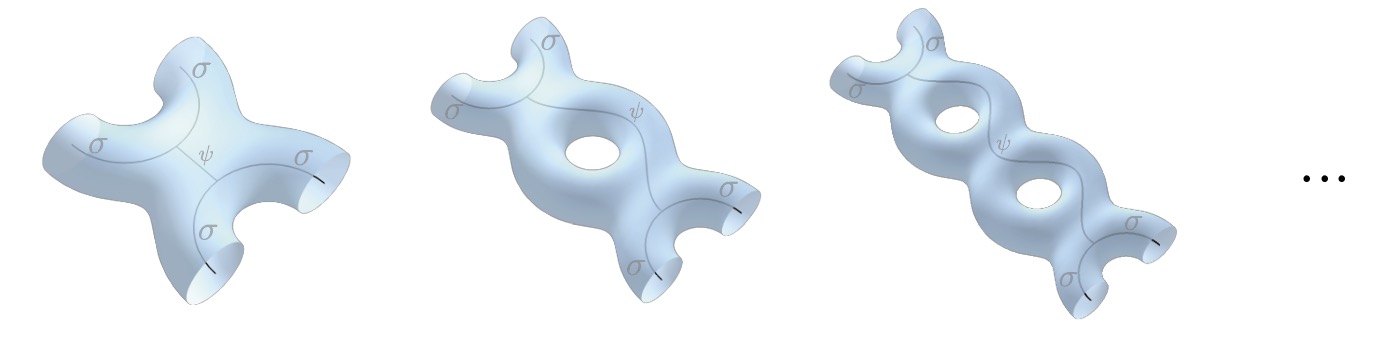}
    \caption{Local configurations of fusion diagrams with some of the $w_i$ labels taking the value $\psi$}
    \label{fig:psi}
\end{figure}

Now, we are ready to construct the operators $|\vac \rangle \Babw $ with some of the $w_i$ labels equal to $\psi$. When some of the $w_i$ labels equal to $\psi$, the anyon diagram associated to $\Kabw$ must be in one of the configurations shown in Figure \ref{fig:psi} in the vicinity of the diagram where the $w_i$ labels take the value $\psi$. In the Ising TQFT, we have the following linear relations between the diagrams
\begin{align}
\raisebox{-0.4 \height}{\includegraphics[width=2cm]{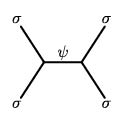}}~~ \propto ~~
\left(
\raisebox{-0.4 \height}{\includegraphics[width=2cm]{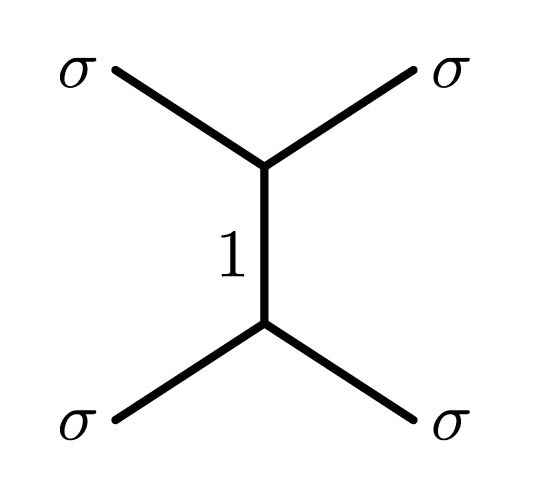}} - \raisebox{-0.4 \height}{\includegraphics[width=2cm]{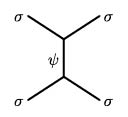}} \right)= \raisebox{-0.4 \height}{\includegraphics[width=2cm]{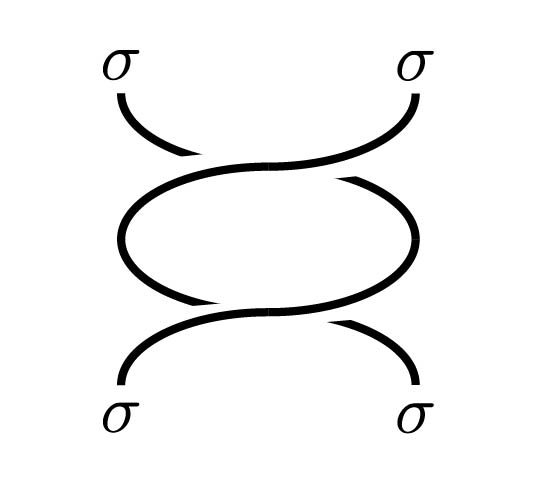}},
\label{eq:FRmove}
\end{align}
which can help us relate an anyon diagram with some of the $w_i$ labels equal to $\psi$ to another diagram with less of the $w_i$ labels equal to $\psi$. For example, the 
leftmost
configuration shown in Figure \ref{fig:psi} obeys 
\begin{align}
\raisebox{-0.4 \height}{\includegraphics[width=3cm]{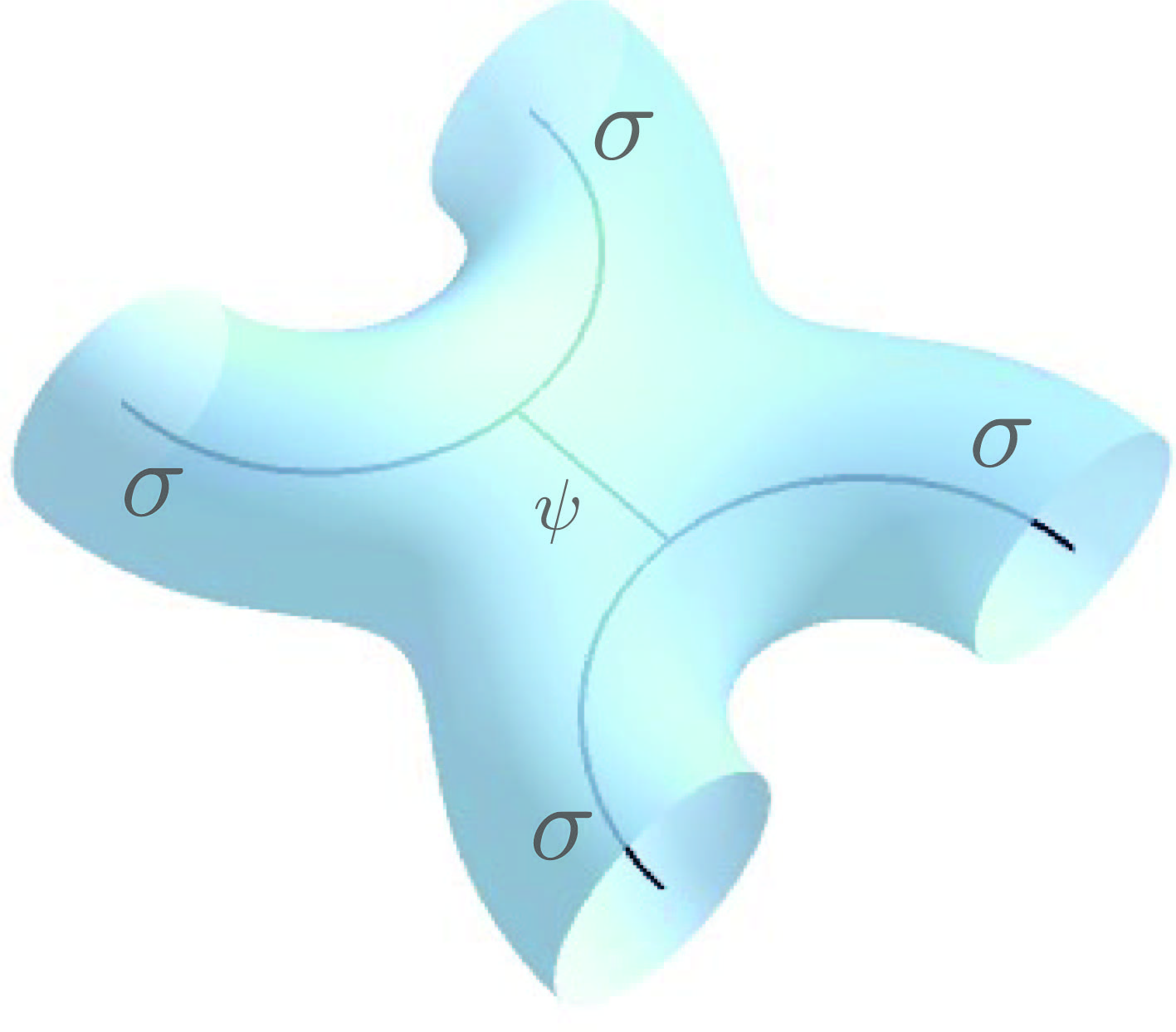}}~~ \propto ~~
\raisebox{-0.4 \height}{\includegraphics[width=3cm]{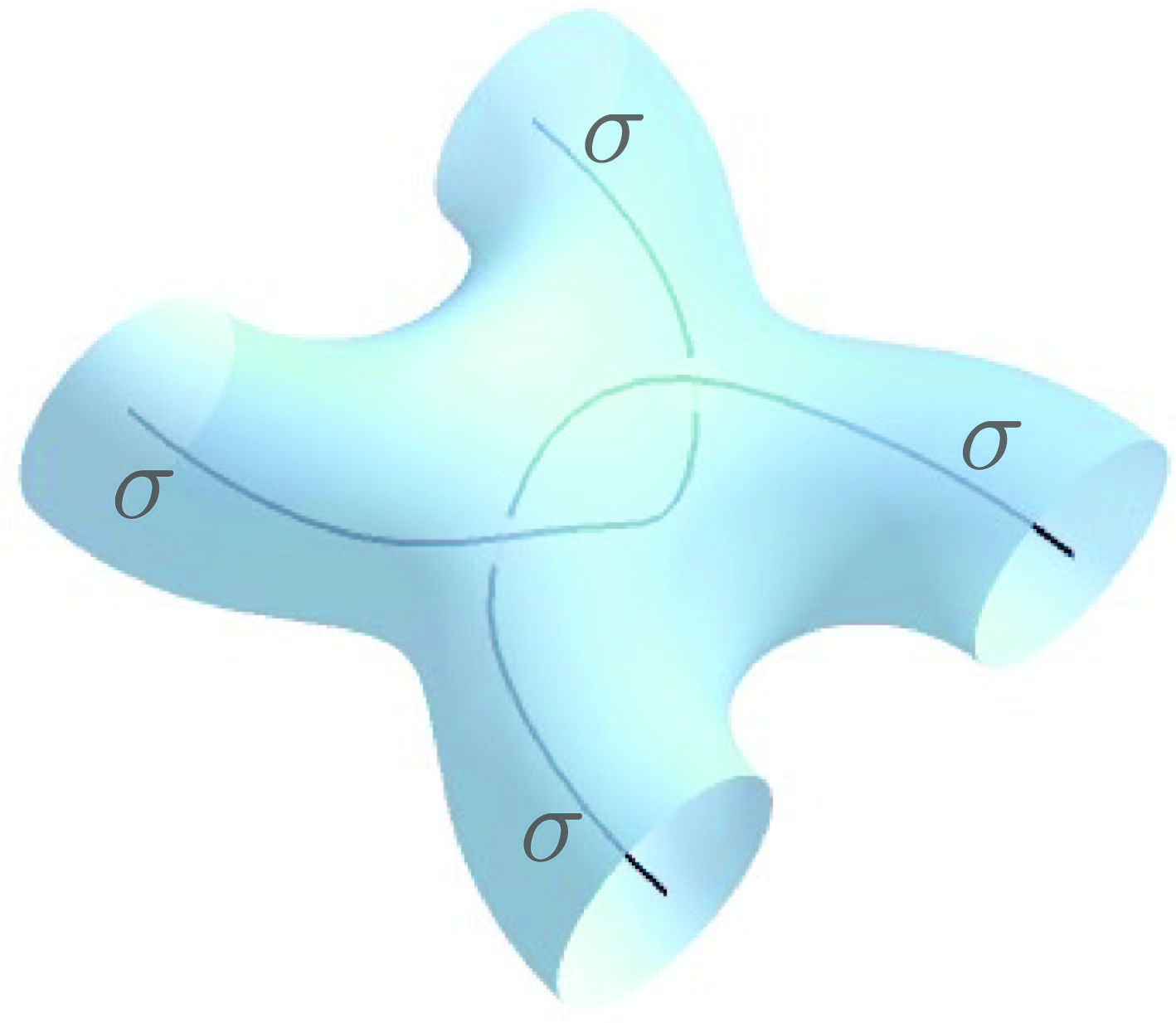}}~~ =~~  \raisebox{-0.4 \height}{\includegraphics[width=3cm]{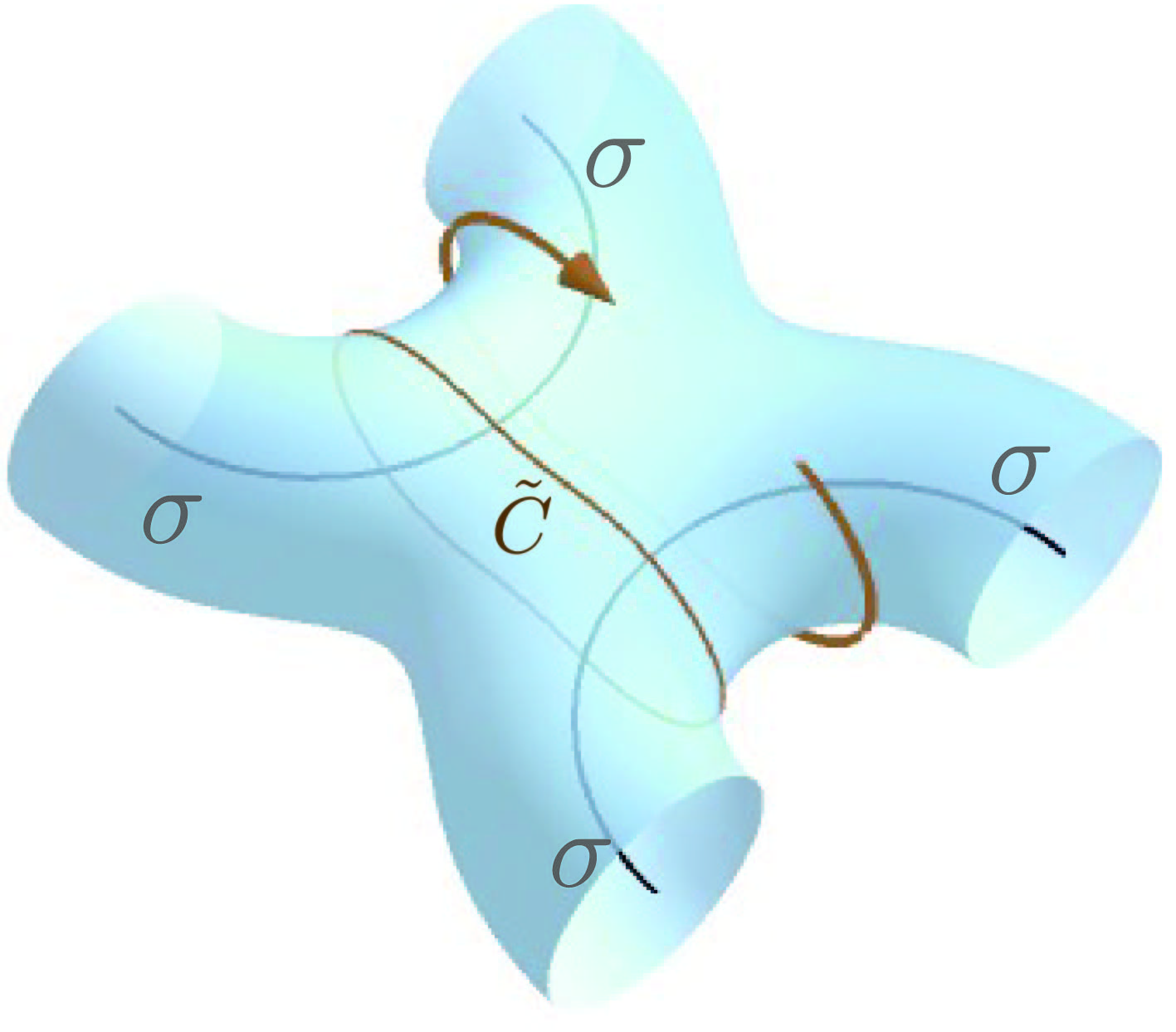}},
\label{eq:FRmove_PsiToOne_1}
\end{align}
where the relation between the first two diagrams is a graphical representation of the relation (\ref{eq:FRmove}). The last equality in \eqref{eq:FRmove_PsiToOne_1} means that a Dehn twist along the loop $\tilde{C}$ can transform 
the rightmost diagram shown in \eqref{eq:FRmove_PsiToOne_1}, before it was acted on by the  Dehn twist, to  the diagram shown in the middle of the same equation. Remember that the Dehn twist along $\tilde{C}$ on 
the surface should be extended into the interior of the handlebody leading to the transformation from the third diagram to the second in (\ref{eq:FRmove_PsiToOne_1}). Thus, Equation 
(\ref{eq:FRmove_PsiToOne_1}) shows an example to use Dehn twists to relate a diagram with a $w_i$ label equal to $\psi$ to another diagram without such a $w_i$ label. A similar relation can also be obtained for the second configuration shown in Figure \ref{fig:psi}:
\begin{align}
\raisebox{-0.4 \height}{\includegraphics[width=3cm]{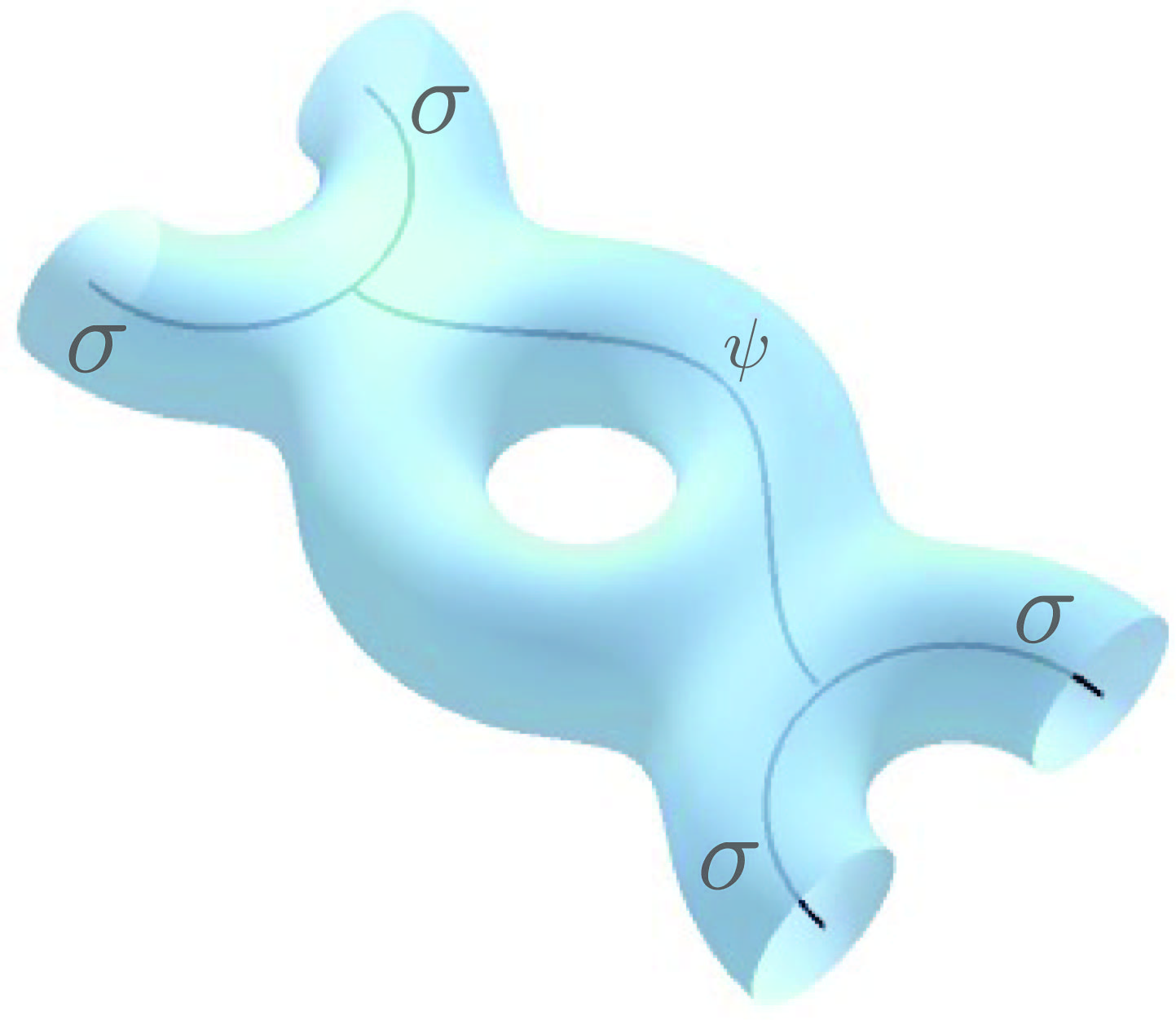}}~~ \propto ~~
\raisebox{-0.4 \height}{\includegraphics[width=3cm]{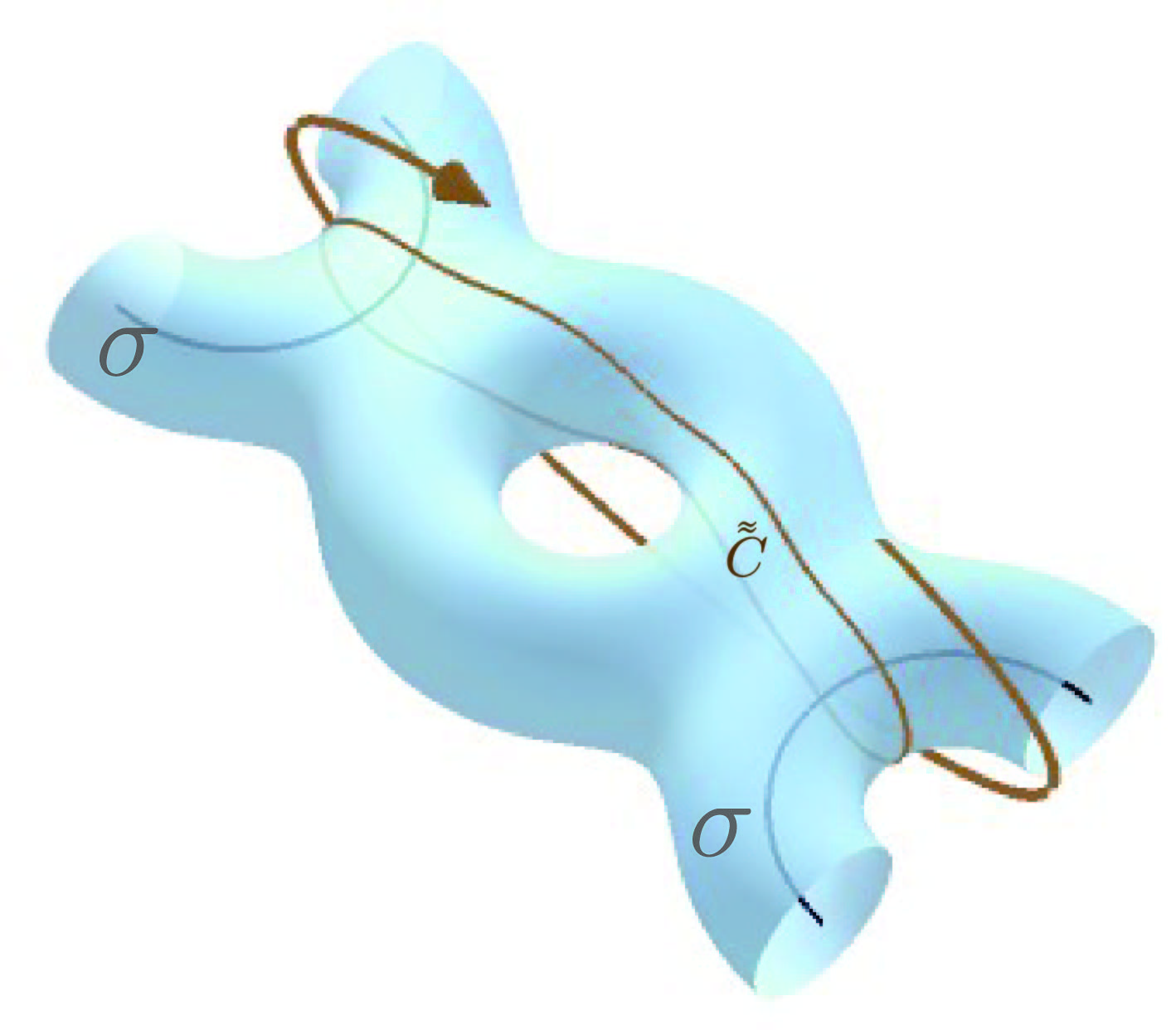}},
\label{eq:FRmove_PsiToOne_2}
\end{align}
where a Dehn twist along $\tilde{\tilde{C}}$ is performed. In fact, similar procedures can be carried out on all of the 
configurations shown in Figure \ref{fig:psi} (and their generalizations that are not depicted). Therefore, all of the states $\Kabw$ with some $w_i$ labels equal to $\psi$ can be obtained from the states the states without such $w_i$ labels, i.e., the states $|\{a_i\}\rangle$, 
by applying one or a sequence of Dehn twists of the type shown above. Consequently, all the operators $|\vac \rangle \Babw $ and $\Kabw \langle \vac| $
can be obtained from multiplying the operators of the form $|\vac\rangle\langle\{a_i\}|$ or $|\{a_i\}\rangle\langle\vac|$ with the unitary operators associated to the proper set of Dehn twists. 

Having constructed all of the operators $|\vac\rangle \Babw$ and $\Kabw\langle\vac|$ (regardless of the value of the $w_i$ labels) within $\mathbb{C}[\rho_g]$, we can simply obtain
via matrix multiplication operators of the more general form $\Kabwp\Babw$, which 
form a complete basis for the full matrix algebra $\mathbb{M}_{N_g}$, within 
$\mathbb{C}[\rho_g]$. At this point, we have completed the proof for $\mathbb{C}[\rho_g] \cong \mathbb{M}_{N_g} $ and, hence, for the irreducibility of the 
(projective) representation $\rho_g$ of the MCG 
for a general $g$.

\subsection{Duality to 2D Ising CFT}
\label{sec:relation}
In Section \ref{sec:genus_g_vacuum_seed}, we proposed the expression (\ref{eq:Zvac_MultiGenus}) for the vacuum seed in terms of the product of the holomorphic and anti-holomorphic vacuum conformal blocks of the 2D Ising CFT. Based on this
proposed vacuum seed, we proved the finiteness of the ``gravitational'' modular sum (\ref{eq:Zgrav_MultiGenus}) in Section \ref{sec:finiteness_modular_sum} and obtained
the final expression (\ref{eq:Zgrav_diagonal_modular_invariance}) of
the gravitational partition function
$Z_{\rm grav}$ up to a multiplicative constant in Section \ref{sec:irreducibility}. We need to emphasize that, in our discussion, the result (\ref{eq:Zgrav_diagonal_modular_invariance}) is purely a consequence of our arguments for the vacuum seed $Z_{\rm vac}$ which
were made from the gravity bulk perspective, as well as of the mathematical results that we proved including the finiteness of $\rho_g(\Gamma_g)$ and the irreducibility of the MCG representation $\rho_g$. In fact, even if $Z_{\rm vac}$ is not of the form (\ref{eq:Zvac_MultiGenus}), as long as it be written as a product of a holomorphic and an anti-holomorphic conformal block 
of the 2D Ising CFT (or even as a sum of  products of this type), 
we can still conclude the finiteness of the modular sum (\ref{eq:Zgrav_MultiGenus}) and further obtain the {\it same}
expression (\ref{eq:Zgrav_diagonal_modular_invariance}) 
for $Z_{\rm grav}$, based on our mathematical results, i.e., the finiteness of $\rho_g(\Gamma_g)$ and the irreducibility of $\rho_g$. 

The right hand side of (\ref{eq:Zgrav_diagonal_modular_invariance}) can also be naturally identified with the (full) partition function of the 2D Ising CFT on the Riemann surface $\Sigma_g$ with period matrix $\Omega$. We therefore conclude that, 
at Brown-Henneaux central charge $c=1/2$, 
and for genus $g$, the full gravitational partition function $Z_{\rm grav} (\Omega, \bar{\Omega})$ with a genus-$g$ asymptotic boundary $\Sigma_g$ is always proportional to the partition function of the 2D Ising CFT $Z_{\rm Ising} (\Omega, \bar{\Omega})$ on $\Sigma_g$: 
\begin{align}
    Z_{\rm grav} (\Omega, \bar{\Omega}) \propto \sum_{i=1}^{N_g} \chi_i(\Omega) \bar{\chi}_i (\bar{\Omega}) = Z_{\rm Ising}(\Omega, \bar{\Omega}).
    \label{eq:ZadsPropToZising}
\end{align}
\begin{figure}
    \centering
    \includegraphics[width=0.4\textwidth]{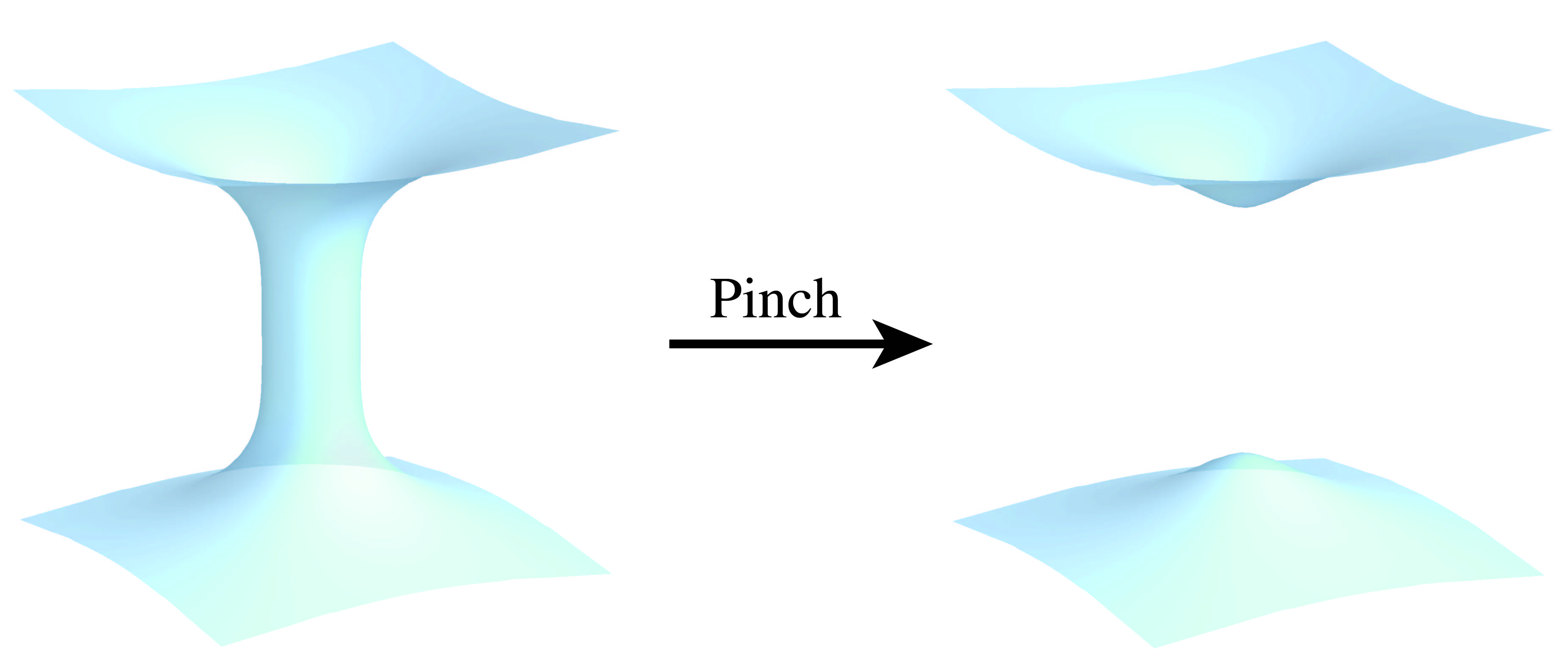}
    \caption{A schematic picture for pinching off a long cylinder.}
    \label{fig:Pinch}
\end{figure}

At this point, we would like to come back to our proposed expression (\ref{eq:Zvac_MultiGenus})
for the vacuum seed. 
In Section \ref{sec:genus_g_vacuum_seed}, we have already provided physical arguments that suggest that (\ref{eq:Zvac_MultiGenus}) is a natural expression for the vacuum seed. Now, we would like to further substantiate this proposal (\ref{eq:Zvac_MultiGenus}) by commenting on the resulting gravitational partition function $Z_{\rm grav}$ (\ref{eq:ZadsPropToZising}). (\ref{eq:ZadsPropToZising}) is a sensible result from the following perspectives. Firstly, the gravitational partition function (\ref{eq:ZadsPropToZising}) for arbitrary genus $g$ is compatible with and is the natural extension of the genus-one result obtained in \cite{Castro}. Secondly, the gravitational partition function (\ref{eq:ZadsPropToZising}), 
in the ``pinching limits'', is self-consistent and is consistent with the genus-one result obtained in \cite{Castro}. The pinching limit we focus on here is the limit of the period matrix $\Omega$ of the asymptotic boundary $\Sigma_g$ 
in which some part of the asymptotic boundary $\Sigma_g$ is stretched into a very long cylinder and eventually
can be effectively viewed as pinched off.  
Figure \ref{fig:Pinch} is a schematic picture for pinching off a long cylinder. In the gravity context, such a pinching limit has 
previously only been investigated, to the best of our knowledge, in
semi-classical gravity \cite{Yin1,Pinching}.
With the gravitational partition function given by (\ref{eq:ZadsPropToZising}), we can now study the pinching limit of strongly coupled gravity 
with Brown-Henneaux central charge $c=1/2$. In the pinching limit, intuitively, we expect the genus of the asymptotic boundary to be effectively reduced by 1. Hence, we expect a reduction 
of a gravitational partition function with a genus-$g$
boundary to one with a genus-$(g-1)$ boundary. 
This physical intuition is indeed consistent with (\ref{eq:ZadsPropToZising}), since the partition function of 2D Ising CFT on a genus-$g$ surface indeed reduces to that on a genus-$(g-1)$ surface in the pinching limit \cite{DVV}.

Starting with a genus-$g$ asymptotic boundary, we can take successive pinching limits such that the  genus of the resulting asymptotic boundaries
eventually reduces to $g=1$. In this case,  (\ref{eq:ZadsPropToZising}) implies that the gravitational partition function eventually 
reduces, up an overall multiplicative constant, to the product of genus-one partition functions of the 2D Ising CFT. 
This result is again consistent with \cite{Castro}. Here, 
we have provided general arguments for the behavior of the gravitational partition function 
in the pinching limits using (\ref{eq:ZadsPropToZising}) and using the behavior of the 2D Ising CFT partition 
in the same limit. In Appendix \ref{app:Ising}, we provide an example of analytic studies of the pinching limit of the gravitational partition function with genus-two asymptotic boundaries. Having provided arguments that substantiate the result (\ref{eq:ZadsPropToZising}) (and thereby its starting point (\ref{eq:Zvac_MultiGenus})), we would like to conclude that based on a natural choice for the vacuum seed,
we establish duality between 3D AdS quantum gravity at Brown-Henneaux central charge $c=1/2$ and the 2D Ising CFT using the all-genus partition functions. 

Besides the gravitational partition function, the mathematical result that $\rho_g$ is an irreducible projective representation of the MCG $\Gamma_g$ of the Riemann surface $\Sigma_g$ 
for any $g$ also has an interesting implication purely 
for the 2D Ising CFT. It was proven in \cite{Cappelli} that, up to a multiplicative constant, there is only one unique modular invariant partition function that can be constructed using 2D Ising CFT conformal blocks on a genus-one surface. To the best of our knowledge, there is no generalization of such a proof to higher-genus surfaces in previous works. Our result that $\rho_g$ is an irreducible representation of $\Gamma_g$ implies that, up to a multiplicative constant, for any fixed genus $g$, there is always 
a unique partition function constructed from Ising-CFT conformal blocks that is invariant under the MCG $\Gamma_g$ action on a genus-$g$ surface.

\subsection{Difficulty in extending beyond the 2D Ising CFT}
\label{sec:difficulty}

Solely based on the consideration of gravitational partition functions with genus-1 asymptotic boundary, Castro et al. \cite{Castro} argued that, for Brown-Henneaux central charge $c<1$, the only 2D CFTs that can be dual to pure Einstein gravity in AdS$_3$ at the corresponding Brown-Henneaux central charge $c$ are the Ising and the Tricritical Ising CFTs of central charges $c=1/2$ and $c=7/10$. Our results obtained in the previous subsections on the all-genus partition functions has established duality between 3D AdS quantum gravity at Brown-Henneaux central charge $c= 1/2$ and the 2D Ising CFT.

However, as we will show later in this subsection, the consideration of higher-genus partition functions at $c= 7/10$ reveals a fundamental difficulty in establishing duality between 3D gravity at Brown-Henneaux central charge $c= 7/10$ and the Tricritical Ising CFT.

At Brown-Henneaux central charge $c= 7/10$, we can follow the same reasoning as in Section \ref{sec:genus_g_vacuum_seed} to argue that the corresponding gravitational vacuum seed at $c=7/10$ with a genus-$g$ asymptotic boundary $\Sigma_g$ should be identified as the vacuum conformal block of the 2D Tricritical Ising CFT on $\Sigma_g$. Hence, the modular sum (\ref{eq:Zgrav_MultiGenus}) at $c= 7/10$ is dictated by the MCG $\Gamma_g$ representation $\rho_g'$ that governs how the holomorphic conformal blocks of the 2D Tricritical Ising CFT transform under the action of $\Gamma_g$. In complete analogy with the connection between the 2D Ising CFT and the 3D Ising TQFT, the information about the MCG representation $\rho_g'$ (which is associated with the holomorphic conformal blocks of the 2D Tricritical Ising CFT) is fully contained in the 3D Tricritical Ising TQFT, which can be mathematically described, equivalently, by 
the chiral Tricritical Ising Modular Tensor Category (MTC) \cite{MooreSeiberg, Kirillov}.

By inspection, this MTC contains a sub-Modular Tensor Category called $\mathrm{Fib}$,
with the so-called Fibonacci Fusion Rules\footnote{The MTC $\mathrm{Fib}$ has only two labels ($=$``simple objects'' or ``particles'' or ``anyons'' types) which, when denoted by $1$ and $x$, possess the Fusion Rules $x\times x=1+ x$, $1\times x=x \times 1 =x$, $1\times 1 = 1$. Strictly speaking, the 
MTC Fib that is needed here is the conjugate of the
MTC Fib typically used in the literature \cite{WangBook}.}
(see, e.g., \cite{WangBook}). It follows from a very general Theorem by M\"{u}ger \cite{Muger} that the Tricritical Ising MTC at $c=7/10$ must then be the tensor product of the sub MTC $\mathrm{Fib}$ and the MTC associated to the chiral Ising CFT. This factorization implies that the MCG representation $\rho_g'$ given by the Tricritical Ising MTC must be a tensor product of an MCG representation given by the MTC $\mathrm{Fib}$ and an MCG representation given by the MTC of the 2D chiral Ising CFT. Each of these MCG representations mentioned here can be viewed as a map from the MCG to a unitary group.

Finally, a fundamental Theorem by Freedman, Larsen and Wang \cite{Freedman} states that the MCG representations given by the MTC
$\mathrm{Fib}$ has an infinite image set.\footnote{In fact, this image set is dense in a unitary group, a result that, as is well known, is related to the fundamental importance of the $\mathrm{Fib}$ MTC for the subject of fault-tolerant quantum computation.} It then immediately follows that the image set of $\rho_g'$ must also be infinite, i.e., 
the set $\rho_g'(\Gamma_g)$, where $\Gamma_g$ is the Mapping Class Group of the genus-$g$ Riemann surface,
is an infinite set. This result implies that at Brown-Henneaux central charge $c=7/10$, the modular sum of the  gravitational partition function in (\ref{eq:Zgrav_MultiGenus}) cannot be defined for genus $g\geq 2$
because the sum occurring in this equation has an infinite number of terms and cannot be naturally regularized, as discussed 
in Footnote \footref{foot:regularization}. 

\acknowledgments

We thank Jing-Yuan Chen, Scott Collier, Xi Dong, Michael H. Freedman, Ori J. Ganor, Thomas Hartman, Alexei Y. Kitaev, Thilo Kuessner, Xiao-Liang Qi, Shu-Heng Shao, Douglas Stanford, Yong-Shi Wu and Xi Yin for helpful discussions on various topics. H.-Y. S. thanks Semeon Artamonov, Filip Kos and Christopher Paciorek for help with parallel computing. 
C.-M. J. is supported by the Gordon and Betty Moore Foundations EPiQS Initiative through Grant GBMF4304. Z.-X. L. and H.-Y. S. are supported in part by the National Science Foundation under Grant No. NSF PHY-1748958 and by the Heising-Simons Foundation. Z. W. is partially supported by NSF grant  FRG-1664351. This work was supported in part by the National Science Foundation under Grant No. DMR-1309667 (A.W.W.L.). This work used the Savio computational cluster resource provided by the Berkeley Research Computing program at the University of California, Berkeley.

\appendix

\section{Genus one modular sum revisited}
\label{app:kernel}
This appendix presents some new results concerning the genus-one case, which aim to explain the mathematical meaning of the factor of eight in \eqref{eq:torusindex}. We will first introduce several necessary concepts.

As discussed in \cite{Bantay}, for any 2D rational CFT $\mathcal{C}$ with a finite set $\mathscr{I}$ of primaries, the field extension $F$ of $\mathbb{Q}$ by adjoining the all matrix elements of the modular $\mathcal{S}$ matrix as in \eqref{eq:modular} \cite{de Boer, Gannon} is a subfield of a cyclotomic field $\mathbb{Q}[\zeta_n]$ for some positive integer $n$, where $\zeta_n\equiv e^{2\pi i/n}$ is the primitive $n^{\textbf{th}}$ root of unity, by the Kronecker-Weber theorem in number theory.
Following the terminology in algebraic number theory, the smallest $n$ for which $F\subseteq \mathbb{Q}[\zeta_n]$ is called the {\it conductor} of $\mathcal{C}$ (also defined in \cite{GukovVafa}), and can be shown \cite{Bantay} to be equal to the {\it order} $N$ of modular $\mathcal{T}$ matrix. For the Ising CFT of interest to us here, the modular $\mathcal{T}$ matrix is listed in \eqref{eq:modular}.

Another important player for us is the {\it kernel} $\mathscr{K}$ of the {\it linear} representation of $SL(2,\mathbb{Z})$, defined as the set of modular transformations represented by the identity matrix,
\begin{equation}
\label{eq:kernel}
    \mathscr{K}=\left\{\gamma\in SL(2,\mathbb{Z})|\,M_{ij}(\gamma)=\delta_{ij}\right\},
\end{equation}
where $i,j\in \mathscr{I}$, and $M_{ij}(\gamma)$ is the representation matrix of $\gamma$ transforming between characters:
\begin{equation}
\label{eq:transformation}
    \chi_i\left(\gamma\cdot\tau\right)=\sum_{j}M_{ij}(\gamma)\chi_j(\tau).
\end{equation}
$\mathscr{K}$ can be shown \cite{Bantay,NgSch} to be a congruence subgroup of level $N$, whose meaning will be clear soon.

Now one can consider the index of the {\it principal} congruence subgroup $\Gamma(N)$ of $SL(2,\mathbb{Z})$ of level $N$ inside the kernel $\mathscr{K}$, i.e., the quantity $|\mathscr{K}:\Gamma(N)|$. $\Gamma(N)$ is defined as\footnote{For the {\it principal congruence subgroup} of a Siegel modular group $Sp(2g,\mathbb{Z})$ of level $N$, the definition is the group of diagonal matrices with entries being 1 mod $N$ \cite{Koecher,Krieg}.}
\begin{equation}
\Gamma(N)=\left\{
\begin{pmatrix}
a & b\\
c & d\\
\end{pmatrix}\in SL(2,\mathbb{Z})\Bigg|\,a,d\equiv1 \,(\text{mod} \,N), \,b,c\equiv0 \,(\text{mod} \,0)
\right\},
\end{equation}
the kernel of the homomorphism $\pi_N:SL(2,\mathbb{Z})\rightarrow SL(2,\Z_N)$ induced by the reduction modulo $N$ homomorphism $\mathbb{Z}\rightarrow\Z_N$. 
With the above definition, a subgroup of $SL(2,\mathbb{Z})$ is called a {\it congruence subgroup} of level $n$ if there exists $N\geq1$ such that it contains the principal congruence subgroup $\Gamma(N)$, and and $n$
is the smallest such $N$.

Bantay then showed \cite{Bantay}, using solely the knowledge of the modular $\mathcal{S}$ matrix, that the conductor of the 2D Ising CFT is $N=48$, and he further proved\footnote{Bantay proved that the index $|\mathscr{K}:\Gamma(N)|$ equals the order of the image of $\mathscr{K}$ under a group homomorphism $\mu_N: SL(2,\mathbb{Z})\rightarrow SL(2,\mathbb{Z})/\Gamma(N)$. \label{foot:Bantay}} that the index $|\mathscr{K}:\Gamma(48)|=64$.

On the other hand, the formula for the index of $\Gamma(N)$ inside $SL(2,\mathbb{Z})$ is
\begin{equation}
|SL(2,\mathbb{Z}):\Gamma(N)|=N^3\prod_{p|N} \left(1-\frac{1}{p^2}\right),
\end{equation}
where the product is (as indicated) over all prime numbers $p$ that divide the level $N$. The above equals 73728 for $N=48$.

Then by the Lagrange's theorem in group theory, we have
\begin{equation}
|SL(2,\mathbb{Z}):\Gamma(48)|=|SL(2,\mathbb{Z}):\mathscr{K}|\cdot|\mathscr{K}:\Gamma(48)|,
\end{equation}
so one would naively expect that the index $|SL(2,\mathbb{Z}):\mathscr{K}|$ for the 2D Ising CFT should be $73728/64=1152$. 

However, there is a subtlety: There exist three distinct ways of lifting a projective representation of $SL(2,\mathbb{Z})$ to a linear representation
(i.e., three distinct  ``linearizations''), see for example \cite{Wang1,TenerWang}. Bantay considered a linear representation in \cite{Bantay}, while we are focusing on projective ones because a TQFT (in our case as discussed in Section \ref{sec:finiteness_modular_sum}, the 3D Ising TQFT, whose algebraic theory is described by the Ising MTC \cite{Wang1}) gives rise to a {\it projective} representation of the mapping class group of a Riemann surface (compare with \eqref{eq:AGMV} and \eqref{eq:translationgen}), partially due to the non-degeneracy axiom for the modular $\mathcal{S}$ matrix of the TQFT \cite{Muger,Turaev,Kirillov}. This is also consistent with the (projective) transformations of Jacobi theta functions under $SL(2,\mathbb{Z})$ mentioned in 
Appendix \ref{app:generators} below. After linearization of the projective representation, the order of the image of the $SL(2,\mathbb{Z})$ generator $T$ changes from $16$ to $48$. Taking into account this factor of three, we finally arrive at the index $384=1152/3$, which agrees with the result from our {\it Mathematica} code.\footnote{The indices of the $SL(2,\mathbb{Z})$ subgroups which preserve {\it only one of the three} characters $\chi_{1,1}$, $\chi_{1,2}$, or $\chi_{2,1}$ are $384$, $384$ and $48$, respectively. 
If the characters are required to be preserved only up to a $U(1)$ phase, then these indices become $24$, $24$ and $3$, respectively.}

Now we notice that $\mathscr{K}$ in \cite{Bantay} is {\it not} the enhanced symmetry group $\Gamma_c^{[g=1]}$ for genus 1 discussed in \cite{Castro}, since the latter is defined as in \eqref{eq:chi11}, in a similar but different way than in \eqref{eq:kernel}, i.e., only preserving the vacuum character up to a $U(1)$ phase. $\Gamma_c^{[g=1]}$ turns out to be an index-24 subgroup of $SL(2,\mathbb{Z})$, consistent with \footref{foot:Bantay}, and in \cite{Castro} Bantay's $\mathscr{K}$ is merely used as an argument justifying the finiteness of the genus one modular sum.

Up to now, our entire discussion is about genus one. Our final remark is that the mathematical meaning of the prefactor 384 in \eqref{eq:384} in the genus two case is 
the analogue of the prefactor $8$ 
for the genus one partition function 
in \eqref{eq:torusindex}.\footnote{The fact that the numerical prefactor ``$384$''
in \eqref{eq:384} is the same number as the above-mentioned {\it index} ``$384$'' $= 1152/3$
is probably a coincidence.}

The factor $384$ arises because the subgroup $\Gamma_c^{[g=2]}$ of the mapping class group $\Gamma_{g=2}$ is the immediate counterpart of the subgroup $\Gamma_c^{[g=1]}$ of the mapping class group $\Gamma_{g=1}=SL(2, Z)$, 
both  subgroups being denoted by $\Gamma_c$ in \eqref{eq:chi11} as well as below \eqref{eq:ModularSum}.

\section{Generators for \texorpdfstring{$Sp(4,\Z)$}{} and the Algorithm}
\label{app:generators}
The group $Sp(4,\Z)$ is minimally generated by $K$ and $L$ with the following representations:
\begin{equation}
\label{eq:KL}
K=\left( \begin{matrix} 1 & 0 & 0 & 0 \\ 1 & -1 & 0 & 0\\ 0 & 0 & 1 & 1\\ 0 & 0 & 0 & -1\end{matrix} \right), \quad
L=\left( \begin{matrix} 0 & 0 & -1 & 0 \\ 0 & 0 & 0 & -1\\ 1 & 0 & 1 & 0 \\ 0 & 1 & 0 & 0\end{matrix}\right).
\end{equation}
They satisfy $K^2=L^{12}=\mathbbm{1}_4$ and 
the following six additional relations \cite{Bender},
\begin{equation}
\begin{split}
&\left(L^2X\right)^2=\left(XL^2\right)^2,
\quad L\left(L^6X\right)^2=\left(L^6X\right)^2L,\quad \left(KL^5\right)^5=\left(L^6X\right)^2,\\
&\left(KL^7KL^5K\right)L=LX, \quad\left(L^2KL^4\right)X=X\left(L^2KL^4\right),\quad \left(L^3KL^3\right)X=X(L^3KL^3),
\end{split}
\end{equation}
with $X\equiv KL^5KL^7K=\mathbbm{1}_2\otimes \sigma_x$, $L^6=\mathbbm{1}_2\otimes \sigma_z$, where $\sigma_i$ denote the standard three Pauli matrices. Its generalization to arbitrary genus $Sp(2g,\mathbb{Z})$ with at most 3 generators and $3g+5$ relations can be found in \cite{Lu}. In the following basis of Riemann theta functions
\begin{equation}
\begin{aligned}
\label{eq:basis}
    &\vartheta^{1/2}\left[\begin{matrix} 0 & 0\\ 0 & 0\end{matrix}\right](\Omega), \,
    \vartheta^{1/2}\left[\begin{matrix} \frac{1}{2} & 0\\ 0 & 0\end{matrix}\right](\Omega),\, 
    \vartheta^{1/2}\left[\begin{matrix} 0 & 0\\ \frac{1}{2} & 0\end{matrix}\right](\Omega),\, 
    \vartheta^{1/2}\left[\begin{matrix} 0 & 0\\ \frac{1}{2} & \frac{1}{2}\end{matrix}\right](\Omega),\,
    \vartheta^{1/2}\left[\begin{matrix} 0 & \frac{1}{2}\\ 0 & 0\end{matrix}\right](\Omega),\\
    &\vartheta^{1/2}\left[\begin{matrix} \frac{1}{2} & 0\\ 0 & \frac{1}{2}\end{matrix}\right](\Omega),\,
    \vartheta^{1/2}\left[\begin{matrix} 0 & \frac{1}{2}\\ \frac{1}{2} & 0\end{matrix}\right](\Omega),\,
    \vartheta^{1/2}\left[\begin{matrix} 0 & 0\\ 0 & \frac{1}{2}\end{matrix}\right](\Omega),\,
    \vartheta^{1/2}\left[\begin{matrix} \frac{1}{2} & \frac{1}{2}\\ 0 & 0\end{matrix}\right](\Omega),\,
    \vartheta^{1/2}\left[\begin{matrix} \frac{1}{2} & \frac{1}{2}\\ \frac{1}{2} & \frac{1}{2}\end{matrix}\right](\Omega),
\end{aligned}
\end{equation}
the projective representations of $K$ and $L$ are:
\begin{equation}
\mathcal{K}=\begin{pmatrix}
1 & 0 & 0 & 0 & 0 & 0 & 0 & 0 & 0 & 0\\
0 & 1 & 0 & 0 & 0 & 0 & 0 & 0 & 0 & 0\\
0 & 0 & 0 & 1 & 0 & 0 & 0 & 0 & 0 & 0\\
0 & 0 & 1 & 0 & 0 & 0 & 0 & 0 & 0 & 0\\
0 & 0 & 0 & 0 & 0 & 0 & 0 & 0 & 1 & 0\\
0 & 0 & 0 & 0 & 0 & 1 & 0 & 0 & 0 & 0\\
0 & 0 & 0 & 0 & 0 & 0 & 0 & 0 & 0 & 1\\
0 & 0 & 0 & 0 & 0 & 0 & 0 & 1 & 0 & 0\\
0 & 0 & 0 & 0 & 1 & 0 & 0 & 0 & 0 & 0\\
0 & 0 & 0 & 0 & 0 & 0 & 1 & 0 & 0 & 0\\
\end{pmatrix},
\quad
\mathcal{L}=\begin{pmatrix}
0 & 0 & 1 & 0 & 0 & 0 & 0 & 0 & 0 & 0\\
1 & 0 & 0 & 0 & 0 & 0 & 0 & 0 & 0 & 0\\
0 & e^{\pi i/8} & 0 & 0 & 0 & 0 & 0 & 0 & 0 & 0\\
0 & 0 & 0 & 0 & 0 & 0 & 0 & 0 & e^{\pi i/8} & 0\\
0 & 0 & 0 & 1 & 0 & 0 & 0 & 0 & 0 & 0\\
0 & 0 & 0 & 0 & 1 & 0 & 0 & 0 & 0 & 0\\
0 & 0 & 0 & 0 & 0 & e^{\pi i/8} & 0 & 0 & 0 & 0\\
0 & 0 & 0 & 0 & 0 & 0 & 1 & 0 & 0 & 0\\
0 & 0 & 0 & 0 & 0 & 0 & 0 & 1 & 0 & 0\\
0 & 0 & 0 & 0 & 0 & 0 & 0 & 0 & 0 & e^{3\pi i/8}\\
\end{pmatrix},
\end{equation}
where $\mathcal{L}$ is of order 24 and $\mathcal{K}$ is of order 2. $Z_{\text{vac}}^{\text{cl}}$ in \eqref{eq:Zvac} is invariant under the action of $\mathcal{K}$, just like the torus vacuum seed $Z_{\text{vac}}$ is invariant under $T$ of $SL(2,\mathbb{Z})$. Additionally, $Z_{\text{vac}}^{\text{cl}}$ is also invariant under $\mathcal{L}^6$.

We ignore the factor det$(C\Omega+D)^{-1}$ in the modular transformation of Riemann theta functions \eqref{eq:AGMV}, because it is expected to be absorbed in the overall quantum factor for the characters \eqref{eq:quantum}, similar to the Dedekind eta function in the torus case \cite{BigYellowBook}. The phase $\epsilon(\gamma)$ in \eqref{eq:AGMV} has no effect because it is an overall factor independent of $\Omega$, which drops out when taking the norm in the expression for $Z_{\text{vac}}^{\text{cl}}.$

Below we present the pseudocode similar to those used in an arbitrary solvable ``word problem'' for the MCG \cite{Primer}. 
$\mathcal{K}[\,\cdot\,]$ or $\mathcal{L}[\,\cdot\,]$ means that $\mathcal{K}$ or $\mathcal{L}$ acts on the period matrix $\Omega$ in all $seed$, $seed1$ and $seed2$. 
\begin{algorithm*}
\caption{}\label{euclid}
\begin{algorithmic}[1]
\State Inititalize $\,\mathcal{K},\,\mathcal{L},\,seed1:=\chi_{1,1}$
\State $seed=\mathcal{K}[seed1]$
\For {$n=0$, $n\leq23$, $n++$}
\State $temp=\mathcal{L}[seed]$
\If {$temp$ $\notin$ $seed$,} do nothing
\Else {\,\,$seed=$Append [$seed$, $temp$]}
\EndIf
\EndFor
\State $seed2=seed$
\If {Length [Intersection [$seed1$, $seed2$]]$<$Length [$seed2$]} 
$seed1=seed2$, and repeat from 2 to 8
\Else {\,\,stop}
\EndIf
\State Print $Z_{\text{grav}}:=$Total [$seed2$]
\end{algorithmic}
\end{algorithm*}

Finally, we comment on the ``translational'' subgroup $\Gamma_{\infty}^{[g=2]}$ of $Sp(4,\mathbb{Z})$. The group gets its name from its genus-one counterpart, where $\Gamma_{\infty}^{[g=1]}$ is generated by the translation $T:\tau\rightarrow \tau+1$. (The superscripts $\cdot^{[g=1]}$ and $\cdot^{[g=2]}$ specify the corresponding genus.) As in the genus one case, the group $\Gamma_\infty^{[g=2]}$ is the ``classical analogue'' of $\Gamma_c^{[g=2]}$. There is no canonical choice for $\Gamma_{\infty}^{[g=2]}$ on genus two surfaces, but one possibility is generated (not necessarily minimally) by
\begin{equation}
\begin{aligned}
\label{eq:Generators}
\tilde{S}&=\left(L^6X\right)^3,\quad \tilde{T}=XKL^6X,\\
T_1&=XT_2X, \quad T_2=L^{-1}XL^{-2}XL^{-2},\quad T_3=L^8KL^4X\tilde{S},
\end{aligned}
\end{equation}
with the same $X$ as before. Here $T_1$, $T_2$ and $T_3$ respectively shift the entries $\Omega_{11}$, $\Omega_{22}$ and $\Omega_{12}$ by 1; each of $\tilde{S}$ and $\tilde{T}$ acting on $\Omega$ as in \eqref{eq:sp2z} performs the conjugation $\Omega\rightarrow M\Omega M^{-1}$, where $M$ denotes
$S$ or $T$ of $SL(2,\mathbb{Z})$ \cite{Yin1}. In the same basis \eqref{eq:basis}, the
10-dimensional projective representations
of the generators listed in \eqref{eq:Generators} are:
\begin{equation}
\label{eq:translationgen}
\tilde{\mathcal{T}}_1=\begin{pmatrix}
0 & 0 & 1 & 0 & 0 & 0 & 0 & 0 & 0 & 0\\
0 & e^{\pi i/4} & 0 & 0 & 0 & 0 & 0 & 0 & 0 & 0\\
1 & 0 & 0 & 0 & 0 & 0 & 0 & 0 & 0 & 0\\
0 & 0 & 0 & 0 & 0 & 0 & 0 & 1 & 0 & 0\\
0 & 0 & 0 & 0 & 0 & 0 & 1 & 0 & 0 & 0\\
0 & 0 & 0 & 0 & 0 & e^{\pi i/4} & 0 & 0 & 0 & 0\\
0 & 0 & 0 & 0 & 1 & 0 & 0 & 0 & 0 & 0\\
0 & 0 & 0 & 1 & 0 & 0 & 0 & 0 & 0 & 0\\
0 & 0 & 0 & 0 & 0 & 0 & 0 & 0 & e^{\pi i/4} & 0\\
0 & 0 & 0 & 0 & 0 & 0 & 0 & 0 & 0 & e^{\pi i/4}\\
\end{pmatrix},
\quad
\tilde{\mathcal{T}}_2=\begin{pmatrix}
0 & 0 & 0 & 0 & 0 & 0 & 0 & 1 & 0 & 0\\
0 & 0 & 0 & 0 & 0 & 1 & 0 & 0 & 0 & 0\\
0 & 0 & 0 & 1 & 0 & 0 & 0 & 0 & 0 & 0\\
0 & 0 & 1 & 0 & 0 & 0 & 0 & 0 & 0 & 0\\
0 & 0 & 0 & 0 & e^{\pi i/4} & 0 & 0 & 0 & 0 & 0\\
0 & 1 & 0 & 0 & 0 & 0 & 0 & 0 & 0 & 0\\
0 & 0 & 0 & 0 & 0 & 0 & e^{\pi i/4} & 0 & 0 & 0\\
1 & 0 & 0 & 0 & 0 & 0 & 0 & 0 & 0 & 0\\
0 & 0 & 0 & 0 & 0 & 0 & 0 & 0 & e^{\pi i/4} & 0\\
0 & 0 & 0 & 0 & 0 & 0 & 0 & 0 & 0 & e^{\pi i/4}\\
\end{pmatrix},\nonumber
\end{equation}
\begin{equation}
\tilde{\mathcal{T}}_3=\begin{pmatrix}
1 & 0 & 0 & 0 & 0 & 0 & 0 & 0 & 0 & 0\\
0 & 0 & 0 & 0 & 0 & 1 & 0 & 0 & 0 & 0\\
0 & 0 & 1 & 0 & 0 & 0 & 0 & 0 & 0 & 0\\
0 & 0 & 0 & 1 & 0 & 0 & 0 & 0 & 0 & 0\\
0 & 0 & 0 & 0 & 0 & 0 & 1 & 0 & 0 & 0\\
0 & 1 & 0 & 0 & 0 & 0 & 0 & 0 & 0 & 0\\
0 & 0 & 0 & 0 & 1 & 0 & 0 & 0 & 0 & 0\\
0 & 0 & 0 & 0 & 0 & 0 & 0 & 1 & 0 & 0\\
0 & 0 & 0 & 0 & 0 & 0 & 0 & 0 & 0 & -i\\
0 & 0 & 0 & 0 & 0 & 0 & 0 & 0 & -i & 0\\
\end{pmatrix},
\,\,
\tilde{\mathcal{S}}=\begin{pmatrix}
1 & 0 & 0 & 0 & 0 & 0 & 0 & 0 & 0 & 0\\
0 & 0 & 0 & 0 & 1 & 0 & 0 & 0 & 0 & 0\\
0 & 0 & 0 & 0 & 0 & 0 & 0 & 1 & 0 & 0\\
0 & 0 & 0 & 1 & 0 & 0 & 0 & 0 & 0 & 0\\
0 & 1 & 0 & 0 & 0 & 0 & 0 & 0 & 0 & 0\\
0 & 0 & 0 & 0 & 0 & 0 & 1 & 0 & 0 & 0\\
0 & 0 & 0 & 0 & 0 & 1 & 0 & 0 & 0 & 0\\
0 & 0 & 1 & 0 & 0 & 0 & 0 & 0 & 0 & 0\\
0 & 0 & 0 & 0 & 0 & 0 & 0 & 0 & 1 & 0\\
0 & 0 & 0 & 0 & 0 & 0 & 0 & 0 & 0 & 1\\
\end{pmatrix},
\,\,
\tilde{\mathcal{T}}=\begin{pmatrix}
1 & 0 & 0 & 0 & 0 & 0 & 0 & 0 & 0 & 0\\
0 & 0 & 0 & 0 & 0 & 0 & 0 & 0 & 1 & 0\\
0 & 0 & 1 & 0 & 0 & 0 & 0 & 0 & 0 & 0\\
0 & 0 & 0 & 0 & 0 & 0 & 0 & 1 & 0 & 0\\
0 & 0 & 0 & 0 & 1 & 0 & 0 & 0 & 0 & 0\\
0 & 0 & 0 & 0 & 0 & 0 & 0 & 0 & 0 & 1\\
0 & 0 & 0 & 0 & 0 & 0 & 1 & 0 & 0 & 0\\
0 & 0 & 0 & 1 & 0 & 0 & 0 & 0 & 0 & 0\\
0 & 1 & 0 & 0 & 0 & 0 & 0 & 0 & 0 & 0\\
0 & 0 & 0 & 0 & 0 & 1 & 0 & 0 & 0 & 0\\
\end{pmatrix}.
\end{equation}

\section{Partition function of Ising CFT at genus two}
\label{app:Ising}

The partition function of the 2D Ising CFT 
can be computed on a Riemann surface $\Sigma_g$ with arbitrary genus $g$ as the square root of that of the 
$\Z_2$-orbifold CFT 
of a free scalar field at central charge $c=1$ with compactification radius
$R=1$ \cite{AGMV,DVV}. It is given by the product of $Z^{\text{qu}}$, representing the quantum fluctuations of the compactified scalar field, and a classical part $Z^{\text{cl}}$. The latter is the partition sum over the classical solutions in $2g$ winding or soliton 
sectors \cite{AGMV} around the $\alpha$ and $\beta$ cycles
depicted in Figure \ref{fig:handlebody} of Section \ref{sec:two} above, 
and turns out to be
given by \cite{DVV}
\begin{equation}
\label{eq:IsingArbitraryGenus}
Z^{\text{cl}}(\Omega,\bar{\Omega})=2^{-g} \sum\limits_{\mathbf{a},\mathbf{b}\in \left(\frac{1}{2}\Z\right)^g} \left| \vartheta \left[\begin{matrix} \mathbf{a}\\ \mathbf{b}\end{matrix}\right] (\Omega)\right|.
\end{equation}

The more subtle quantum factor is \cite{AGMV,Freed,BN}
\begin{equation}
\label{eq:quantum}
Z^{\text{qu}}(\Omega,\bar{\Omega})=\left(\frac{\det '(-\Delta_G)}{\int_{\Sigma_g} \sqrt{h} \det \left(\text{Im} \Omega\right)}\right)^{-1/4}.
\end{equation}
Here $\Delta_{G}$ is defined as
\begin{equation}
    \Delta_G\equiv-\frac{1}{\sqrt{G}}\partial_{\mu}\sqrt{G}G^{\mu\nu}\partial_{\nu}.
\end{equation}
It is the scalar Laplacian on real functions\footnote{$\Delta_G$ equals the natural covariant Laplacians $\Delta^{\pm}_0$ on $\mathbb{T}^n$, the space of all weight $(n,0)$ tensor fields on $\Sigma_g$ \cite{DHokerPhong1}. Generally $\Delta^{+}_n=-2\nabla^z_{n+1}\nabla^n_z=2\bar{\partial}_{n+1}\bar{\partial}^{\dagger}_{n+1}$ and $\Delta^{-}_n=-2\nabla^{n-1}_z\nabla^z_n=2\bar{\partial}^{\dagger}_n\bar{\partial}$, where the covariant derivatives are
\begin{equation*}
\begin{aligned}
    &\nabla^n_z:\mathbb{T}^n\rightarrow\mathbb{T}^{n+1},\quad\nabla^n_z\left(T(dz)^n\right)\equiv\left(G_{z\bar{z}}\right)^n\frac{\partial}{\partial z}\left(\left(G^{z\bar{z}}\right)^nT\right)(dz)^{n+1};\\
    &\nabla^z_n:\mathbb{T}^n\rightarrow\mathbb{T}^{n-1}, \quad\nabla^z_n\left(T(dz)^n\right)\equiv G^{z\bar{z}}\frac{\partial}{\partial \bar{z}}T(dz)^{n-1},
\end{aligned}
\end{equation*}
This subtlety is explained here, because in the original papers, the numerator in (\ref{eq:quantum}) is $\det '(-\nabla^2)$ \cite{AGMV,BN} or $\det '\Delta^{\pm}_0$ \cite{DHokerPhong2,DHokerPhong3}, instead of $\det '\Delta_G$.
}, i.e., the Laplace-Beltrami operator, and $G$ is the metric on 
$\Sigma_g$. The prime in $\det '$ indicates regularization by omitting zero modes of $\Delta_G$.
For genus one, with the standard metric $|d\sigma_1+\tau d\sigma_2|^2$, the entire 
expression in \eqref{eq:quantum} is simply $Z^{\text{qu}}(\tau,\bar{\tau})=1/\left(\sqrt{2}|\eta(\tau)|\right)$, 
as it appears in standard texts, such as e.g., \cite{BigYellowBook}. 
For genus $g>1$, the determinant alone is evaluated as \cite{Fried,DHokerPhong1,DHokerPhong2,DHokerPhong3}
\begin{equation}
\label{eq:determinant}
    \det\text{$'$}\Delta_G=
    \zeta_S'(1)\exp\left\{(g-1)\left[\ln 2\pi-1/2+4\zeta'(-1)\right]\right\}\approx \zeta_S'(1)e^{0.6762(g-1)},
\end{equation}
where $\zeta_S(s)$ and $\zeta(s)$ are Selberg and Riemann zeta functions,  respectively. The Selberg zeta function for $\Sigma_g$ is defined as 
\begin{equation}
    \zeta_S(s)=\prod_{p\text{ primitive}}\prod^{\infty}_{k=1}\left[1-e^{-(s+k)l(p)}\right],
\end{equation}
where the primitive $p$'s are the simple closed oriented geodesics on $\Sigma_g$ annd $l(p)$ is the hyperbolic length of $p$.

Since the Ising CFT can also be expressed in terms of
the CFT of a single non-interacting Majorana fermion species, 
the classical part in \eqref{eq:IsingArbitraryGenus} is simply proportional to the summation over the partition function for the free Majorana
fermion theory of the corresponding spin structure. For example in the case of torus, we have \cite{BigYellowBook}
\begin{equation}
\begin{split}
& 2\sqrt{\eta(\tau)} ~\chi_{1,1}(\tau)=
2\sqrt{\eta(\tau)} ~\chi_{1}(\tau)=
\vartheta^{1/2} \left[\begin{matrix} 0\\ 0\end{matrix}\right](\tau)+\vartheta^{1/2} \left[\begin{matrix} 0\\ 1/2\end{matrix}\right](\tau),\\
& \sqrt{2\eta(\tau)} ~\chi_{1,2}(\tau) =
\sqrt{2\eta(\tau)} ~\chi_{\sigma}(\tau) =
\vartheta^{1/2} \left[\begin{matrix} 1/2\\ 0\end{matrix}\right](\tau),\\
&2\sqrt{\eta(\tau)} ~\chi_{2,1}(\tau)=
2\sqrt{\eta(\tau)} ~\chi_{\psi}(\tau)=
\vartheta^{1/2} \left[\begin{matrix} 0\\ 0\end{matrix}\right](\tau)-\vartheta^{1/2} \left[\begin{matrix} 0\\ 1/2\end{matrix}\right](\tau).\\
\label{eq:TorusChiTheta}
\end{split}
\end{equation}

\begin{figure}[htbp]
\centering
\includegraphics[width=0.3\textwidth]{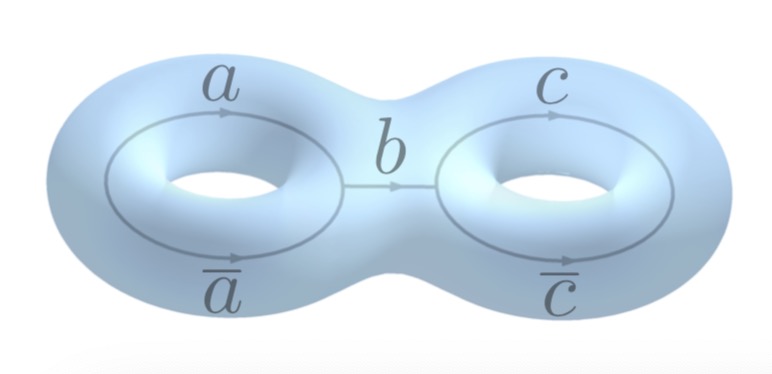}
\caption{All possible admissible label sets are $\{a, b, c\}=\{1,1,1\},$ $\{\psi,1,1\},$ $\{1,1,\psi\},$ $\{\psi,1,\psi\},$ $\{\sigma,1,1\},$ $\{1,1,\sigma\},$ $\{\sigma,1,\psi\},$ $\{\psi,1,\sigma\},$ $\{\sigma,1,\sigma\},$ $\{\sigma,\psi,\sigma\}.$}
\label{fig:anyon}
\end{figure}

At genus $g=2$, there 
turn out to be 
ten holomorphic conformal blocks of the Ising theory. As shown in Figure \ref{fig:anyon}, the three primary fields $a, b, c \in \{1,\sigma,\psi\}$ satisfy the following fusion rules,
\begin{equation}
a\times \bar{a}\rightarrow b,\quad c\times \bar{c}\rightarrow b,
\end{equation}
where the overbar denotes the anti-particle
(and all particles $1, \sigma, \psi$ are their own anti-particle).

There are sixteen $g=2$ Riemann theta functions corresponding to the different possible choices of characteristic vectors $\mathbf{a}$ and $\mathbf{b}$, where $\mathbf{a}, \mathbf{b} \in
(\frac{1}{2}\mathbb{Z}_2)^2$ - compare with Section \ref{sec:math} above. Only the ten even ones are non-vanishing, which are listed in \eqref{eq:basis}. In Table \ref{tab:basis}, we present the matrix of basis change\footnote{Table \ref{tab:basis} is the result of an educated guess based on
(\ref{eq:Factorization}) below, and its content passed all the consistency checks to our best knowledge. Perhaps it could be derived by considering six-point functions of twist operators of conformal dimension $\frac{c}{6}\left(2-\frac{1}{2}\right)=\frac{1}{8}$ in the orbifold CFT Ising$^{\otimes2}/\mathbb{Z}_2$ on the Riemann sphere in the spirit of \cite{Witten,CardyMaloneyMaxfield}.}
from the ``free Majorana fermion basis'' of square-roots of theta functions (right part of Table)  to the classical parts of the basis of the  genus-two Ising characters (left part of Table).

\begin{table}[hbtp]
\centering
    \begin{tabular}{ | l | l |}
    \hline  &\\[-10pt]
    $4\chi_{111}^{\text{cl}}$ & $\vartheta^{1/2}\spin{0}{0}{0}{0}+\vartheta^{1/2}\spin{0}{0}{0}{1/2}+\vartheta^{1/2}\spin{0}{0}{1/2}{0}+\vartheta^{1/2}\spin{0}{0}{1/2}{1/2}$\\[15pt] \hline
     &\\[-10pt]
    $4\chi_{\psi 11}^{\text{cl}}$ & $\vartheta^{1/2}\spin{0}{0}{0}{0}+\vartheta^{1/2}\spin{0}{0}{0}{1/2}-\vartheta^{1/2}\spin{0}{0}{1/2}{0}-\vartheta^{1/2}\spin{0}{0}{1/2}{1/2}$\\[15pt] \hline
     &\\[-10pt]
    $4\chi_{11\psi}^{\text{cl}}$ & $\vartheta^{1/2}\spin{0}{0}{0}{0}-\vartheta^{1/2}\spin{0}{0}{0}{1/2}+\vartheta^{1/2}\spin{0}{0}{1/2}{0}-\vartheta^{1/2}\spin{0}{0}{1/2}{1/2}$\\[15pt] \hline
     &\\[-10pt]
    $4\chi_{\psi 1\psi}^{\text{cl}}$ & $\vartheta^{1/2}\spin{0}{0}{0}{0}-\vartheta^{1/2}\spin{0}{0}{0}{1/2}-\vartheta^{1/2}\spin{0}{0}{1/2}{0}+\vartheta^{1/2}\spin{0}{0}{1/2}{1/2}$\\[15pt] \hline
    \end{tabular}
    
    \begin{tabular}{|l|l|l|l|}
    \hline & & &\\[-10pt]
    $2\sqrt{2} \chi_{\sigma 11}^{\text{cl}}$ & $\vartheta^{1/2}\spin{1/2}{0}{0}{0}+\vartheta^{1/2}\spin{1/2}{0}{0}{1/2}$
    & $2\sqrt{2} \chi_{\sigma 1\psi}^{\text{cl}}$ & $2\vartheta^{1/2}\spin{1/2}{0}{0}{0}-\vartheta^{1/2}\spin{1/2}{0}{0}{1/2}$\\[15pt] \hline
    & & &\\[-10pt]
    $2\sqrt{2} \chi_{11 \sigma}^{\text{cl}}$ & $\vartheta^{1/2}\spin{0}{1/2}{0}{0}+\vartheta^{1/2}\spin{0}{1/2}{1/2}{0}$
    & $2\sqrt{2} \chi_{\psi 1\sigma}^{\text{cl}}$ & $\vartheta^{1/2}\spin{0}{1/2}{0}{0}-\vartheta^{1/2}\spin{0}{1/2}{1/2}{0}$\\[15pt] \hline
    \end{tabular}
    
    \begin{tabular}{|l|l|l|l|}
    \hline & & &\\[-10pt]
    $2\chi_{\sigma 1\sigma}^{\text{cl}}$ & $\vartheta^{1/2}\spin{1/2}{1/2}{0}{0}$
    & $2\chi_{\sigma\psi\sigma}^{\text{cl}}$ & $\vartheta^{1/2}\spin{1/2}{1/2}{1/2}{1/2}$ \\[15pt] \hline
    \end{tabular}
    \caption{The correspondence between 
    (the classical parts of) Ising characters (left) and free fermion characters (right).}
    \label{tab:basis}
\end{table}

The table can be understood intuitively in the pinching limit, where the off-diagonal entries of the period matrix
$\Omega$ vanish. When $\Omega_{12}\rightarrow 0$, all of the above characters except $\chi_{\psi\sigma\psi}$ 
factorize into a product of two genus-one characters: 
\begin{equation}
\chi_{\mu 1 \nu}(\Omega)\rightarrow \chi_{\mu}(\Omega_{11})\chi_{\nu}(\Omega_{22}),
\label{eq:Factorization}
\end{equation}
with $\mu, \nu \in \{1,\sigma,\psi\}.$ (For simplicity, we use the notations $\chi_{1,1}\equiv \chi_1,$ $\chi_{1,2}\equiv\chi_\sigma,$ and $\chi_{2,1}\equiv\chi_\psi.$)  The factorization is not possible for $\chi_{\psi\sigma\psi}$ because when the particle $b$ in Figure \ref{fig:anyon} is non-trivial, the character is intrinsically genus-two and cannot be viewed as disjoint union of two genus-one components, even from a topological point of view. For the other nine sectors, \eqref{eq:Factorization} can be traced back to the factorization of Jacobi theta functions in such a limit:
\begin{equation}
\vartheta\spin{a_1}{a_2}{b_1}{b_2}(\Omega)\rightarrow \vartheta\left[\begin{matrix} a_1\\ b_1\end{matrix}\right](\Omega_{11})~ \vartheta\left[\begin{matrix} a_2\\ b_2\end{matrix}\right](\Omega_{22}). 
\end{equation}

\section{Genus two long-cylinder limit}
\label{app:AlgCurve}
In this appendix, we provide some details regarding the low-temperature or the long-cylinder limit of the Ising and the $c=3l/2G_N=1/2$ gravity partition functions on genus two. 
As reviewed in \eqref{eq:curve}, a genus-two Riemann surface with a $\Z_2$ time-reflection symmetry can be constructed as a complex curve by the ``replica trick''
on two copies of real lines with six branch points, i.e., three finite intervals \cite{Riemann-Hilbert, CoserTagliacozzoTonni}. For computational convenience, we choose an alternative but equivalent expression other than \eqref{eq:curve}:
\begin{equation}
\label{eq:curves}
y(z)^2=u(z)v(z),\quad u(z)=\prod_{i=1}^{3}(z-x_{2i-2}),\quad v(z)=\prod_{i=1}^{2}(z-x_{2i-1}),
\end{equation}
where $y,z\in \mathbb{C}^2$, and we have used a conformal map to fix three of the six branch points as cross ratios: 
\begin{equation}
    x(z)\equiv\frac{(u_1-z)(u_2-u_3)}{(u_1-u_2)(z-u_3)},
\end{equation}
such that $x(u_1)=0,$ $x(u_2)=1,$ and  $x(u_3)\mapsto$ infinity, which we denote as $z_\infty$. For simplicity we have denoted
$x(u_n)\equiv x_{2n-2},$ $x(v_n)\equiv x_{2n-1}$, $n=1,2,3$.

This curve has a non-normalized basis of holomorphic 1-forms:
\begin{equation}
\label{eq:form}
\omega_i=\frac{z^{i-1}}{y(z)}dz,\quad i=1,2.
\end{equation}
Given the canonical homology basis $\{\alpha_i,\beta_j\}$ on the Riemann surface as in (\ref{eq:intersection}), two 2-by-2 non-symmetric matrices can be defined on the surface
\begin{equation}
    \mathcal{A}_{j,i}\equiv\oint_{\alpha_i}\omega_j,\quad \mathcal{B}_{j,i}\equiv\oint_{\beta_i}\omega_j.
\end{equation}
The corresponding period matrix of the surface can then be expressed as
\begin{equation}
    \Omega=\mathcal{A}^{-1}\cdot \mathcal{B},
\end{equation}
separating the contributions from integrals along the $\alpha$ and $\beta$ cycles. (Notice that here we used a different normalization on $\omega$ than the one in \eqref{eq:normalization}.)

Next, we do a basis transformation by decomposing $\{\alpha_i,\beta_i\}$ into auxiliary cycles $\{\alpha_i^{\text{aux}},\beta_i^{\text{aux}}\}$,
\begin{equation}
\label{eq:decompose}
    \alpha_i=\sum_{k=1}^i\alpha^{\text{aux}}_k,\quad \beta_i=\beta^{\text{aux}}_i.
\end{equation}
Correspondingly, one can define following the matrices, which are simply integrals of the one-forms (\ref{eq:form}) along the auxiliary cycles
\begin{equation}
    \mathcal{A}_{j,i}=\sum_{k=1}^i \left(\mathcal{A}^{\text{aux}}\right)^{j,k},\quad \mathcal{B}_{j,i}=\left(\mathcal{B}^{\text{aux}}\right)^{j,k}.
\end{equation}
and finally \cite{Riemann-Hilbert,CoserTagliacozzoTonni},
\begin{equation}
\begin{split}
\left(\mathcal{A}^{\text{aux}}\right)_{j,i}&\equiv\oint_{\alpha^{\text{aux}}_i}\omega_j=-2(-1)^{3-i}\mathscr{F}_j\mid^{x_{2i-1}}_{x_{2i-2}},\\
\left(\mathcal{B}^{\text{aux}}\right)_{j,i}&\equiv\oint_{\beta^{\text{aux}}_i} \omega_b=-2i(-1)^{3-i}\mathscr{F}_j\mid^{x_{2i}}_{x_{2i-1}},\quad i=1,2.
\end{split}
\end{equation}
Here $\mathscr{F}_j\mid^b_a$ can be expressed in terms of the fourth Lauricella function $F^{(3)}_D$ \cite{Exton}, a generalization of the hypergeometric function $_2F_1$,
\begin{equation}
\begin{split}
    \mathscr{F}_j \mid^b_a&=\int^1_0 dt~ \frac{(b-a)[(b-a)t+a]^{j-3/2}}{\prod_{k=2}^3\left|(b-a)t-(x_{2k-2}-a)\right|^{1/2}\prod_{k=1}^{2}|(b-a)t-(x_{2k-1}-a)|^{1/2}}\\
    &=\frac{\pi a^{j-3/2}}{\prod^3_{\substack{k=2\\ x_{2k-2}\neq a}}|x_{2k-2}-a|^{1/2}\prod^2_{\substack{l=1\\ x_{2k-2}\neq a }}|x_{2l-1}-a|^{1/2}}F^{(3)}_D\left(\frac{1}{2},\frac{3}{2}-j,\frac{1}{2},\frac{1}{2};1;\mathbf{q}^{(a,b)}\right),
\end{split}
\end{equation}
where the 3-dimensional vector $\mathbf{q}^{(a,b)}$ has components:
\begin{equation}
    \mathbf{q}^{(a,b)}_{\xi}\equiv\frac{b-a}{x_{\xi}-a}, \quad\xi\in\{0,1,2,3,4\} \backslash \{\eta|x_{\eta}\neq a,b\},
\end{equation}
and $F^{(3)}_D$ has the integral representation:
\begin{equation}
    F^{(3)}_D\left(a,b_1,b_2,b_3;c;q_1,q_2,q_3\right)=\frac{\Gamma(c)}{\Gamma(a)\Gamma(c-a)}\int_0^1dt\frac{t^{a-1}(1-t)^{c-a-1}}{\prod_{j=1}^3(1-q_jt)^{b_j}}.
\end{equation}

Taking all $(x_{2i-1}-x_{2i-2})\equiv\epsilon_i$ to be small for $i\in \{1,2,3\}$, which is required by the long-cylinder limit, we obtain
\begin{equation}
\mathcal{A}=-\frac{2\pi}{\sqrt{z_{\infty}}} \left(\begin{matrix} 1 & 0 \\ 0 & -1 \end{matrix}\right),\quad
\mathcal{B}=\frac{2i}{\sqrt{z_\infty}} \left( \begin{matrix} \log \epsilon_1\epsilon_2 & -\log\epsilon_2\\ \log\epsilon_2 & -\log \epsilon_2\end{matrix}\right),
\end{equation}
and the period matrix is 
\begin{equation}
\label{eq:limit}
    \Omega=\frac{i}{\pi} \left( \begin{matrix} -\log\epsilon_1\epsilon_2 & \log\epsilon_2 \\ \log\epsilon_2 & -\log\epsilon_2 \end{matrix}\right).
\end{equation}
Inserting this into the equations \eqref{eq:IsingArbitraryGenus} and \eqref{eq:Zvac}, we find that they match each other at leading order, which further justifies our expression for $Z_{\text{vac}}^{\text{cl}}$.  The subleading terms will not agree, because the contribution of other sectors will enter.

One remark is that, in this appendix we have used a non-rotating, i.e., purely imaginary period matrix for convenience. Adding an angular potential complicates the calculations but does not affect the match between the low temperature limits of $Z_{\text{vac}}^{\text{cl}}$ and $Z^{\text{cl}}$, which is robust against arbitrarily large angular momenta due to the cancellation between fast oscillating phases in Riemann theta functions. For a review of the rotating case in general, see the following Appendix \ref{app:selection}.

For genus greater than 2 with $\mathbb{Z}_2$ symmetry, one can allow for more branch points and take two copies, and follow the general treatment in \cite{CoserTagliacozzoTonni} to obtain $\Omega$ similarly.

\section{Superselection sectors of angular momenta}
\label{app:selection}

In this section, we explain the nature of rotation of BTZ black holes at genus one, which is not usually discussed in the literature. The lesson will be general enough to extend to higher genus.

It is well-known that given a modular parameter $\tau$ on a torus, which is the asymptotic boundary of the BTZ black hole, its temperature is $\text{Im}\,\tau$, and the angular momentum/potential is $\text{Re}\,\tau$, and if $\text{Re}\,\tau=0$, then it is not rotating. Then what if we shift the purely imaginary $\tau$ by an integer? Apparently it becomes rotating. However, the modulus $\tau$ of the boundary torus is only defined up to $SL(2,\mathbb{Z})$ transformations \cite{MaloneyWitten}, so the torus boundary and hence the asymptotic AdS$_3$ stays the same, which is non-rotating. As a result, we say that both $\tau$ and the shifted $\tau$ are in same superselection sector of rotation, which obtains its name due to the following reasons.

Generically, different $\tau$'s on the upper half plane $\mathbb{H}$ are not connected by $SL(2,\mathbb{Z})$ transformations. For example, take $\tau_1=i$ and $\tau_2=1/3+i/2$. For them to be connected, we need some $\gamma\in SL(2,\Z),$ such that
$$
\gamma\tau=\frac{ai+b}{ci+d}=1/3+i/2=\tau_2
$$
for some $\gamma\in SL(2,\mathbb{Z})$.

However, this is not possible, because this requires
$$
\frac{bd+ac}{c^2+d^2}=\frac{1}{3},\quad\frac{ad-bc}{c^2+d^2}=\frac{1}{2}.
$$
The second equation implies that $c,d=\pm1$. Substituting them into the first equation, we obtain $bd+ac=2/3$, which is impossible.

A more obvious example is to consider $\tau_1=i$ and $\tau_3$ with an irrational real or imaginary part. Hence we say that disconnected $\tau$'s belong to different superselection sectors, or mathematically speaking, they are in different conformal classes, i.e., they are different points in the moduli space of the boundary torus.

Our description of BTZ angular momentum is consistent with the phase diagram for 3D quantum gravity (not necessarily pure or Einstein) shown in Figure 3b in \cite{MaloneyWitten}. Based on the standard tessellation of $\mathbb{H}$ by $SL(2,\mathbb{Z})$ fundamental regions, this phase diagram is a subtessellation obtained by erasing curves which can be crossed without changing the dominant geometry $M_{c,d}$, so all degree 6 vertices become fixed points of $SL(2,\mathbb{Z})$ of order 3. Rotating and non-rotating BTZ black holes can coexist in the same phase, since dominant geometries $M_{c,d}$ for them can have the same 2-tuple $(c,d)$, e.g., all $\text{Im}\,\tau\geq1$ saddles belong to one single phase, where $M_{1,0}$ dominates.

For genus two, in a different geometrical limit than the one in Appendix \ref{app:AlgCurve} (e.g., when two regions where three cylinders join each other are folded around the axis perpendicular to the $\mathbb{Z}_2$-symmetry plane in an opposite way\footnote{Simply twisting cylinders along the axis perpendicular to the $\mathbb{Z}_2$-symmetry plane, or tilting them with respect to the same plane will not introduce rotation.}), $\Omega$ develops a real part and the spacetime rotates, but our $Z_{\vac}^\text{cl}$ will stay the same.
Analytic continuation of a rotating asymptotic AdS$_3$ into the Euclidean signature requires a more complicated version of Schottky double \cite{Krasnov2}, and there is no longer time-reversal symmetry with respect to the $t=0$ slice. However, as long as the doubling remains, one can calculate $\Omega$ using the same replica trick for $\mathbb{Z}_2$ symmetry as in Appendix \ref{app:AlgCurve}.

\section{Extended Property F of Ising Theory}
\label{app:propertyf}

Unitary $(2+1)$-TQFTs are 
well-captured, physically, by anyon models, or, mathematically, by unitary modular tensor categories. 
The {\it extended property F conjecture} asserts that all representations of mapping class groups (MCGs) from a TQFT would have finite images if the total quantum dimension $D^2=\sum_i d_i^2$ is an integer. For the original non-extended {\it property F conjecture} on braid groups instead of MCGs, see \cite{ERW,Naidu}. In this appendix, we prove the Ising TQFT extended case.

The Ising theory has three labels or anyon types $\{1,\sigma,\psi\}$.  The same fusion rule can be realized by $8$ different anyon models     \cite{kitaev2006anyons} with chiral central charges $c=\frac{a}{2}, a=1,3,...,15$, where $c=\frac{1}{2}$ for Ising TQFT. 
The results in this appendix apply to all $8$ theories. 

The Ising TQFT can be constructed explicitly using Temperley-Lieb algebras and Jones-Wenzl (JW) projectors with $A=ie^{\pm \frac{2\pi i}{16}}$.  The three anyon types $\{1,\sigma,\psi\}$ then correspond to the JW projectors $\{p_i\}, i=0,1,2$.
For the notations and terminologies, see for example \cite{WangBook}. 
The mapping class group representations are explicitly described in \cite{Bloomquist}.

To understand the representations of MCGs $\Gamma_g$ from the Ising TQFT, we will use four different bases of the Hilbert spaces $V_{\textrm{iTQFT}}(\Sigma_g)$: the defining basis $\{e_{\bf b}^{\bf a}\}$, the standard basis $\{v_{\bf k}^{\bf i}\}$, the geometric basis $\{u_{\bf k}^{\bf i}\}$, and the spin basis $\{\omega_{\bf l}^{\bf m}\}$---the last three bases are defined and used in \cite{Wright}, and the defining basis is used in \cite{Bloomquist}. The defining basis and standard basis consist of labeled fusion graphs using $\{1,\sigma,\psi\}$. The defining basis is the one as in Figure \ref{fig:ConformalBlock}, while the standard basis can be obtained from Figure \ref{fig:ConformalBlock} by performing $F$-moves at all $w_i$-labeled edges. The geometric basis consists of skeins of simple closed curves; the spin basis consists of even spin structures (those with Arf invariant 0) or even quadratic enhancements of the intersection forms. 

The four bases can be changed to each other through the following explicit formulae \cite{Wright}.  From the geometric to the standard,
$v_{\bf k}^{\bf i}=\sum_{{\bf i'}\leq {\bf i}}\alpha_{\bf i'}u_{\bf k}^{\bf i'},$
where $\alpha_{\bf i'}=\left(-[2]_A\right)^{\frac{g-\sum\left(i_n-i_n'\right)}{2}}$, where $[k]_A\equiv\frac{A^{2k}-A^{-2k}}{A^2-A^{-2}}$ is the quantum integer, and $A=ie^{-2\pi i/16}$ for Ising TQFT. From the standard to the spin, 
$\omega_{\bf l}^{\bf m}=\sum_{\bf k}\alpha_{\bf k}v_{\bf k}^{\bf i},$
where $i_n=m_nl_n+1$, $k_n=2$ if $m_n=1$, $k_n=1$ or $3$ if $m_n=0$, and $\alpha_{\bf k}=(-1)^l 2^{\frac{\sum m_n -g}{2}}, l=\sum \delta_{3,k_n}l_n$.  To go from the defining to the standard, we first note that the label $w_1$ in Figure \ref{fig:ConformalBlock} would be either $1$ or $\psi$.  Inductively, we obtain a change of basis by applying $F$ moves.

Any self-diffeomorphism $f$ of a surface $\Sigma_g$  induces an action on the $\mathbb{Z}_2$-homology group $H_{1}(\Sigma_g; \mathbb{Z}_2)$ of $\Sigma_g$.  The images of all such $\{f\}$ form $Sp\left(2g, \mathbb{Z}_2\right)$.  The kernel consists all diffeomorphims that fix $H_{1}(\Sigma_g; \mathbb{Z}_2)$, which form a subgroup $D_g$ of $\Sigma_g$.  It is proved in \cite{Wright} that $D_g$ is generated by all squares of Dehn twists on simple closed curves.

Any simple closed curve $s$ on $\Sigma_g$ defines a function from the set of spin structures $\omega$ of $\Sigma_g$ to $\mathbb{Z}_2$ by sending $\omega$ to $\omega(s)$.
To prove the finiteness of all representations, it is convenient to use the spin basis.  In this basis, the square of Dehn twist on any simple closed curve $s$ is a diagonal matrix with non-zero entries $\left(-A^6\right)^{\omega(s)}$.  Since $A^6$ is of order $8$, the image of $D_g$ is an abelian group inside $\mathbb{Z}_8^N$ for some $N$.

It follows that we have an exact sequence:

$$1\rightarrow \rho_{\textrm{iTQFT}}(D_g) \rightarrow \rho_{\textrm{iTQFT}}(\Gamma_g) \rightarrow Sp(2g, \mathbb{Z}_2) \rightarrow 1,$$
where $\rho_{\textrm{iTQFT}}(D_g)$ is a subgroup of $\mathbb{Z}_8^N$ for some $N$.

\section{Towards a Formulation of Bekenstein-Hawking entropy in strongly coupled \texorpdfstring{\ads}{}}
\label{sec:discussion}

In this section, we use the proposed duality to compute a gravitational entropy. The resulting expressions, reported in (\ref{eq:GeneralizedBekensteinHawking}) at the end of this section, resemble the form of the universal subleading correction to the entanglement entropy
of the ground states of long-range entangled topological phases in $(2+1)$ dimensions \cite{KitaevPreskill,LevinWen}. 

For this purpose, we consider the genus-one case
and use the fact that the gravitational partition function equals that of the modular invariant 2D Ising CFT at the asymptotic boundary. We then use Cardy's method \cite{Cardy1,Cardy2, Carlip} to extract a variant of the familiar expression for the entropy. Our suggested expressions are listed in
(\ref{eq:GeneralizedBekensteinHawking}) below. We first briefly review familiar manipulations of
the modular invariant 2D CFT partition function for general central charge $c$, and specialize to $c=1/2$ at a suitable point below when we exhibit the new features.

The partition function can be written as
\begin{equation}
Z\left(\tau,\bar{\tau}\right)=\text{Tr}_{\mathcal{H}}~e^{2\pi i \tau \left(L_0-c/24\right)}e^{-2\pi i \bar{\tau}\left(\bar{L}_0-c/24\right)}\equiv \mathcal{Z}(\tau,\bar{\tau})e^{-2\pi ic\tau/24}e^{2\pi ic\bar{\tau}/24},
\label{eq:mathcalZ}
\end{equation}
where ${\cal H}$ denotes the Hilbert space of the CFT (i.e., a choice of pairs of holomorphic and anti-holomorphic primaries), and, when denoting the eigenvalues of $L_0$ and $\overline{L}_0$ as $\Delta$ and $\overline{\Delta}$, the quantity $\mathcal{Z}(\tau,\overline{\tau})$ is related to the density of states $\rho\left(\Delta,\overline{\Delta}\right)$ of the CFT by 
\begin{equation}
\mathcal{Z}\left(\tau,\bar{\tau}\right)=\sum_{\Delta,\overline{\Delta}} \rho(\Delta,\overline{\Delta})e^{2\pi i \Delta \tau}
e^{-2\pi i\bar{\Delta}\bar{\tau}}.
\label{eq:DefDOS}
\end{equation}
We can extract the density of states $\rho$ from the partition function by contour integration via the inverse Laplace transformation going from the canonical to the microcanonical ensemble
\begin{equation}
\rho\left(\Delta,\overline{\Delta}\right)=\frac{1}{(2\pi i)^2}\left(\int^{i\epsilon+\infty}_{i\epsilon-\infty}d\tau\right)\left(\int^{i\epsilon+\infty}_{i\epsilon-\infty}d\bar{\tau}\right) q^{-1-\Delta}\bar{q}^{-1-\overline{\Delta}}Z\left(q,\bar{q}\right),
\end{equation}
where $q=e^{2\pi i \tau}$ and $\bar{q} = e^{2 \pi i {\bar{\tau}}}$.
Using modular invariance $Z(\tau,\bar{\tau})=Z\left(-1/\tau,-1/\bar{\tau}\right)$, as well as the definition
of $\mathcal{Z}(\tau, \bar{\tau})$ from (\ref{eq:mathcalZ}), we obtain
\begin{equation}
\begin{split}
\mathcal{Z}(\tau,\bar{\tau})
&= e^{\frac{2\pi i c}{24}\tau}e^{-\frac{2\pi i c}{24}\bar{\tau}}
Z(\tau,\bar{\tau})
=e^{\frac{2\pi i c}{24}\tau}e^{-\frac{2\pi i c}{24}\bar{\tau}}
Z(-1/\tau,-1/\bar{\tau})\\
&= e^{\frac{2\pi i c}{24}\tau}e^{\frac{2\pi i c}{24}\frac{1}{\tau}}
e^{-\frac{2\pi i c}{24}\bar{\tau}}e^{-\frac{2\pi i c}{24}\frac{1}{\bar{\tau}}}
\mathcal{Z}(-1/\tau,-1/\bar{\tau}),
\end{split}
\end{equation}
and we can rewrite the density of states as
\begin{equation}
\rho\left(\Delta,\overline{\Delta}\right)=
\int_{i\epsilon-\infty}^{i\epsilon+\infty} d\tau \int_{i\epsilon-\infty}^{i\epsilon+\infty} d\bar{\tau}
~e^{-2\pi i \Delta\tau}
e^{2\pi i \overline{\Delta}\bar{\tau}}
e^{2\pi ic\tau/24}e^{-2\pi ic\bar{\tau}/24}e^{2\pi ic/24\tau}e^{-2\pi ic/24\bar{\tau}}\mathcal{Z}(-1/\tau,-1/\bar{\tau}).
\label{eq:DOSSaddlePointEvaluation}
\end{equation}
The asymptotic form of the density of states for large $\Delta$ and $\bar{\Delta}$, of interest to us here, is then obtained from
(\ref{eq:DOSSaddlePointEvaluation}) by steepest descend: Assuming first that
$\mathcal{Z}(-1/\tau,-1/\bar{\tau})$ varies slowly near the 
saddle point 
(a fact that we subsequently check to be correct), 
one finds the saddle point $\tau_*$, $\bar{\tau}_*$ to be located at
\begin{equation}
    \tau_*\approx i\sqrt{\frac{c}{24\Delta}},
    \quad
    \bar{\tau}_*\approx i\sqrt{\frac{c}{24 \bar{\Delta}}},
    \end{equation}
where $\Delta/c\gg 1$, $\bar{\Delta}/c\gg 1$ 
was used\footnote{Thus $(-1/\tau_*)\to i\infty $ and 
$(-1/\bar{\tau}_*)\to i\infty$, implying that 
$\mathcal{Z}(-1/\tau,-1/\bar{\tau})\to 1$ varies slowly, in agreement with the assumption made above.}, implying
$|\tau_*|, |\bar{\tau}_*| \ll 1$.
Substituting back into the integral above yields the ``Cardy formula''
\begin{equation}
    \log \rho\left(\Delta,\overline{\Delta}\right)
    \sim 2\pi c
    \left (
    \sqrt{\frac{\Delta}{6c}}+\sqrt{\frac{\bar{\Delta}}{6 c}}\right ),
    \qquad \left({\rm when} \ \frac{\Delta}{c} \gg 1, \frac{\bar{\Delta}}{c} \gg 1\right).
\label{eq:DOSCardyFormula}
\end{equation}

Now we discuss the Ising case with $c=1/2$.
For convenience we make the identification 
$\left\{\chi_{1,1}, \chi_{1,2}, \chi_{2,1}\right\}=\{\chi_{1}, \chi_{\sigma}, \chi_{\psi}\}$. 
With $\tau=i \left(\beta/L\right)$, the partition function in (\ref{eq:mathcalZ}) describes the quantum partition function of thermal \ads, where the spatial cycle has circumference $L$. 
In the low-temperature limit (small $q$, $\tau\to i \infty$), the gravitational system is dominated by the thermal \ads solution, i.e., $Z(\tau,\bar{\tau})\sim|\chi_{1,1}(\tau)|^2$. In the opposite high-temperature limit, the black hole solutions dominate. Specifically, the BTZ saddle point can be obtained from the thermal \ads saddle by an $S$ modular transformation $\tau\rightarrow -1/\tau$.
Considering the high-temperature ($\beta\to 0$) limit $\tau\to 0$,
where $-1/\tau=i (L/\beta) \to i \infty$,
we obtain
\begin{eqnarray}
\nonumber
&&
|\chi_{1}(\tau)|^2 + |\chi_{\sigma}(\tau)|^2 + |\chi_{\psi}(\tau)|^2=Z_{\text{Ising}}(\tau) = 
Z_{\text{Ising}}(-1/\tau) \\
&& = 
|\chi_{1}(-1/\tau)|^2 + |\chi_{\sigma}(-1/\tau)|^2 + |\chi_{\psi}(-1/\tau)|^2
\sim |\chi_{1}(-1/\tau)|^2, \ (-1/\tau \to i \infty).
\label{eq:CardyS}
\end{eqnarray}
Now we re-write the first line
using the action of the modular transformation on the characters 
\begin{equation}
\chi_a(\tau) = \sum_{b=1,\sigma,\psi}
\mathcal{S}_{a,b} \ \chi_b(-1/\tau).
\end{equation}
The {\it normalized} modular matrices are
\begin{equation}
\label{eq:modular}
\mathcal{S}=\frac{1}{2} \left( \begin{matrix} 1  & \sqrt{2} & 1 \\ \sqrt{2} & 0 & -\sqrt{2}\\ 1 & -\sqrt{2} & 1\end{matrix}\right),\quad \mathcal{T}= e^{-2\pi i/48}\left( \begin{matrix} 1 & 0 & 0 \\ 0 & e^{2\pi i/16} & 0 \\ 0 & 0 & -1 \end{matrix}\right).
\end{equation}
Collecting the leading terms in the limit $-1/\tau \to i \infty$, 
\begin{equation}
|\chi_a(\tau)|^2\sim |\mathcal{S}_{a,1}|^2  \ |\chi_1(-1/\tau)|^2
+ \dots,
\label{eq:LeadingTerms}
\end{equation}
the first line of \eqref{eq:CardyS} then reads in this limit 
\begin{equation}
\begin{split}
&
\qquad  \qquad \qquad
|\chi_{1}(\tau)|^2 + |\chi_{\sigma}(\tau)|^2 + |\chi_{\psi}(\tau)|^2  \\
\sim~& \frac{d_{1}^2}{
\mathcal{D}^2}\ |\chi_1(-1/\tau)|^2
+
\frac{d_{\sigma}^2}{
\mathcal{D}^2}\ |\chi_1(-1/\tau)|^2
+
\frac{d_{\psi}^2}{\mathcal{D}^2}
 |\chi_1(-1/\tau)|^2, \ \ \ \ \   (-1/\tau \to i \infty)
\end{split}
\label{eq:ThreeEntropies}
\end{equation}
where we have made use of the relationship of the quantum dimensions $d_a=\mathcal{S}_{1,a}/\mathcal{S}_{1,1}$, and the total quantum dimension $\mathcal{D}^2=\sum_a {d_a}^2=1/(\mathcal{S}_{1,1})^2$
of the Ising TQFT, with the modular $\mathcal{S}$-matrix (from the Verlinde formula).  \eqref{eq:ThreeEntropies} suggests that the three summands in the second
line arise from the corresponding three summands in the first line.
Using \eqref{eq:mathcalZ}, \eqref{eq:DefDOS} and \eqref{eq:DOSCardyFormula}, we have
\begin{equation}
\begin{split}
& Z(\tau, \bar{\tau}) = \mathcal{Z}(\tau,\bar{\tau})
\ e^{-\frac{2\pi i c}{24} \tau}\ e^{-\frac{2\pi i c}{24}\bar{\tau}}
\\ 
 =~& Z(-1/\tau, -1/\bar{\tau}) = \mathcal{Z}(-1/\tau,-1/\bar{\tau})
\ e^{-\frac{2\pi i c}{24}{(-1/\tau)}}\ e^{-\frac{2\pi i c}{ 24}{(-1/\bar{\tau})}}
\end{split}
\end{equation}
which, in the limit $-1/ \tau \to i \infty$, yields
\begin{eqnarray}
\sum_{\Delta,\bar{\Delta}}
\rho\left(\Delta,\bar{\Delta}\right) e^{2\pi i \Delta \tau}
e^{2\pi i \bar{\Delta}\bar{\tau}}
\sim e^{-\frac{2\pi i c}{24}{(-1/\tau)}}\ e^{-\frac{2\pi i c}{24}{(-1/\bar{\tau})}}
\sim |\chi_1(-1/\tau)|^2,
\ \  (-1/ \tau \to i \infty), \ \ \ \ \ 
\label{eq:HighTempLimitDOSPartitionFunction}
\end{eqnarray}
and we need these expressions here with $c=1/2$.
Comparison with (\ref{eq:ThreeEntropies}) suggests that we can identify three different densities of states,
\begin{eqnarray}
\rho_{1}\left(\Delta,\bar{\Delta}\right)\equiv \frac{d_{1}^2}{\mathcal{D}^2} \rho\left(\Delta,\bar{\Delta}\right),
\,\,
\rho_{\sigma}\left(\Delta,\bar{\Delta}\right)\equiv \frac{d_{\sigma}^2}{\mathcal{D}^2} \rho\left(\Delta,\bar{\Delta}\right),
\,\,
\rho_{\psi}\left(\Delta,\bar{\Delta}\right)\equiv \frac{d_{\psi}^2}{\mathcal{D}^2} \rho\left(\Delta,\bar{\Delta}\right)
\label{eq:ResultThreeEntropies}
\end{eqnarray}
in the regime of large $\Delta/c$ and $\bar{\Delta}/c$. Taking the logarithm of \eqref{eq:ResultThreeEntropies}, we arrive at 
\begin{equation}
S_a= \left\{ \log \rho\left(\Delta, \bar{\Delta}\right)- \log D^2 \right\}+\log d_a^2,
\qquad {\rm where} \ a=\sigma, \psi.
\label{eq:GeneralizedBekensteinHawking}
\end{equation}
Following the interpretation in \cite{Witten} that non-trivial primaries in the dual CFT correspond to black holes, the expression \eqref{eq:GeneralizedBekensteinHawking} suggests that the different types of black holes labeled by $\sigma$ and $\psi$ can be distinguished by a subleading constant term
in their entropy, apart from the extensive contribution (the term in curly brackets in (\ref{eq:GeneralizedBekensteinHawking}))
arising from boundary gravitons dressing the black hole. Here, a black hole dressed by boundary gravitons corresponds to descendant states in the dual CFT. The term in curly brackets in (\ref{eq:GeneralizedBekensteinHawking}) is
independent of the labels $a$ and, hence, is universal.

We note that, as already mentioned above, these expressions
resemble the form of the universal subleading correction to the entanglement entropy of the ground states of long-range entangled topological phases of matterin $(2+1)$ dimensions \cite{KitaevPreskill,LevinWen}. Earlier studies of connections between ``topological entanglement entropies'' and Bekenstein-Hawking entropy of BTZ black holes, from different perspectives, include \cite{Verlinde,LuoSun}.

\end{document}